\documentclass[11pt, a4paper, hidelinks]{report}
\usepackage{xcolor}
\usepackage[utf8]{inputenc}
\usepackage{pgfplots}
\usepackage{braket}
\usepackage[numbers,sort&compress]{natbib}
\usepackage[font=small]{caption}
\usepackage[labelformat=empty, position=top]{subcaption}
\usepackage[export]{adjustbox}
\usepackage{graphicx}
\usepackage{float}
\bibpunct{[}{]}{,}{n}{}{;}
\usepackage{amsfonts}
\usepackage{amsmath}

\pgfplotsset{compat=1.18}  % Ensure compatibility with recent versions

\usepackage{enumitem, multirow, hyperref, graphicx, siunitx, scalerel, amsmath, array, longtable}
\usepackage{tikz}
\usepackage{amsmath}
\usepackage[font={small}]{caption}
\usepackage[font={small}]{subcaption}
\usepackage{graphicx}
\usepackage{caption}
\usepackage{subcaption}

\graphicspath{ {images/} }

\AtBeginDocument{%
 \abovedisplayskip=15pt plus 3pt minus 4pt
 \abovedisplayshortskip=15pt plus 3pt minus 4pt
 \belowdisplayskip=15pt plus 3pt minus 6pt
 \belowdisplayshortskip=15pt plus 3pt minus 4pt
}

\usepackage[
    a4paper,
    top = 20mm,
    inner = 28mm,   % left margin on an odd page
    outer = 28mm,   % right margin on an odd page
  ]{geometry}

\usepackage{parskip}
\usepackage{titlesec}

\titlespacing\section{0pt}{12pt plus 4pt minus 2pt}{0pt plus 2pt minus 2pt}
\titlespacing\subsection{0pt}{12pt plus 4pt minus 2pt}{0pt plus 2pt minus 2pt}
\titlespacing\subsubsection{0pt}{12pt plus 4pt minus 2pt}{0pt plus 2pt minus 2pt}

\begin{document}

% Frontmatter - - - - - - - - - - - - - - - - - - - - 
% Title Page
\begin{titlepage}
    
    \begin{figure}
        \subfloat{\includegraphics[width=0.35\textwidth]{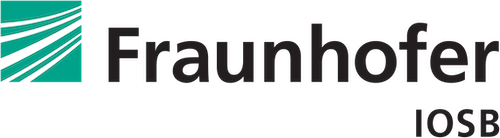}} 
        \hfill
        \subfloat{\includegraphics[width=0.25\textwidth]{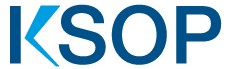}}
        \\[1.5cm]
    \end{figure}

    \begin{center}
        \Huge
        \textbf{Spatial-mode demultiplexing for quantum-inspired superresolution experiments} \\[10ex]
        
        \LARGE
        \textbf{Nickolay Erin Titov} \\[5ex]

        \Large
        Karlsruhe Institute of Technolgy, Karlsruhe, Germany \\
        Fraunhofer IOSB, Ettlingen, Germany \\[1.5ex]
        
        \vfill
        
        \textbf{Supervisors:} Dr. Szymon Gladysz, Prof. Uli Lemmer, Dr. Giacomo Sorelli\\[1.5ex]
        \textbf{Submitted:} 24.07.2025\\[5.0ex]
        
        \Large
        Submitted in partial fulfillment of the\\ requirements for the degree\\of Master of Science \\[3.5ex]
        
    \end{center}
    
\end{titlepage}
\clearpage
\pagenumbering{roman}   % Roman Numerals in Frontmatter

\vspace*{\fill}

Ich versichere wahrheitsgemäß, die Arbeit selbstständig verfasst, alle benutzten Hilfsmittel vollständig und genau angegeben und alles kenntlich gemacht zu haben, was aus Arbeiten anderer unverändert oder mit Abänderungen entnommen wurde sowie die Satzung des KIT zur Sicherung guter wissenschaftlicher Praxis in der jeweils gültigen Fassung beachtet zu haben.

\medskip % small gap between paragraphs

I herewith declare that the present thesis is original work written by me alone, that I have indicated completely and precisely all aids used as well as all citations, whether changed or unchanged, of other theses and publications, and that I have observed the KIT Statutes for Upholding Good Scientific Practice, as amended.

\bigskip

\noindent
Signed:\raisebox{-0.3\height}{\includegraphics[height=2cm]{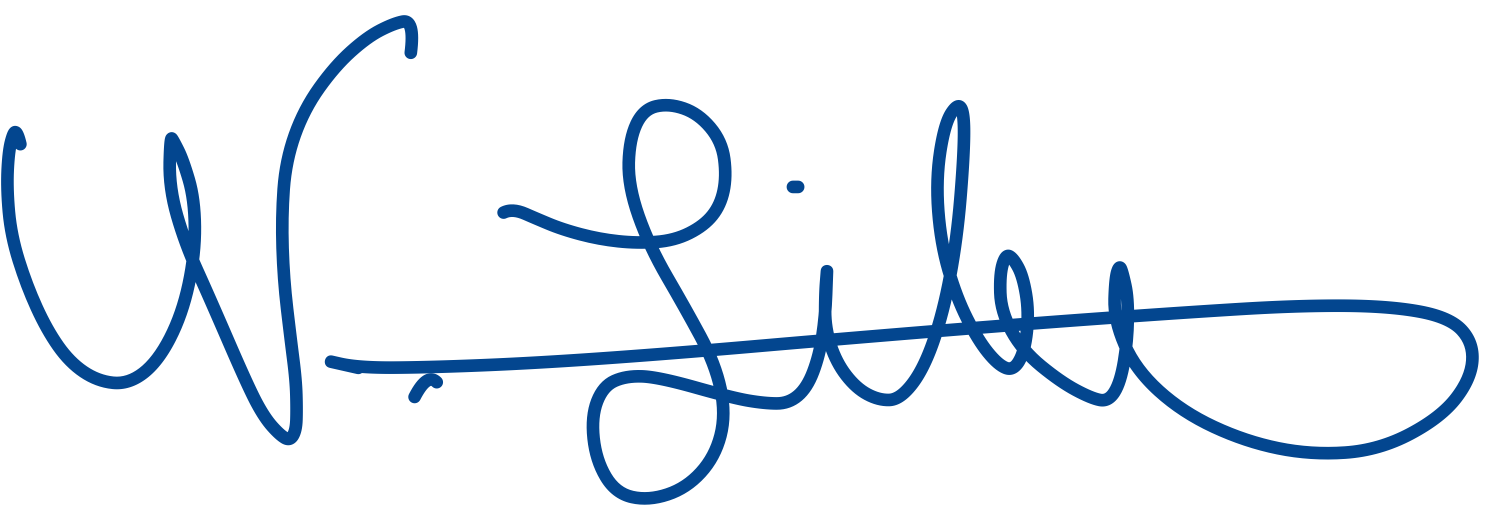}}

\vspace{2pt}

\noindent
Nickolay Erin Titov

\clearpage

\section*{Abstract}
\vspace{0.75cm}
Conventional optical imaging is limited by diffraction, preventing discrimination of closely spaced incoherent sources. Inspired by quantum parameter estimation, this thesis explores spatial-mode demultiplexing (SPADE) as a method to overcome this limit by projecting light onto an orthonormal mode basis. In this work, we design and experimentally validate a practical implementation of spatial‐mode demultiplexing via Multi‐Plane Light Conversion (MPLC).

\clearpage

\listoffigures

\clearpage

\tableofcontents
\clearpage

\chapter*{List of Abbreviations}
{\renewcommand{\arraystretch}{1.2}
\begin{longtable}{cl}
    SPADE     &   Spatial-mode Demultiplexing \\
    LG      &   Laguerre-Gaussian \\
    HG     &   Hermite-Gaussian \\
    BS     &   Beam Splitter \\
    MPLC      &   Multi-plane Light Conversion \\
    SLM     &  Spatial Light Modulator \\
    WFM     & Wavefront Matching Method \\
    ASM & Angular Spectrum Method  \\
    CGH & Computer Generated Holograms\\
    GUI & Graphical User Interface\\
\end{longtable}}
\clearpage

% - - - - - - - - - - - - - - - - - - - - - - - - - - - - -

% Chapters - - - - - - - - - - - - - - - - - - - - - - - -
\pagenumbering{arabic}  % Begin numbering after title page

\chapter{Introduction}
Optical imaging plays a central role across diverse fields such as astronomy, semiconductor inspection, and biology. Despite decades of development, classical imaging systems remain fundamentally constrained by diffraction, which limits their ability to resolve closely spaced incoherent sources. This resolution barrier appears in various criteria, such as that proposed by Rayleigh, and it is not just a theoretical concern; it directly impacts the achievable information gain, biological detail, and sensing precision in practical applications \cite{SalehTeich2007, Pawley2006}.

In response, a wide range of super-resolution techniques have been developed, particularly in microscopy and astronomy. In fluorescence microscopy, methods such as STED (Stimulated Emission Depletion) microscopy \cite{Hell1994}, PALM (Photo-Activated Localization Microscopy), STORM (Stochastic Optical Reconstruction Microscopy), and SIM (Structured Illumination Microscopy) \cite{Betzig2006, Dickson1997} have substantially improved resolution by manipulating emission dynamics or exploiting interference and computational reconstruction. In astronomy, the technique of aperture synthesis measures the spatial coherence between signals from multiple telescopes, combining them to synthesize an image with a resolution far exceeding that of any single instrument \cite{thompson2008interferometry}.

While various superresolution strategies have made significant progress, they are often constrained by specific experimental or physical conditions. A broadly applicable approach should satisfy several key criteria. First, far-field operation is essential for non-invasive imaging in fields such as astronomy, remote sensing, and biomedical imaging, where placing detectors or probes near the object is infeasible. Second, linear-optical systems are preferable because they avoid the complexity and power requirements of nonlinear techniques like STED, making them compatible with weak signals and non-interacting systems. Third, passivity ensures that the method does not actively perturb the source, which is critical for fragile biological samples or remote thermal sources. Meeting all these conditions defines a desirable superresolution strategy that is simple, robust, and widely applicable. However, no existing technique fully satisfies all these criteria in a single framework. This thesis investigates a new, quantum-inspired superresolution approach based on spatial-mode demultiplexing (SPADE). Recent developments in quantum parameter-estimation theory have revealed that the classical resolution limit is not fundamental, but rather a consequence of sub-optimal measurements \cite{Tsang2016}. By projecting the optical field onto an orthonormal spatial-mode basis instead of recording its intensity distribution, it is possible to saturate the fundamental limit and achieve resolution beyond classical limits.

Translating SPADE from theory to practice presents an optical engineering challenge: implementing a high-fidelity, low-loss, and compact mode demultiplexer. Multi-Plane Light Conversion (MPLC), introduced by Morizur et al. in 2010, offers a scalable solution for implementation of SPADE using cascaded phase masks and free-space propagation \cite{Morizur2010}. Early MPLC systems handled 3–15 modes with 7–20 planes \cite{Labroille2014,Zhang2023}, and Boucher et al. later clarified their fundamental limits via filtered-random-matrix theory \cite{Boucher2021}. A major advance came in 2017, when Fontaine et al. proposed a separable "magic mapping" that drastically reduced the number of planes \cite{Fontaine2017}, enabling sorters for 325 modes with only 7 phase masks and, By 2021, a 1035-mode HG multiplexer using just 14 planes \cite{Fontaine2021}. We adopt and adapt this technology to a quantum-inspired metrological context.

The main goal of this thesis is to design, simulate, and experimentally validate a MPLC setup that performs the SPADE measurement to potentially overcome diffraction limit for resolving incoherent point sources. This thesis covers topics from the theoretical foundation of quantum-inspired super-resolution to its practical implementation. Chapter 2 establishes the necessary language of quantum optics to describe light in terms of spatial modes. Chapter 3 then frames the two-source resolution problem within quantum metrology, formally demonstrating why SPADE is the optimal measurement strategy to overcome the diffraction limit. Chapter 4 translates this theory into a practical engineering design, presenting the simulation and optimization of a MPLC system for mode sorting. Finally, Chapter 5 details the experimental realization and performance characterization of this MPLC setup, providing a proof-of-concept for high-fidelity spatial mode sorting.

\chapter{Quantum Optics}
This chapter provides a brief introduction to the classical and quantum descriptions of light, establishing notation and concepts used throughout this thesis. We first review Maxwell’s equations and modes of optical beams, then develop the quantization procedure that promotes these modes to quantum operators. Finally, we discuss photon statistics to prepare for our focus on SPADE in later chapters.

\section{Classical Electromagnetic Waves}
As a foundation for quantum optics, we first revisit the classical description of electromagnetic waves provided by Maxwell’s equations, which determine how the electric field $\vec{E}(\vec{r},t)$ and magnetic field $\vec{B}(\vec{r},t)$ evolve in space and time. When free charges and currents are absent, Maxwell’s equations reduce to the following differential form \cite{Griffiths2017}:
\begin{align}
    &\nabla \cdot \vec{E}(\vec{r},t) = 0, \quad \text{(Gauss's law for electricity)} \\
    &\nabla \cdot \vec{B}(\vec{r},t) = 0, \quad \text{(Gauss's law for magnetism)} \\
    &\nabla \times \vec{E}(\vec{r},t) = -\frac{\partial \vec{B}(\vec{r},t)}{\partial t}, \quad \text{(Faraday's law)} \\
    &\nabla \times \vec{B}(\vec{r},t) = \mu_0 \epsilon_0 \frac{\partial \vec{E}(\vec{r},t)}{\partial t}, \quad \text{(Maxwell--Ampère law)}
\end{align}
where \(\epsilon_0\) is the permittivity and \(\mu_0\) is the permeability of free space. In free space, the wave equation for the electric field is:
\begin{align}
\nabla^2\vec{E}(\vec{r},t)
-\frac{1}{c^2}\frac{\partial^2}{\partial t^2}\vec{E}(\vec{r},t)=0.
\end{align}  
where c is speed of light in vacuum.
Any electric field \( \vec{E}(\vec{r}, t) \) is a function of space and time and it can be expressed using a Fourier integral:
\begin{align}
\vec{E}(\vec{r}, t) = \int_{-\infty}^{\infty} \vec{E}_\omega(\vec{r}, \omega) e^{-i \omega t} \, d\omega.
\end{align}
The positive frequency component of this field is defined as:
\begin{align}
\vec{E}^{(+)}(\vec{r}, t) = \int_{0}^{\infty} \vec{E}_\omega(\vec{r}, \omega) e^{-i \omega t} \, d\omega.
\end{align}
Similarly, the negative frequency component is given by:
\begin{align}
\vec{E}^{(-)}(\vec{r}, t) = \int_{-\infty}^{0} \vec{E}_\omega(\vec{r}, \omega) e^{-i \omega t} \, d\omega = \int_{0}^{\infty} \vec{E}_\omega^*(\vec{r}, \omega) e^{i \omega t} \, d\omega.
\end{align}

Having reviewed the classical field equations, we now introduce the concept of optical modes, which serve as the basis for both classical and quantum descriptions of light.

\section{Modes of the Electromagnetic Field}
We limit our study to a finite volume \( V \) which is much larger than the physical system. A mode of the electromagnetic field is a vector field \( \vec{u}(\vec{r}, t) \) that satisfies Maxwell’s equations along with the orthonormality condition \cite{Fabre2020ModesStates}:
\begin{align}
\frac{1}{V} \int_V d\vec{r} \, \vec{u_m}(\vec{r}, t)\vec{u_n}^*(\vec{r}, t) = 1.
\end{align}
Since Maxwell’s equations are linear, any linear combination of their solutions is also a solution. The vector field is confined in a finite volume V, hence we can find a discrete set of orthonormal solutions, denoted by \(\vec{u}_m(\vec{r}, t)\), that can be used as a basis to express any other solution that satisfies the same boundary conditions.

If the set  of \(  \vec{u}_m(\vec{r}, t) \) is complete, then any electromagnetic field can be written as a sum of these modes:
\begin{align}
\vec{E}(\vec{r}, t) = \sum_m \varepsilon_m \vec{u}_m(\vec{r}, t),
\end{align}
where the complex coefficients \( \varepsilon_m \) represent the amplitude of each mode and completely define the state of the classical electromagnetic field.

\subsection{Spatial Modes}
We confine our attention to electromagnetic fields in the form of laser beams. We consider optical fields formed by the superposition of plane waves with wave vectors concentrated around a central wave vector \( \vec{k}_0 \), and with frequencies narrowly distributed around a central frequency \( \omega_0 = c|\vec{k}_0| \).

To simplify notation, we chose the z-axis as the propagation direction and limit our discussion to linearly polarized electromagnetic field. In free space, the wave equation for the electric field yields solutions of the form:
\begin{align}
\vec{E}(\vec{r},t)
=\,A(x,y,z,t)\,e^{i(k_0 z - \omega_0 t)}\vec{\epsilon}
\end{align}
where $\vec{\epsilon}$ is the polarization in xy-plane and $e^{i(k_0 z - \omega_0 t)}$ represents the carrier plane wave. We assume that a mode of the field $A(x,y,z,t)$ is separable in space and time \cite{phd_clemen}:
\begin{align}
A(x,y,z,t)
= u(x,y,z)\;g(t - z/c).
\end{align}
Since our focus is on the spatial mode structure of paraxial beams, we will omit the temporal envelope $g(t - z/c)$ and analyze only the spatial part $u(x, y, z)$. This simplification is valid for quasi-monochromatic beams, where temporal effects are decoupled from the spatial propagation.

Then substituting into the wave equation gives the intermediate form:
\begin{align}
\frac{\partial^2u}{\partial x^2}
+\frac{\partial^2u}{\partial y^2}
+\frac{\partial^2u}{\partial z^2}
-2ik_0\frac{\partial u}{\partial z}
=0.
\end{align}
Under the paraxial conditions, where the beam’s transverse spatial profile changes gradually along the propagation direction compared to the wavelength and the variations arising from the transverse width of beam:
\begin{align}
\bigl|\frac{\partial^2u}{\partial z^2}\bigr|\ll\bigl|2k_0\frac{\partial^2u}{\partial z^2}\bigr|,\quad
\bigl|\frac{\partial^2u}{\partial z^2}\bigr|\ll\bigl|\frac{\partial^2u}{\partial x^2}\bigr|,\quad
\bigl|\frac{\partial^2u}{\partial z^2}\bigr|\ll\bigl|\frac{\partial^2u}{\partial y^2}\bigr|.
\end{align}
the \(\frac{\partial^2u}{\partial z^2}\) term may be neglected.  Altogether, we get the paraxial wave equation:
\begin{align}
\frac{\partial^2u}{\partial x^2} + \frac{\partial^2u}{\partial y^2}
- 2ik_0\frac{\partial u}{\partial z}= 0.
\label{paraxial}
\end{align}
Accordingly, the solutions to Eq. \ref{paraxial} describe the transverse spatial structure of the beam, known as spatial modes. Two important families of these solutions are the Hermite–Gaussian (HG) and the Laguerre Gaussian (LG) modes.

\subsubsection{Hermite--Gaussian Modes}

Hermite-Gaussian (HG) modes, are solutions to the paraxial wave equation in Cartesian coordinates \((x,y,z)\). They are characterized by two indices, $m$ and $n$, which correspond to the mode orders in the horizontal ($x$) and vertical ($y$) directions, respectively. At the beam waist where the beam diameter reaches its minimum value, the transverse field amplitude is given by the product of Hermite polynomials and a Gaussian envelope:
\begin{equation}
    u_{m,n}(x, y, 0) \propto H_m\left(\frac{\sqrt{2}x}{w_0}\right) H_n\left(\frac{\sqrt{2}y}{w_0}\right) \exp\left(-\frac{x^2 + y^2}{w_0^2}\right),
\end{equation}
where $H_m$ and $H_n$ are Hermite polynomials of order $m$ and $n$, and $w_0$ is the beam waist radius.

\begin{figure}[h]
    \centering
    \begin{subfigure}[h]{0.31\textwidth}
        \centering
        \includegraphics[width=\textwidth]{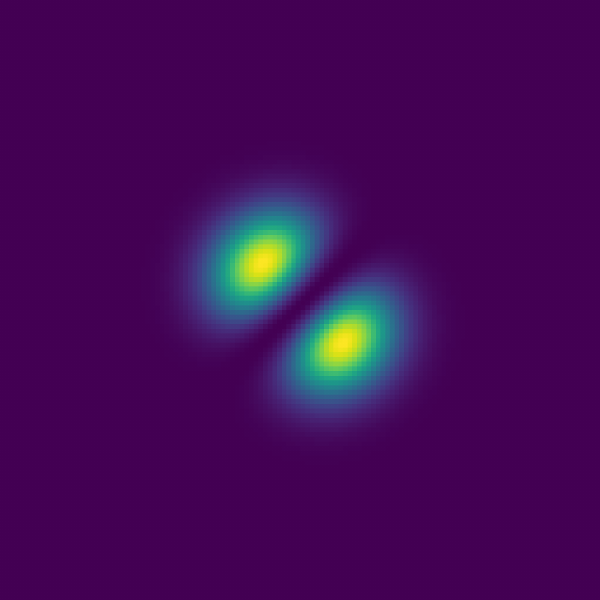}
        \caption{HG$_{10}$}
        \label{HG10}
    \end{subfigure}
    \hfill
    \begin{subfigure}[h]{0.31\textwidth}
        \centering
        \includegraphics[width=\textwidth]{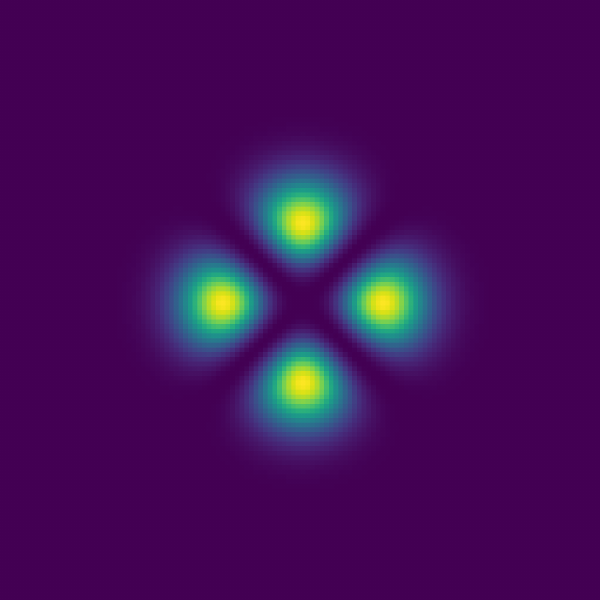}
        \caption{HG$_{11}$}
        \label{HG11}
    \end{subfigure}
    \hfill
     \begin{subfigure}[h]{0.31\textwidth}
        \centering
        \includegraphics[width=\textwidth]{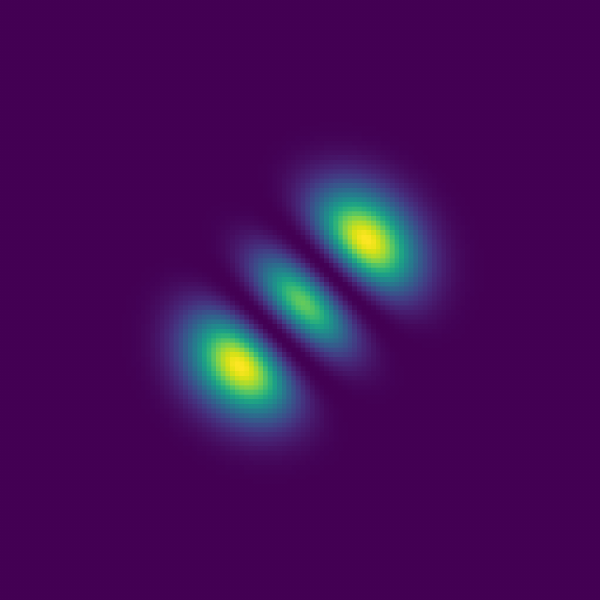}
        \caption{HG$_{20}$}
        \label{HG02}
    
    \end{subfigure}
    
    \caption[HG mode examples]{Examples of HG modes intensity profiles}
    \label{hg_modes_example}
\end{figure}
The intensity distribution of Hermite-Gaussian (HG) modes is characterized by having \( m \) nodes horizontally and \( n \) nodes vertically. For indices  (\( m = n = 0 \)), the mode simplifies to a Gaussian beam, known as the fundamental mode~\cite{HG_mode}. Fig. \ref{hg_modes_example} displays the intensity profiles of several low-order HG modes with for example, the two-lobed pattern of HG$_{10}$ and the four-lobed HG$_{11}$ illustrating their increasing nodal structure. Any paraxial field can be decomposed into a linear combination of HG modes, making them an essential basis in laser cavity theory and beam propagation analysis~\cite{Siegman,KogelnikLi}.

Due to their orthogonality and completeness, HG modes form the basis for SPADE-based spatial mode demultiplexing techniques. These modes are used throughout this work to represent and sort spatial modes of light in both simulation and experiment.

\subsubsection{Laguerre--Gaussian Modes}

\begin{figure}[h]
    \centering
    \begin{subfigure}[h]{0.31\textwidth}
        \centering
        \includegraphics[width=\textwidth]{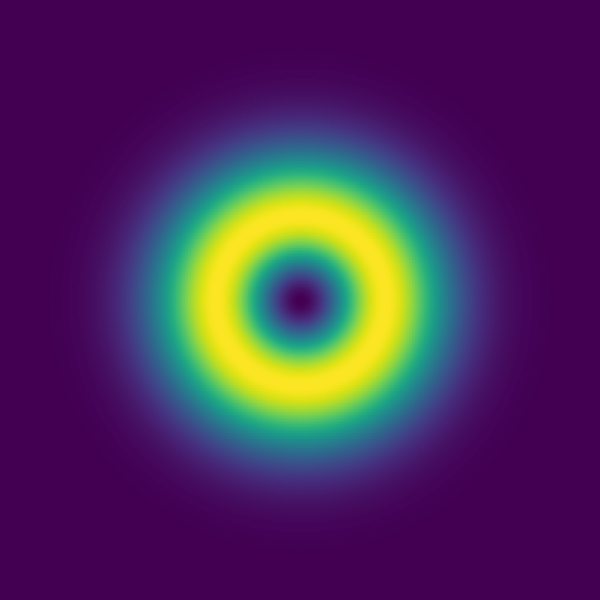}
        \caption{LG$_{01}$}
        \label{LG01}
    \end{subfigure}
    \hfill
    \begin{subfigure}[h]{0.31\textwidth}
        \centering
        \includegraphics[width=\textwidth]{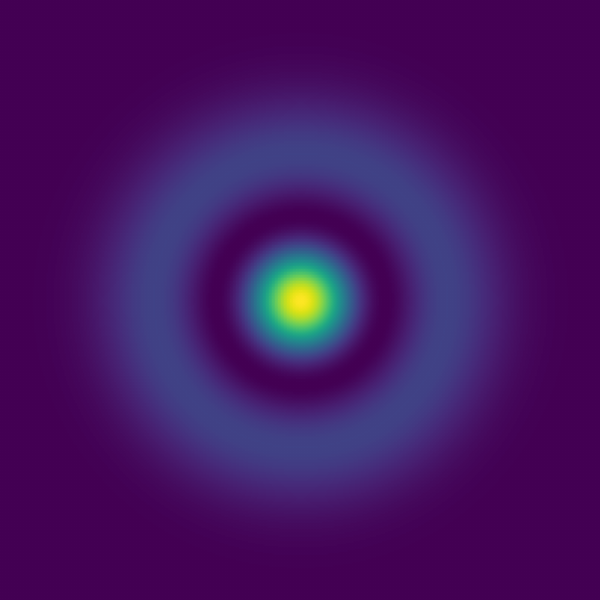}
        \caption{LG$_{10}$}
        \label{LG10}
    \end{subfigure}
    \hfill
     \begin{subfigure}[h]{0.31\textwidth}
        \centering
        \includegraphics[width=\textwidth]{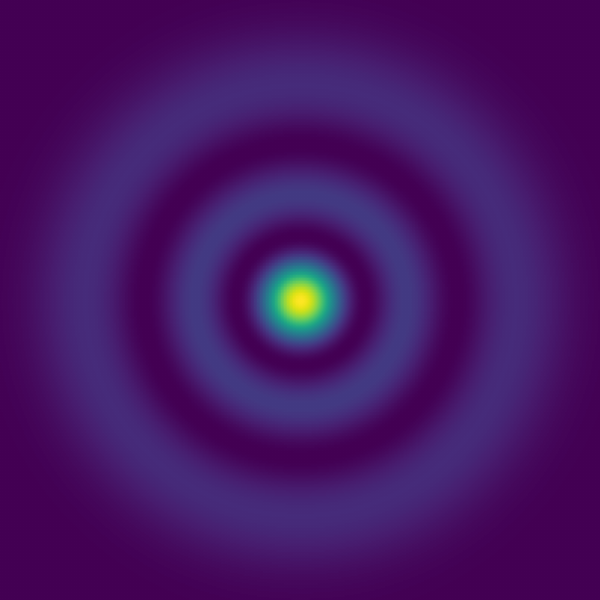}
        \caption{LG$_{20}$}
        \label{HG20}
    \end{subfigure}
    
    \caption[LG mode examples]{Examples of LG modes intensity profiles}
    \label{lg_modes_ex}
\end{figure}

In cylindrical coordinates \((r,\phi,z)\), the paraxial wave equation admits the Laguerre–Gaussian (LG) family of modes \(u_{p,l}(\vec r,z)\), labeled by radial index \(p\ge0\) and azimuthal index $l\in \mathbb{Z}$.  Their normalized transverse profile is:  
\begin{equation}
    u_{p,l}(r, \phi, 0) \propto \left( \frac{\sqrt{2}r}{w_0} \right)^{|l|} L_p^{|l|}\left( \frac{2r^2}{w_0^2} \right) \exp\left( -\frac{r^2}{w_0^2} \right) \exp(il\phi),
\end{equation}
where $L_p^{|l|}$ are the generalized Laguerre polynomials. The phase factor $\exp(il\phi)$ imparts an azimuthal dependence that creates a helical phase front around the beam axis. Some representative intensity patterns for \(u_{p,l}\) are plotted in Fig. \ref{lg_modes_ex} , illustrating the doughnut-shaped profile for \(l \neq 0\) and the central peak for \(l = 0\). These LG modes carry orbital angular momentum \(l\hbar\) per photon and exhibit a doughnut‐shaped intensity for \(l\neq0\) \cite{Yang2022}. 

Once the mode structure is defined, we can proceed to quantize the electromagnetic field and understand how individual photons emerge from this formalism.

\section{Quantization of the Electromagnetic Field}
The quantization of the electromagnetic field allows us to describe light as a collection of photons distributed across spatial modes. This formalism is crucial for our analysis. It provides the foundation for understanding the spatial mode decomposition and transformations that are central to both SPADE and its practical implementation with MPLC.

We start the quantization by expressing the classical Hamiltonian of the electromagnetic field in terms of vector potential $\vec{A}(\vec{r},t)$, a field whose time derivative and curl give the physical fields via: 
\begin{align}
\vec{E}(\vec{r},t) = -\frac{\partial \vec{A}(\vec{r},t)}{\partial t}, 
\qquad
\vec{B}(\vec{r},t) = \nabla \times \vec{A}(\vec{r},t).
\label{eq:EB_from_A}
\end{align}
In Eq. \ref{eq:EB_from_A}, we impose  Coulomb gauge condition $\nabla \cdot \vec{A} = 0$ which, in a region free of charges and currents,  eliminates any longitudinal component of the vector potential. Physically, this restriction leaves only the two transverse polarization modes as the true degrees of freedom of the free electromagnetic field,  simplifying both the mode expansion and the subsequent quantization procedure \cite{Griffiths2017}.

We follow the quantization procedure of the electromagnetic field primarily as presented in \cite{Grynberg2010}, with modifications to the notation to suit our formalism. By confining $\vec A(\vec{r},t)$ to a finite volume $V$ with periodic boundary conditions, the solutions can be decomposed into a discrete set of modes:
\begin{align}
\vec A(\vec r,t) = \sum_{\vec k}\Bigl(\vec A_{-\vec k}\,e^{-i(\vec k\cdot\vec r - \omega_{\vec{k}}t)} + \vec A_{\vec k}\,e^{i(\vec k\cdot\vec r - \omega_{\vec{k}}t)}\Bigr).
\end{align}

Since $\vec A(\vec r,t)$ is real, it implies $\vec A_{-\vec k} = (\vec A_{\vec k})^*$.

As we did in spatial modes we confine our attention to linearly polarized electromagnetic field. It follows that we may decompose the vector potential into modes of wavevector $\vec k$ and polarization $\vec\epsilon$ by writing $\vec A_{-\vec k} = A_{\vec{k}}\vec{\epsilon}$:
\begin{align}
\vec A(\vec r,t)
&= \frac{1}{\sqrt{V}}
  \Bigl(A_{\vec k}\,e^{\,i(\vec k\cdot\vec r - \omega_k t)} + A_{\vec k}^*\,e^{-i(\vec k\cdot\vec r - \omega_k t)}\Bigr)\,
  \vec{\epsilon}.
\end{align}
In the absence of charges, the classical Hamiltonian for the free EM field is given by the energy stored in the fields:
\begin{equation}
H_{\text{class}} 
= \frac{1}{2} \int_V \left( \epsilon_0 |\vec{E}(\vec{r},t)|^2 + \frac{1}{\mu_0} |\vec{B}(\vec{r},t)|^2 \right) \, d^3r,
\label{eq:EM_classical_H}
\end{equation}
Substituting Eq.~\eqref{eq:EB_from_A} into Eq.~\eqref{eq:EM_classical_H}, we obtain
\begin{equation}
H_{\text{class}} 
= \frac{1}{2} \int_V \left( \epsilon_0 \left| -\frac{\partial \vec{A}(\vec{r},t)}{\partial t} \right|^2 
  + \frac{1}{\mu_0} \left| \nabla \times \vec{A}(\vec{r},t) \right|^2 \right) d^3r.
\label{eq:EM_classical_H2}
\end{equation}

By adopting the vector-potential formulation, the classical Hamiltonian can be written as follows (for a thorough derivation, see Appendix \ref{appendix_qem}):
\begin{align}
H_{\text{class}} = 2 \varepsilon_0 \omega_{\vec{k}^2} |A_{\vec{k}}|^2 
= 2 \varepsilon_0 \omega_{\vec{k}^2} \left( |A_{\vec{k}}^R|^2 + |A_{\vec{k}}^I|^2 \right).
\end{align}
where \( A_{\vec{k}} = A^R_{\vec{k}} + i A^I_{\vec{k}} \). Note that if we define \( A_{\vec{k}}(t) = A_{\vec{k}} e^{-i\omega_{\vec{k}} t} \), then:
\begin{align}
\dot{A}^R_{\vec{k}} &= \omega_{\vec{k}} A^I_{\vec{k}}, \quad 
\dot{A}^I_{\vec{k}} = -\omega_{\vec{k}} A^R_{\vec{k}}.
\end{align}
and thus:
\begin{align}
\frac{\partial H}{\partial A^R_{\vec{k}}} &= 4\varepsilon_0 \omega_{\vec{k}^2} A^R_{\vec{k}} 
= -4\varepsilon_0 \omega_{\vec{k}} \dot{A}^I_{\vec{k}}. \\
\frac{\partial H}{\partial A^I_{\vec{k}}} &= 4\varepsilon_0 \omega_{\vec{k}^2} A^I_{\vec{k},} 
= 4\varepsilon_0 \omega_{\vec{k}} \dot{A}^R_{\vec{k},}.
\end{align}
This implies that \( A^R_{\vec{k}} \) and \( A^I_{\vec{k}} \) form a pair of canonically conjugate variables, up to a proportionality factor. Accordingly, we can define the corresponding conjugate position and momentum variables as follows:
\begin{align}
Q_{\vec{k}} &= 2\sqrt{\varepsilon_0} A^R_{\vec{k}}. \\
P_{\vec{k}} &= 2 \omega_{\vec{k}} \sqrt{\varepsilon_0} A^I_{\vec{k}}.
\end{align}
respectively. Clearly, these satisfy:
\begin{align}    
\left\{
\begin{aligned}
\dot{Q}_{\vec{k}} &= P_{\vec{k}}. \\
\dot{P}_{\vec{k}} &= -\omega_{\vec{k}}^2 Q_{\vec{k}}.
\end{aligned}
\right.
\qquad
\left\{
\begin{aligned}
\frac{\partial H}{\partial Q_{\vec{k}}} &= -\dot{P}_{\vec{k}}. \\
\frac{\partial H}{\partial P_{\vec{k}}} &= \dot{Q}_{\vec{k}}.
\end{aligned}
\right.
\end{align}

as would be the case for a harmonic oscillator. Consequently, also the Hamiltonian will be identical to that of a harmonic oscillator:
\begin{align}
H = \frac{1}{2} \left( P_{\vec{k}}^2 + \omega_{\vec{k}}^2 Q_{\vec{k}}^2 \right).
\end{align}
We can now quantize the electromagnetic field just as one would quantize the harmonic oscillator. We promote $P_{\vec{k}}$ and $Q_{\vec{k}}$ to quantum operators $\hat{P}_{\vec{k}}$ and $\hat{Q}_{\vec{k}}$ which satisfy the canonical commutation relations:
\begin{align}
[\hat{Q}_{\vec{k}}, \hat{P}_{\vec{k}'}] &= i\hbar \delta_{\vec{k},\vec{k}'}  \\
[\hat{Q}_{\vec{k}}, \hat{Q}_{\vec{k}'}] &= [\hat{P}_{\vec{k}}, \hat{P}_{\vec{k}'}] = 0
\end{align}
so that:
\begin{align}
\hat{H} = \frac{1}{2} \left( \hat{P}_{\vec{k}}^2 + \omega_{\vec{k}}^2 \hat{Q}_{\vec{k}}^2 \right).
\end{align}
We introduce the ladder operators:
\begin{align}
\left\{
\begin{aligned}
a_{\vec{k}}^\dagger &= \sqrt{\frac{\hbar}{2\omega_{\vec{k}}}} \left( \omega_{\vec{k}} \hat{Q}_{\vec{k}} - i \hat{P}_{\vec{k}} \right). \\
a_{\vec{k}} &= \sqrt{\frac{\hbar}{2\omega_{\vec{k}}}} \left( \omega_{\vec{k}} \hat{Q}_{\vec{k}} + i \hat{P}_{\vec{k}} \right).
\end{aligned}
\right.
\quad \Longrightarrow \quad
\left\{
\begin{aligned}
\hat{Q}_{\vec{k}} &= \sqrt{\frac{\hbar}{2\omega_{\vec{k}}}} \left( a_{\vec{k}}^\dagger + a_{\vec{k}} \right). \\
\hat{P}_{\vec{k}} &= i \sqrt{\frac{\hbar \omega_{\vec{k}}}{2}} \left( a_{\vec{k}}^\dagger - a_{\vec{k}} \right).
\end{aligned}
\right.
\end{align}
With these new operators, the Hamiltonian turns into the familiar quantum harmonic oscillator:
\begin{align}
\hat{H} = \hbar \omega_{\vec{k}} \sum_{\vec{k}} \left( a_{\vec{k}}^\dagger a_{\vec{k}} + \frac{1}{2} \right).
\label{quantum_H}
\end{align}
With this picture, each mode enters the Hamiltonian \eqref{quantum_H} as a separate harmonic oscillator and the quantized mode excitations are interpreted as photons, each carrying energy \(\hbar\omega_{\vec{k}}\). 

After quantization, the operators for the electric and magnetic fields are obtained from annihilation and creation operators as follows:
\begin{align}
\hat{E}(\vec{r}, t) &= i \sqrt{\frac{\hbar \omega_{\vec{k}}}{2 \varepsilon_0 \mathcal{V}}} 
\sum_{\vec{k}} 
\left( e^{i(\vec{k} \cdot \vec{r} - \omega_{\vec{k}} t)} a_{\vec{k} } - 
e^{-i(\vec{k} \cdot \vec{r} - \omega_{\vec{k}} t)} a^\dagger_{\vec{k}} \right) \vec{\epsilon}. \\
\hat{B}(\vec{r}, t) &= i \sqrt{\frac{\hbar}{2 \varepsilon_0 \omega_{\vec{k}} \mathcal{V}}} 
\sum_{\vec{k}}  
\left( e^{i(\vec{k} \cdot \vec{r} - \omega_{\vec{k}} t)} a_{\vec{k}} - 
e^{-i(\vec{k} \cdot \vec{r} - \omega_{\vec{k}} t)} a^\dagger_{\vec{k}} \right) 
\left( \vec{k} \times \vec{\epsilon} \right).
\end{align}
In this formalism, the electric field operator \(\hat{E}^{(+)}(\vec{r}, t)\) can be simplified as: 
\begin{align}
    \hat{E}^{(+)}(\vec{r}, t) 
    &= i \sum_{m,\lambda} 
    \sqrt{\frac{\hbar \omega_m}{2 \varepsilon_0 V}} \,
    \hat{a}_{m,\lambda} \, 
    \vec{\epsilon}_{m,\lambda} \,
    e^{i(\vec{k}_m \cdot \vec{r} - \omega_m t)} 
    \label{quantum_mode} \\
    \vec{u}_{m,\lambda}(\vec{r}, t) 
    &= e^{i(\vec{k}_m \cdot \vec{r} - \omega_m t)} \, 
    \vec{\epsilon}_{m,\lambda}, 
    \qquad 
    \mathcal{E}_m^{(1)} = \sqrt{\frac{\hbar \omega_m}{2 \varepsilon_0 V}}.
\end{align}

As shown in Eq. \ref{quantum_mode} the annihilation operator is associated to mode $\vec{u_m}(\vec{r},t)$. We now examine how a unitary transformation \( U \), which maps the mode basis \( \vec{u}_m(\vec{r}, t) \) onto the basis \(\vec{v}_l(\vec{r}, t) \), influences the corresponding annihilation operators.

A unitary transformation \( U \) with complex coefficients can relate the two bases and it is defined by \cite{Fabre2020ModesStates}:
\begin{align} 
U_{m\ell} = \frac{1}{V} \int_V d^3\vec{r} \, \vec{u}^*_m(\vec{r}, t) \cdot \vec{v}_l(\vec{r}, t).
\end{align}
Now the mode functions can be expressed in terms of each other using the unitary matrix:
\begin{align} 
    \vec{u}_m(\vec{r}, t) = \sum_m U_{ml} \vec{v}_l(\vec{r}, t),
\end{align}
which leads to the corresponding transformation for the photon creation operators:
\begin{align} 
    \hat{b}^\dagger_l = \sum_m U_{lm} \hat{a}^\dagger_m.
\end{align}
Here, the operator \(\hat{b}^\dagger_l\) corresponds to the new mode basis \(\vec{v}_l(\vec{r}, t)\). The photon annihilation operator \(\hat{b}_l\) is related by Hermitian conjugation:
\begin{align} 
    \hat{b}_l = (\hat{b}^\dagger_l)^\dagger = \sum_m U^*_{lm} \hat{a}_m.
\end{align}
We can define a given mode either via the annihilation operator \(\hat{b}_l\) or through its spatial-temporal structure \(\vec{u}_l(\vec{r}, t)\). The unitarity of \(U_{lm}\) ensures that the canonical commutation relations are preserved:
\begin{align} 
    [\hat{b}_l, \hat{b}^\dagger_{l'}] = \delta_{ll'}, \qquad [\hat{b}_l, \hat{b}_{l'}] = 0.
\end{align}
The total energy of the field,
\begin{align}  
    \hat{H} = \sum_m \hbar \omega_m \left( \hat{a}^\dagger_m \hat{a}_m + \frac{1}{2} \right) 
    = \sum_l \hbar \omega_l \left( \hat{b}^\dagger_l \hat{b}_l + \frac{1}{2} \right).
\end{align}
is also invariant under this mode transformation.

\section{ Quantum States}

Before introducing quantum states, we recall that a quantum state is represented by a vector, ket, in a Hilbert space, which is a complete complex vector space equipped with an inner product $\langle\phi\mid\psi\rangle$ that is linear, conjugate‐symmetric ($\langle\phi\mid\psi\rangle=\langle\psi\mid\phi\rangle^*$), positive‐definite ($\langle\psi\mid\psi\rangle\ge0$) \cite{Loudon2000}. Physically, the ket $\lvert\psi\rangle$ encodes the states of the field, while the inner product determines transition amplitudes and probabilities.

In our modal decomposition each spatial mode $u_m(\vec r)$ becomes an independent quantum harmonic oscillator with its own Hilbert space $\mathcal{H}_m$, and the full field state lives in the tensor‐product space
\begin{align}
\mathcal{H} &= \bigotimes_{m}\mathcal{H}_m.
\end{align}

In the following subsections we will introduce the notions of pure and mixed states and discuss their relevance for quantum‐inspired superresolution.

\subsection{Pure States}
A pure state in quantum mechanics represents a situation in which we have maximal information about the system. Such a state is completely described by a state vector \( |\psi\rangle \), which encodes all that can be known about the system. For example, knowing \( |\psi\rangle \) allows one to predict the time evolution of the system by solving the Schrödinger equation. However, even though the state \( |\psi\rangle \) embodies full information about the system, this complete description still yields inherently probabilistic predictions when physical properties are measured. Now, we introduce two important pure states, namely Fock and coherent states. Their importance stems from the fact that Fock states $\lvert n_m\rangle$ have a definite photon number in each spatial mode enabling precise photon counting statistics while coherent states $\lvert\alpha_m\rangle$ accurately model laser beams.

\subsubsection{Fock States}
Fock states are eigenstates of the photon number operator $\hat n_m=\hat a_m^\dagger\hat a_m$ which satisfies $\hat n_m\lvert n_m\rangle = n_m\lvert n_m\rangle$ and are a natural basis for $\mathcal{H}$:
\begin{align}
\lvert n_1, n_2, \dots\rangle &= \bigotimes_{m}\lvert n_m\rangle,
\end{align}
 For example, the single‐photon excitation in mode $m$ is $\lvert1_m\rangle=\hat a_m^\dagger\lvert0\rangle$, and a two‐mode, one‐photon state $\lvert1_{m_1},1_{m_2}\rangle$ carries exactly one photon in each of the modes $m_1$ and $m_2$.

The Fock states $\lvert n_m\rangle$ form a basis of the single‐mode Hilbert space, so that any pure state in mode $m$ can be written as \cite{MandelWolf1995}:
\begin{align}
\lvert\psi\rangle_m = \sum_{n_m} c_{n_m}\,\lvert n_m\rangle,\quad \sum_{n_m}\lvert c_{n_m}\rvert^2 = 1.
\end{align}
The Fock states satisfies:
\begin{align}
\hat a_m\,\lvert n_1,\dots,n_m,\dots\rangle 
= \sqrt{n_m}\,\lvert n_1,\dots,n_m-1,\dots\rangle,\\
\hat a_m^\dagger\,\lvert n_1,\dots,n_m,\dots\rangle 
= \sqrt{n_m+1}\,\lvert n_1,\dots,n_m+1,\dots\rangle.\\
\langle n_1,n_2,\dots\lvert \hat H \rvert n_1,n_2,\dots\rangle 
= \sum_m \hbar\omega_m\Bigl(n_m+\tfrac12\Bigr).
\end{align}

\subsubsection{Vacuum State}

The vacuum state $\lvert0\rangle = \lvert n_1=0,\,n_2=0,\,\dots\rangle$ is the state in which every mode has zero photons.  It has the minimum energy, as seen by
\begin{align}
\langle0|\hat H|0\rangle &= \sum_m \tfrac12\,\hbar\omega_m.
\end{align}
It has a zero mean field:
\begin{align}
\langle0|\hat E^{(+)}(\vec r,t)|0\rangle
&= \sum_m \mathcal{E}_m\,u_m(\vec r,t)\,\langle0|\hat a_m|0\rangle
= 0.
\end{align}
However, it has non-zero fluctuations:
\begin{align}
\langle0|\hat E^2(\vec r,t)|0\rangle
&= \langle0|\hat E^{(+)}(\vec r,t)\,\hat E^{(-)}(\vec r,t)|0\rangle \\
&= \sum_{m,n}\mathcal{E}_m\,\mathcal{E}_n^*\,u_m(\vec r)\,u_n^*(\vec r)
\,e^{-i\omega_m t}e^{i\omega_n t}\,\langle0|\hat a_m\,\hat a_n^\dagger|0\rangle \\
&= \sum_m \bigl|\mathcal{E}_m\,u_m(\vec r)\bigr|^2.
\end{align}

\subsubsection{Coherent State}
Coherent states are generated by an ideal laser. Coherent states $\lvert\alpha_m\rangle$ are defined as the eigenstates of the annihilation operator for mode $m$,
\begin{align}
\hat a_m \,\lvert\alpha_m\rangle &= \alpha_m \,\lvert\alpha_m\rangle.
\end{align}
where $\alpha_m$ is a complex number.  In the Fock basis of mode $m$, they expand as \cite{MandelWolf1995}:
\begin{align}
\lvert\alpha_m\rangle 
&= e^{-\tfrac{|\alpha_m|^2}{2}}
  \sum_{n=0}^{\infty}\frac{\alpha_m^n}{\sqrt{n!}}\;\lvert n_m\rangle.
\label{coherent_state_eq}
\end{align}
so that the photon‐number distribution is Poissonian with mean $|\alpha_m|^2$, and the factor $e^{-|\alpha_m|^2/2}$ ensures $\langle\alpha_m\mid\alpha_m\rangle=1$.
The fields expectation in the coherent state is
\begin{align}
\langle\alpha_m|\hat E_m^{(+)}(\vec r,t)|\alpha_m\rangle
&= \mathcal{E}_m\,\alpha_m\,u_m(\vec r,t).
\end{align}

In addition to reproducing the classical field amplitude, coherent states exhibit several mathematically and physically appealing properties:
Each coherent state is normalized, yet not mutually orthogonal. The inner product is given by
\begin{equation}
\langle \alpha_m | \alpha_n \rangle = \exp\left(-\frac{1}{2}(|\alpha_m|^2 + |\alpha_n|^2 - 2\alpha_m^* \alpha_n)\right),
\end{equation}
and satisfies \( |\langle \alpha_m | \alpha_n \rangle|^2 = e^{-|\alpha_m - \alpha_n|^2} \), indicating a quasi-orthogonality in the limit \( |\alpha_m - \alpha_n| \gg 1 \). They also form an complete basis for the Hilbert space, satisfying the resolution of identity~\cite{ScullyZubairy}:
\begin{equation}
\frac{1}{\pi} \int d^2\alpha_{mn}\, |\alpha_m\rangle\langle \alpha_n| =\mathbb{I}.
\end{equation}

\subsection{Mixed States}
Mixed states capture a situation where there is classical uncertainty about which pure state the system is in. They can arise from imperfect state preparation or as a result of decoherence, where a system becomes entangled with an environment.The description of mixed states can be handled by the density operator formalism.

\subsubsection{Density Operator}
The density operator $\rho$ represents the state of a quantum system. It is a positive semidefinite operator with unit trace \cite{Karuseichyk2024}:
\begin{align}
\rho \geq 0 \quad \text{and} \quad \operatorname{Tr}(\rho) = 1.
\end{align}
For a pure state $\lvert \psi \rangle$, the density operator is given by
\begin{align}
\rho = \lvert \psi \rangle \langle \psi \rvert,
\end{align}
If a quantum system can be in the pure states \(  |\psi_k\rangle  \) with corresponding probabilities \( p_k \) (where \( p_k \geq 0 \) and \(\sum_k p_k=1\)), the density operator is defined as:
\begin{align}
\rho = \sum_{k} p_{k} \rho_k.
\end{align}

\subsubsection{Thermal State}
In many practical situations such as black‐body emitters, incandescent lamps, or stellar radiation, the optical field is not prepared in a pure Fock or coherent state but rather in a thermal state, that is a classical mixture of photon‐number eigenstates. In the Fock basis \( \ket{n_m}\) of mode \(m\), the density operator of a single-mode thermal field is given by:
\begin{align}
\hat{\rho}_{\mathrm{m}}
&= \sum_{n=0}^{\infty}P(n)\,\lvert n_m\rangle\langle n_m\rvert
\end{align}
with the Bose–Einstein photon number distribution:
\begin{align}
    P(n)
&=\frac{\langle\hat{n}_m\rangle^{n}}
     {\bigl(1+\langle\hat{n}_m\rangle\bigr)^{\,n+1}}
\end{align}

where \(\langle\hat{n}_m\rangle = \mathrm{Tr}\{\hat{n}_m\,\hat{\rho}_{\mathrm{m}}\}\) denotes the mean photon number in mode \(m\).  By construction, \(\hat{\rho}_{\mathrm{m}}\) contains no off-diagonal coherences in the \(\lvert n_m\rangle\) basis and therefore describes an incoherent mixture of Fock states.

\section{Photon Statistics}
Quantum optics seeks to explore what happens when we treat a light beam as a succession of photons rather than as a continuous classical wave. At first glance, the appearance of discrete pulses from a detector seems to offer proof that the incoming light consists of individual energy quanta, photons, and that variations in the count rate directly mirror the statistical properties of that photon stream. However, the issue is more subtle. Therefore, it is crucial to distinguish between the statistical behavior inherent to the photodetection mechanism and the intrinsic photon statistics of the light itself. In this section we focus on intrinsic photon statistics of the light. For a more thorough development one can see \cite{quantum_fox}.

The photon flux, \(\Phi\), is defined as the mean number of photons traversing the beam’s cross‐section per unit time:
\begin{align}
  \Phi = \frac{IA}{\hbar \omega} \equiv \frac{P}{\hbar \omega}
  \quad\bigl[\text{photons}\,\mathrm{s}^{-1}\bigr].
  \label{eq:flux}
\end{align}
where \(A\) is the beam area, \(P\) its optical power, I is the intensity and \(\hbar\omega\) the photon energy.

Photon‐counting detectors are specified by their quantum efficiency. \(\eta\): the ratio of registered counts to incident photons.  Over a counting interval \(T\), the mean number of counts is
\begin{align}
  N(T) \;=\; \eta\,\Phi\,T
  \;=\;\eta\,\frac{P\,T}{\hbar \omega}.
\end{align}
and the corresponding average count rate is
\begin{align}
  R = \frac{N(T)}{T}
    = \eta\,\Phi
    = \eta\,\frac{P}{\hbar \omega}
    \quad\bigl[\text{counts}\,\mathrm{s}^{-1}\bigr].
    \label{eq:rate}
\end{align}
Although, Eq. \ref{eq:flux} and \ref{eq:rate} describe mean properties of the beam, even a perfectly stable photon flux exhibits fluctuations in short‐time count records. Since photons are indivisible quanta, each finite segment must actually contain an integer number of photons.  Assuming their positions along the beam are uniformly random gives rise to statistical deviations above and below these mean values. Thus, even a perfectly steady beam exhibits short‐time count‐rate fluctuations, a direct consequence of the light’s discrete nature. Indeed, as shown in Eq. \ref{coherent_state_eq}, a coherent state \(\lvert \alpha_m \rangle\) expands in the Fock basis into number states \(\lvert n_m \rangle\) with photon-number probabilities that follow a Poisson distribution having both mean and variance equal to \(\lvert \alpha_m \rvert^2\). These fluctuations are described by the photon statistics of the light. In fact, perfectly coherent light of constant intensity exhibits Poissonian photon statistics.

Poisson distributions are uniquely specified by their mean value \(n\); in particular their variance equals their mean. The standard deviation for the fluctuations of the photon number above and below the mean value is therefore given by:
\begin{align}
   \Delta n = \sqrt{n}
\end{align}
From a classical standpoint, a perfectly coherent beam at constant intensity is the most stable optical field imaginable. It therefore provides a natural benchmark: based on the standard deviation \(\Delta n\) of their photon‐number distributions relative to \(\sqrt{n}\), light sources can be classified as sub‐Poissonian (\(\Delta n<\sqrt{n}\)), Poissonian (\(\Delta n=\sqrt{n}\)), or super‐Poissonian (\(\Delta n>\sqrt{n}\)).

It is straightforward to identify light sources that exhibit super-Poissonian statistics.  Whenever there are classical fluctuations in intensity, the photon-number variance exceeds that of a beam with constant intensity.  Since a perfectly stable beam gives Poissonian statistics, any classical light with time-varying power will necessarily show super-Poissonian photon-number distributions.

While non-classical sources such as squeezed or antibunched light can in principle achieve sub-Poissonian photon-number variances \cite{Walls1983}, our focus will remain on the Poissonian regime. Indeed, whereas most quantum metrology approaches improve measurement sensitivity by employing exotic non-classical states, we do not rely on such resources. Instead of modifying the photon statistics at the source, we pursue a quantum-inspired enhancement by optimally sorting the photons into transverse modes. As we will demonstrate, the SPADE alone permits us to approach the ultimate quantum limits of superresolution without requiring exotic light sources.

\chapter{Quantum Metrology}

This chapter provides the theoretical foundation for precision estimation tasks that arise in optical imaging, specifically in the context of resolving closely spaced incoherent sources. We introduce the formalism of quantum metrology to quantify the ultimate limits of precision and compare classical and quantum-inspired strategies. Although we work entirely with classical light, our analysis is inspired by quantum bounds and shows how spatial-mode sorting allows us to approach these limits using practical measurement schemes.

\section{Introduction}
Measurement of phenomena in our surroundings lies at the core of physics. As technology advances and allows us to explore more complex systems, the demand for precise measurements continues to grow \cite{Karuseichyk2024}. Metrology, the science of measurement, encompasses the development of measurement standards, techniques, and protocols that ensure accuracy and consistency across diverse fields \cite{czichos_springer_2011}. In any metrological framework, it is essential to define clearly what is being measured, describe how the measurement is performed, and acknowledge the uncertainties and limitations associated with the chosen method.

In the context of optical metrology, we discuss two kinds of estimation strategies; Intensity and mode-based measurements. Intensity-based schemes rely on the spatial intensity distribution recorded (for example, on a camera or an array of detectors) to form an estimator for the parameter of interest. Mode-based schemes, by contrast, involve projecting or sorting the incoming optical field into an orthogonal spatial-mode basis (for example, HG modes), with the resulting mode counts serving as the basis for parameter estimation.

Guided by this classification, we will begin by introducing the formal framework of parameter estimation theory, providing the tools to quantify any measurement strategy’s precision before moving on to analyses of intensity-based schemes.

\section{Parameter Estimation Theory}

Often the quantities of interest within a system cannot be measured directly; instead, we perform indirect measurements by observing related, accessible variables and then apply a known mathematical model to infer the true values. The process of deducing unknown model parameters from those measurements and the model is known as parameter estimation \cite{VanTrees2001,Kay1993}.

\subsection{General Parameter Estimation Scheme}

\begin{figure}[ht]
  \centering
  \includegraphics[width=0.9\textwidth]{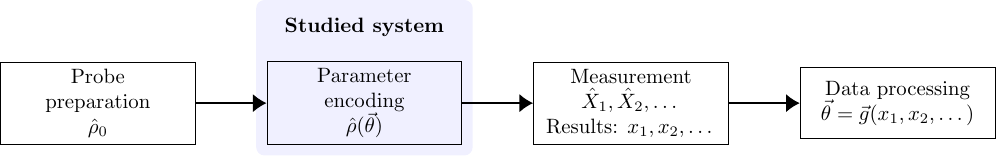}
  \caption{Quantum parameter estimation scheme}
  \label{fig:general_sheme}
\end{figure}

The process of parameter estimation can be divided into distinct stages shown in Fig. \ref{fig:general_sheme}. Specifically, the goal is to find the unknown parameters $\vec{\theta} = (\theta_1, \theta_2, \dots)$ of a given system. 

First, we prepare a probe and let it interact with the system. Its state then evolves into $\hat{\rho}(\vec{\theta})$, which now contains information about those parameters. Next, we perform a series of measurements $\hat{X}_1, \hat{X}_2, \dots$ on the probe, obtaining outcomes $x_1, x_2, \dots$. In the final stage, these results are fed into a classical data-analysis algorithm $g(x)$ to produce an estimator of the unknown parameter \cite{Karuseichyk2024}. 

The central aim of parameter estimation theory is to design such an estimator so that its predicted values match the true parameters as closely as possible, thus reducing estimation error.

\subsection{Classical Parameter Estimation Theory}
Classical parameter estimation typically involves proposing an estimator. There are various systematic methods to derive good estimators, but regardless of the method used, we require ways to assess an estimator’s performance. Important criteria include bias, variance and mean squared error which will be defined shortly. A central goal is often to find an estimator that is unbiased and has the smallest possible variance among a class of estimators. Classical estimation theory provides tools like the Fisher information to quantify how much information the data carry about unknown parameters, and the Cramér–Rao inequality to establish lower bounds on the variance of estimators. These tools let us assess how good an estimator can be in principle, and whether a given estimator is optimal \cite{Fisher1922,Casella2002,Rao1945}.

\subsubsection{Bias and Variance of an Estimator}

Let $\vec{\tilde{\theta}}$ denote an estimator of the true parameter vector $\vec{\theta}$.  We define its bias by\cite{wasserman2005all}:
\begin{align}
\vec{B}(\vec{\tilde{\theta}})
\;=\;
\langle \vec{\tilde{\theta}}
- {\vec{\theta}}\rangle.
\label{eq:bias}
\end{align}
Here, the angle brackets $\langle\cdot\rangle$ denote the statistical expectation value taken over all realizations of the estimator. An estimator is called unbiased if 
\begin{align}
    \vec{B}(\vec{\tilde{\theta}}) = 0.
\end{align}
Unbiased estimators are consistent, in that they converge to the true value $\vec{\theta}$ as the sample size grows.The estimator’s precision is quantified by its variance,
\begin{align}
\Delta^2\vec{\tilde{\theta}}
\;=\;
\langle \vec{\tilde{\theta}}^2 \rangle
\;-\;
\langle \vec{\tilde{\theta}}\rangle^2.
\label{eq:variance}
\end{align}
An optimal unbiased estimator is one that minimizes the variance thereby achieving the highest possible sensitivity to changes in the true parameter.
\subsubsection{Fisher Information}

We limit our work with the estimation of a single real parameter \(\theta\).  Let
\(p(\vec x\mid\theta)\) be the probability density for observing the outcome
\(\vec x\) when the true parameter value is \(\theta\).  This model satisfies the
normalization condition:
\begin{align}
\int \mathrm{d}\vec x\,p(\vec x\mid\theta) &= 1,
\end{align}
and the Fisher information which quantifies how sensitively the distribution
depends on \(\theta\) can be written as\cite{Fisher1922}:
\begin{align}
\mathcal{F}(\theta)
= \int \mathrm{d}x\,\frac{1}{p(\vec x\mid\theta)}
   \biggl(\frac{\partial\,p(\vec x\mid\theta)}{\partial\theta}\biggr)^{2}.
\end{align}
The role of Fisher information becomes especially clear once one invokes the Cramér–Rao bound, which sets a fundamental lower limit on the variance of any unbiased estimator \cite{Cramer1946,Rao1945,Kay1993}.  In particular, if the estimation procedure is repeated independently N times, then
\begin{align}
\Delta^2(\tilde\theta) \;\ge\; \frac{1}{N\,\mathcal{F}(\theta)} \,.
\end{align}
Here \(N\) is the number of independent measurements (or “probes”) used in the estimation protocol, and \(\mathcal{F}(\theta)\) denotes the Fisher information for a single trial.

\subsection{Quantum Parameter Estimation Theory}
In the previous section, we focused on estimating a parameter $\theta$ from measurement outcomes with the highest possible sensitivity. However, it is also possible to identify the fundamental limit to how precisely $\theta$ can be estimated—a limit determined solely by the state of the probe after the parameter has been encoded, independent of the specific measurement setup.

The most general framework for describing quantum measurements is through positive operator-valued measures (POVMs). A POVM consists of a set of Hermitian, non-negative operators \( E_y(x)\ \) that satisfy the normalization condition \cite{phd_clemen}:
\begin{align}
\int d\vec x\, E_y(\vec x) = \mathbb{1}.
\end{align}
The conditional probability of obtaining outcome \( x \) for a given parameter value \( \theta \) is expressed as
\begin{align}
p(\vec x|\theta) = \mathrm{Tr} \left[ E_y(\vec x)\, \hat{\rho}(\theta) \right].
\end{align}
The fundamental bound on the precision of estimating the parameter \( \theta \) is given by maximizing the classical Fisher information \( \mathcal{F} \) over all possible POVMs \( \{E_y(x)\} \). This bound is known as the quantum Cramér-Rao bound (QCRB) and reads:
\begin{align}
\Delta^2 \hat{\theta} \geq \frac{1}{N\, \mathcal{F}[\hat{\rho}(\theta), \{E_y(\vec x)\}]} \geq \frac{1}{N \mathcal{F}_Q[\hat{\rho}(\theta)]},
\label{QCRB_EQ}
\end{align}
where N is the number of independent repetitions of the experiment.
Note that, since we have introduced  POVMs we switch from the  Fisher information $\mathcal{F}(\theta)$ representation to more explicit $\mathcal{F}[\hat{\rho}(\theta), \{E_y(\vec x)\}]$  Fisher information representation so that all subsequent expressions refer to the information contributed by a single use of the probe. \( \mathcal{F}_Q[\hat{\rho}(\theta)] \) denotes the quantum Fisher information (QFI), defined by:
\begin{align}
\mathcal{F}_Q[\hat{\rho}(\theta)] = \max_{\{E_y(\vec x)\}}\, \mathcal{F}[\hat{\rho}(\theta), \{E_y(\vec x)\}] = \mathrm{Tr} \left[ \hat{\rho}(\theta)\, \hat{L}_\theta^2 \right].
\label{QFI}
\end{align}
The operator \( \hat{L}_\theta \), known as the symmetric logarithmic derivative (SLD), satisfies the equation:
\begin{align}
\frac{\partial \hat{\rho}(\theta)}{\partial \theta} = \frac{1}{2} \left( \hat{\rho}(\theta) \hat{L}_\theta + \hat{L}_\theta \hat{\rho}(\theta) \right).
\label{eq:sld}
\end{align}
As described in Eq. \ref{QFI}, the quantum Fisher information represents the maximum classical Fisher information achievable over all possible POVMs. Hence, it acts as an upper limit on precision and sets the ultimate sensitivity for estimating the parameter \( \theta \), regardless of the specific measurement apparatus used.
\begin{align}
    \mathcal{F}[\hat{\rho}(\theta), E_y(\vec{x})] 
    &\leq \operatorname{Tr}[\hat{\rho}(\theta)\, \hat{L}_\theta^2] = \mathcal{F}_Q[\hat{\rho}(\theta)].
    \label{eq:fq_result}
\end{align}
Notice that the right-hand side of the inequality is independent of the chosen POVM. Moreover, one can show that there exists at least one POVM that saturates Eq. \ref{eq:fq_result}, namely the POVM constructed from the eigenstates of \(\hat L_\theta\) \cite{PezzeSmerzi2014}. However, this measurement is in general hard to compute, it depends explicitly on the unknown parameter \(\theta\), and it is even harder to implement physically. Consequently, practical implementations typically seek measurement schemes that can saturate the quantum Cramér–Rao bound for all values of \(\theta\); a notable example is spatial mode demultiplexing (SPADE), which achieves the quantum Fisher information limit across the entire range of source separations, as we will see in the mode based measurement section. 

\section{Direct Intensity Measurements}
\begin{figure}[h]
    \centering
    \begin{tikzpicture}
        % Define the colors
        \definecolor{darkblue}{rgb}{0.1,0.1,0.6}
        \definecolor{lightblue}{rgb}{0.2,0.7,0.9}
    
        % Axes
        \draw[->] (-4,0) -- (3.5,0) node[right] {\textit{x}};

        % Gaussian curves (made taller by increasing the amplitude)
        \draw[thick,darkblue] plot[domain=-3.5:3,samples=100] (\x,{2*exp(-(\x+1)^2)});
        \draw[thick,lightblue] plot[domain=-2.5:2.5,samples=100] (\x,{2*exp(-(\x-0.3)^2)});
        
        % Mean points
        \draw[dashed] (-1,0) -- (-1,{2*exp(-0)}) ;
        \draw[dashed] (0.3,0) -- (0.3,{2*exp(-0)}) ;
         \draw[dashed] (-1,0) node[below] {\textcolor{black}{$X_1$}};
         \draw[dashed] (0.3,0) node[below] {\textcolor{black}{$X_2$}};
    
        % Distance labels
        \draw[<->] (-1,2.3) -- (0.3,2.3) node[midway, above] {\textcolor{black}{$\theta$}};
        
        % Wavefunction labels
        \node[left, darkblue] at (-2.4,0.3) {\textcolor{darkblue}{$u_1(x)$}};
        \node[right, lightblue] at (2.0,0.3) {\textcolor{lightblue}{$u_2(x)$}};

    \end{tikzpicture}
    \caption[Dual-source PSF wavefunctions]{Two photonic wavefunctions on the image plane originate from point sources at \( X_1 \) and \( X_2 \), separated by \( \theta \), and are shaped by the point-spread functions.}
\label{fig:wavefunctions}
\end{figure}
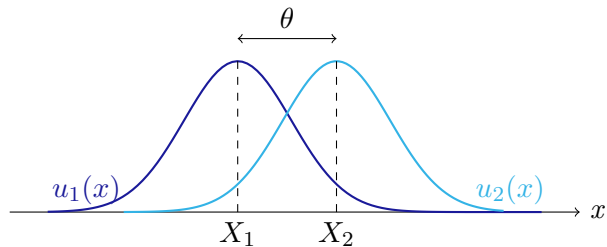
In a direct intensity measurement, photon counts are collected spatially in the image plane. These measurements directly reflect the spatial intensity I distribution, which is proportional to the time‐averaged square of the electric‐field amplitude \cite{SalehTeich2007}:
\begin{align}
    I \propto \langle E(t)^2 \rangle.
\end{align}
On the other hand, SPADE that we will discuss later performs intensity measurements after projecting onto a mode basis.

The main advantages of direct detection are its simplicity and speed: the detector’s output current or voltage can be directly related to optical power via a known responsivity, with minimal post‐processing. Nevertheless, direct intensity measurements exhibit limitations when used to resolve closely spaced sources. 

A natural benchmark for these limitations is the resolution power of diffraction‐limited systems such as telescopes or microscopes. Resolution power is defined as the smallest angular or spatial separation between two point sources that can still be distinguished. The classical Rayleigh criterion, introduced by Lord Rayleigh, specifies this minimum angular separation \(\theta_{\text{R}}\) for a circular aperture of diameter \(D\) \cite{SalehTeich2007}:
\begin{align}
    \theta_{\text{R}} = 1.22 \,\frac{\lambda}{D}.
\end{align}

Note that Rayleigh’s criterion is a purely heuristic rule based on the overlap of intensity profiles. This scenario, with two overlapping wavefunctions on the image plane, is depicted in Fig. \ref{fig:wavefunctions}. This limitation motivates the adoption of an estimation-theoretic framework using Fisher information to rigorously quantify and optimize resolution beyond this heuristic bound.

To illustrate the limitations of direct intensity measurement, we now consider the typical problem of resolving two incoherent, point‐like sources in a diffraction‐limited optical system from a estimation theory point of view. At optical frequencies, where the photon energy $\hbar \omega$ greatly exceeds the thermal energy $k_B T$, thermal sources emit an average photon number $\bar{n} \ll 1$ per coherence interval much smaller than unity, so that the field on the image plane in each interval can be approximated by a statistical mixture of vacuum and single‐photon components \cite{Zmuidzinas2003,Tsang2016}:
\begin{align}
    \rho = (1 - \bar{n})\,\rho_0 + \bar{n}\,\rho_1 + O(\bar{n}^2).
\end{align}
where \(\rho_0\) denotes the vacuum state and the one‐photon contribution is given by:
\begin{align}
    \rho_1 \approx \frac{1}{2}\bigl(\ket{1,u_1}\bra{1,u_1} + \ket{1,u_2}\bra{1,u_2}\bigr).
    \label{one-photon}
\end{align}
Here, $\ket{1, u_1}$ and $\ket{1, u_2}$ denote single-photon states corresponding to the spatial modes $u_1(x)$ and $u_2(x)$, which are the image-plane wavefunctions of two equally bright, incoherent point sources separated by a distance $\theta$ in the object plane. These wavefunctions are given by $u_1(x) = u_0\!\left(x + \tfrac{\theta}{2}\right)$ and $u_2(x) = u_0\!\left(x - \tfrac{\theta}{2}\right)$, where $u_0(x)$ is the normalized point-spread function of the imaging system, typically modeled as a Gaussian $u_0(x) \propto \exp\!\left(-\frac{x^2}{w^2}\right)$.

\begin{figure}[h]   % or [htbp]
  \centering
    \resizebox{1\textwidth}{!}{\input{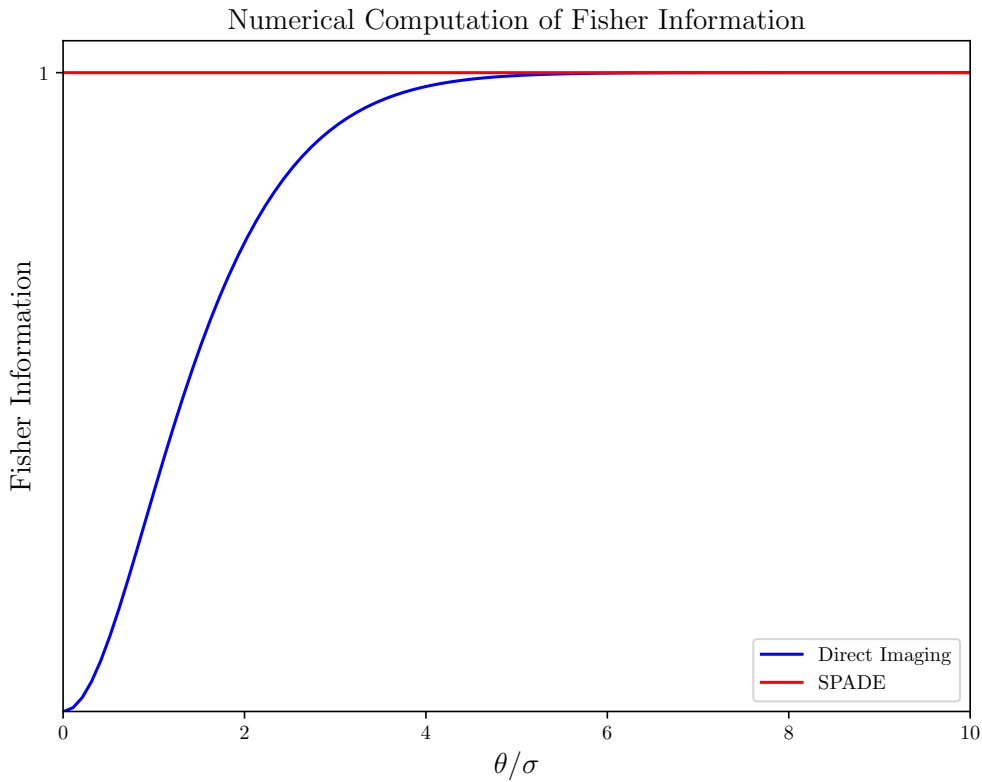}}
  \caption[Fisher information vs separation]{Fisher information as a function of source separation for a Gaussian point-spread function. The horizontal axis shows separation in units of the PSF width \(\sigma\), and the vertical axis shows Fisher information normalized by \(\displaystyle \frac{N}{4\sigma^2}\).}
  \label{fig:fisher_info_custom_2}
\end{figure}

Upon detection of a photon, the resulting intensity‐based probability density reads
\begin{align}
    \Lambda(x) =\operatorname{Tr}[\hat{E}_x \rho]= \frac{1}{2}\bigl|u_1(x)\bigr|^2 + \frac{1}{2}\bigl|u_2(x)\bigr|^2,
\end{align}
and the Fisher information for estimating the source separation \(\theta\) is
\begin{align}
    \mathcal{F}^{direct}(\theta)
    = N \int_{-\infty}^{\infty} \! dx \,\frac{1}{\Lambda(x)}\Bigl(\frac{\partial \Lambda(x)}{\partial\theta}\Bigr)^2.
\end{align}
For a Gaussian point-spread function \cite{Pawley2006}, the derivative \( \frac{\partial \Lambda(x)}{\partial \theta} \) vanishes at \( \theta = 0 \) while $\Lambda(x)$ remains non zero in regions of x where the derivative vanishes, leading a vanishing $\mathcal{F}^{direct}$. Hence, for separations \(\theta \ll \theta_{\text{R}}\) we lost all the information and separation between two sources becomes impossible to estimate leading to the so‐called “Rayleigh’s curse” as can be seen in Fig. \ref{fig:fisher_info_custom_2}. 

Conventional direct intensity measurements record only the spatial intensity distribution. By contrast, SPADE performs projective measurements in a tailored mode basis, allowing one to recover this more information and achieve the quantum Fisher information limit.

\section{Mode-based Measurements (SPADE and Quantum Limit)}
We now describe spatial mode demultiplexing (SPADE), the key measurement technique used in this thesis to implement mode-based estimation. SPADE employs a projection of the incoming field onto an orthonormal spatial‐mode basis in order to extract information about the source separation~$\theta$. For the one‐photon state \(\rho_1\) of Eq. \ref{one-photon}, the probability of detecting the photon in mode \(q\) is then:
\begin{align}
  P(q\mid \theta)
  = \operatorname{Tr}\bigl[\ket{1,u_q}\bra{1,u_q}\,\rho_1\bigr]
= \frac{1}{2}\bigl|\braket{1,u_q|1,u_1}\bigr|^2
    + \frac{1}{2}\bigl|\braket{1,u_q|1,u_2}\bigr|^2.
  \label{eq:P_mode}
\end{align}
where \(\ket{1,u_q}\) denotes single photon states into a complete set of orthonormal modes (for example, the HG functions of width \(\sigma\)).

For a Gaussian point-spread function, the projection probabilities in Eq.~\eqref{eq:P_mode} can be computed analytically. Each quantity $\braket{1,u_q|1,u_{1,2}}$ represents an overlap integral between two spatial modes:
\begin{align}
\braket{1,u_q|1,u_{1,2}} = \int_{-\infty}^{\infty} u_q^*(x)\, u_{1,2}(x)\, dx.
\end{align}
This overlap tells us how similar the displaced input mode \(u_{1,2}(x)\) is to the detection mode \(u_q(x)\); its squared magnitude gives the probability of detecting the photon in mode \(q\). The displaced source modes \(u_{1,2}(x)\) can be interpreted as Gaussian beams shifted slightly to the left or right by a small parameter \(\theta\). Crucially, these overlaps can be calculated exactly because both the source and detection modes are Gaussian-like functions with well-known integral identities. As shown in Ref.~\cite{Tsang2016}, this leads to simple analytical expressions for the detection probabilities:
\begin{equation}
    P(q\mid \theta) \approx \exp(-Q) \frac{Q^q}{q!}, \quad Q \equiv \frac{\theta^2}{16\sigma^2}.
\end{equation}
This formula remains valid even for sources with unequal intensities. 
The classical Fisher information for the HG basis measurement over \( M \) intervals is:
\begin{equation}
    \mathcal{F}^{SPADE}(\theta) = N \sum_{q=0}^{\infty} P(q\mid \theta) \left( \frac{\partial}{\partial \theta} \ln P(q\mid \theta) \right)^2 = \frac{N}{4\sigma^2}=\mathcal{F}_Q,
\end{equation}
which is independent from separation. Moreover, $\mathcal{F}^{SPADE} = \mathcal{F}_Q$ which means that SPADE is the optimal measurement for separation estimation. The $\mathcal{F}^{SPADE}$ (or equally the $\mathcal{F}_Q$) remains finite as \(\theta\to0\) and equals the quantum Fisher information. Hence mode‐based estimation overcomes the vanishing‐information “Rayleigh’s curse” of intensity‐based methods and attains the ultimate precision bound for all separations \cite{Tsang2016}.

In summary, intensity-based measurements suffers from Rayleigh’s curse, meaning its ability to estimate the separation between sources diminishes significantly as their wavefunctions overlap. When the sources move closer together, the classical Fisher information goes to zero. As a result, the Cramér–Rao bound diverges, and separation estimates become extremely unreliable. In contrast, the quantum Fisher information is constant no matter how close the sources get. Therefore, by using the SPADE measurement, one can avoid Rayleigh’s curse and achieve a significant improvement in separation estimation. As we demonstrate in later chapters, MPLC offers a physically realizable method to implement SPADE by performing the required unitary transformation from the Gaussian mode basis to the HG mode basis.

\chapter{Multi-Plane Light Conversion}
Having established in Chapter 3 that SPADE can theoretically overcome the Rayleigh limit by projecting light onto an orthonormal basis, the challenge now lies in physically implementing such a projection. This chapter introduces  MPLC as a practical method to achieve this and details the design and simulation of an MPLC-based mode sorter. We begin by establishing the mathematical formalism underlying MPLC. We then introduce the numerical algorithm used to compute the optimal phase-mask profiles based on wavefront matching. This is followed by a detailed discussion of the design and simulation framework, including its discretization and computational considerations. Finally, we present and analyze simulation results that quantify conversion fidelity while also evaluating the system's robustness to wavelength variation, beam misalignment, propagation distance offsets, and beam-size deviations.

\section{Multi-Plane Light Conversion}

One of the optical techniques to implement the principle of SPADE is Multi-Plane Light Conversion. This optical method employs a series of phase-modulation planes to manipulate light beams as they propagate through free space. By sequentially modulating the wavefront with these phase masks, an MPLC can implement an arbitrary unitary transformation between an input set and an output set of light modes. In simpler terms, MPLC can convert one orthogonal basis of spatial modes into another orthogonal basis (for example, converting a set of separate Gaussian beams into a set of HG modes) while preserving orthogonality and energy. This capability makes MPLC a powerful tool for beam shaping and mode multiplexing in a variety of applications.

\subsection{Theoretical Considerations}

Optical systems that manipulate the spatial profile of a beam are conventionally destructive: they use phase control plus attenuation to reshape a field at the expense of total intensity. By contrast, any unitary spatial transform preserves photon number. It is proved that by interleaving local phase manipulations with Fourier transforms, any unitary mapping of the transverse field is achievable \citep{Morizur2010}. A mathematical proof of the universality of this approach is as follows.

Without loss of generality, assume a monochromatic, linearly polarized beam of wavelength $\lambda$, propagating along the $z$–axis, which has transverse profile at $z=0$ given by:
\begin{align}
E(x,y) = A(x,y)\,e^{\,i\phi(x,y)}
= \sum_{m,n\ge0} a_{mn}\,u_{mn}(x,y).
\end{align}
where the transverse profile can be decomposed in $u_{mn}$ basis with complex coefficients $a_{mn}$.

Any linear optical system is fully specified by how it maps each basis mode $u_{mn}$ to a superposition of $u_{mn}$ modes. If the system is lossless, its matrix $U$ in this basis is unitary. Simple elements like lenses and mirror realize only a small subgroup of $U$: for instance, they cannot convert $HG_{00}$ into $HG_{10}$ since those are eigenmodes of all Gaussian systems \citep{Morizur2010}. This means that another type of elements are needed, such as phase plates.  

However, phase plates only apply local phase shifts and therefore leave the beam’s intensity profile unchanged. Nevertheless, when the beam is composed of a superposition of $u_{mn}$ modes, the distinct Gouy phase shifts that each mode accumulates during propagation change their relative phases, and this phase rebalancing alters the total intensity distribution of the output beam \citep{Morizur2010}.

The two elementary operations available for MPLC are the phase modulation $U_{\text{phase}}$ and the Fourier transform $U_{\text{FT}}$. Let the subgroup $H$ of $U$  be generated by these operations, that is, 
\begin{align} 
H=\langle U_{\text{phase}},\, U_{\text{FT}} \rangle,
\end{align}
and $U_{ij}$ be the subgroup of $U$ that contains all the matrices of the form:

\begin{align}
T_{ij}(\theta) = 
\begin{pmatrix}
1      &        &        &        &        &        &   \\
       & \ddots &        &        &        &        &   \\
       &        & \cos\theta &        & \sin\theta &        &   \\
       &        &        & 1      &        &        &   \\
       &        & -\sin\theta &        & \cos\theta &        &   \\
       &        &        &        &        & \ddots &   \\
       &        &        &        &        &        & 1 \\
\end{pmatrix}.
\end{align}

where the $sin(\theta)$ and $cos(\theta)$ terms are in the $ith$ row and $jth$ column.
It can be shown that it is possible to build $T_{ij}(\theta)$ by interleaving $U_{\rm phase}$ and $U_{\rm FT}$ \cite{BorevichKrupetskii1981}. This means that all the $T_{ij}(\theta)$ are in $H$ hence all $U_{ij}$ are in U. Moreover, the full unitary group is generated by these $T_{ij}(\theta)$ together with $U_{phase}$ and since $H$ is a group it contains all these successions. Hence $H$ is $U$. It follows that any lossless mode optical transform can be exactly decomposed into a finite sequence of phase plane and Fourier transforms.  
\subsection{The Wavefront Matching Algorithm}\label{WFM}

In the previous section, we showed that any unitary transformation between spatial modes can, in principle, be implemented using a sequence of phase modulations and Fourier transforms. In practical optical systems, Fourier transforms are typically implemented through free-space propagation. As a result, combining phase modulations with free-space propagation enables the realization of arbitrary unitary transformations. In this section, we describe how to compute the required phase profiles using an inverse-design algorithm known as Wavefront Matching (WFM), which we adapt specifically for the MPLC architecture.

WFM is an iterative optimization technique that designs the phase profiles \(\{\phi_m(x)\}_{m=1}^M\) for a sequence of \(M\) discrete phase planes. The goal is to transform a given set of orthonormal input modes \(\{u_0^{(i)}(x)\}\) into a desired set of output modes \(\{v_M^{(i)}(x)\}\) with high fidelity. The physical propagation between phase planes is modeled using a free-space propagation operator (the Angular Spectrum Method). Each phase plane modulates the beam by applying a spatially varying phase shift, and the design aims to align the forward-propagated input wavefronts with the backward-propagated target wavefronts at each plane.

The algorithm proceeds as follows:

\begin{enumerate}
    \item \textbf{Initialization}: All phase profiles \(\phi_m(x)\) are initialized to zero.
    
    \item \textbf{Forward Propagation}: Each input mode \(u_0^{(i)}(x)\) is sequentially propagated through the current set of phase masks to generate intermediate fields \(\{u_m^{(i)}(x)\}\) at each plane \(m = 1, \dots, M\).
    
    \item \textbf{Backward Propagation}: Each target output mode \(v_M^{(i)}(x)\) is propagated in reverse through the conjugate of the current optical system, yielding backward fields \(\{v_m^{(i)}(x)\}\) at each phase plane.
    
    \item \textbf{Phase Update}: At each plane \(m\), the phase \(\phi_m(x)\) is updated to locally maximize the overlap between the forward and backward fields. The corresponding update rule is:
    \[
    \phi_m(x) = \arg\left[\sum_i u_m^{(i)}(x)\, v_m^{(i)*}(x)\right],
    \]
    which aligns the summed forward and backward fields in phase to reinforce constructive interference.
    
    \item \textbf{Iteration}: Steps 2--4 are repeated until convergence, typically measured by an increase in mode fidelity or a decrease in phase update norm.
\end{enumerate}

To evaluate performance, we define the fidelity matrix between the output fields \(\{u_M^{(i)}(x)\}\) and the target modes \(\{v_M^{(j)}(x)\}\) as:
\begin{align}
F_{ij} = \left| \int u_M^{(i)}(x)\, v_M^{(j)*}(x)\, dx \right|^2.
\end{align}
High fidelity corresponds to large diagonal values and minimal off-diagonal values, indicating low cross-talk and successful implementation of the unitary transformation.

This forward-backward wavefront matching cycle enables practical design of MPLC systems with high mode selectivity and low insertion loss, even in regimes involving closely spaced or overlapping spatial modes.

\section{Design and Simulation Framework}
This section introduces the simulation framework used to design MPLC. We begin by describing the underlying propagation model and physical constraints, followed by the sampling requirements of the propagation model, tilted-beam configurations, and finally the numerical results evaluating system performance.
\subsubsection{Angular Spectrum Method: Capabilities and Practical Constraints}

Accurately modeling light propagation in MPLC systems requires accounting for both near-field and wide-angle diffraction effects. The near field refers to the regime close to the source or optical element, where the light field exhibits complex spatial variations that are strongly influenced by the aperture shape and phase features. This regime occurs at propagation distances smaller than the Rayleigh length, \(z_R = \pi w_0^2 / \lambda\), where the field has not yet converged into its far-field pattern \cite{SalehTeich2007}. In contrast, far-field propagation describes the regime where the beam has developed a well-defined angular distribution and can be approximated by spherical or plane waves. 

Additionally, many beams in MPLC undergo wide-angle propagation, where significant energy exists at large transverse wavevector components. The Angular Spectrum Method (ASM) is particularly well suited to this setting, as it preserves all spatial frequency components and accurately captures both near-field evolution and large-angle diffraction, making it ideal for simulating free-space multi-plane systems.

\begin{figure}[h]
    \centering
    \includegraphics[width=0.4\textwidth,height=0.4\textwidth]{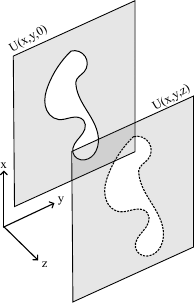}
    \caption[Field propagation of optical field]{Illustration of the free-space propagation of an optical field $U(x,y,0)$ along the z-axis to the observation plane $U(x,y,z)$.}
    \label{fig:asm_prop}
\end{figure}
To simulate such complex propagation accurately, ASM treats the field as a superposition of plane waves and computes their evolution in Fourier space. The method begins by performing a two-dimensional spatial Fourier transform of the input field $U(x,y,0)$ at the initial plane \cite{matsushima2020computerholography}:
\begin{align}
U(k_x, k_y, 0) = \mathcal{F}\{U(x,y,0)\} = \iint_{-\infty}^{\infty} U(x,y,0)\, e^{-i(k_x x + k_y y)}\,dx\,dy,
\end{align}
where $k_x$ and $k_y$ are the transverse spatial frequency components. This operation decomposes the optical field into its constituent plane wave components. Each component then accumulates a phase shift as it propagates over a distance $z$ in a homogeneous medium:
\begin{align}
U(k_x, k_y, z) = U(k_x, k_y, 0)\, e^{-i z \sqrt{k^2 - k_x^2 - k_y^2}},
\label{kernel}
\end{align}
where $k = 2\pi/\lambda$ is the total wave number and the square-root term represents the longitudinal wavevector $k_z$. For evanescent waves, $k_z$ becomes imaginary, resulting in exponential decay.

The propagated field in the spatial domain is then recovered by applying the inverse Fourier transform:
\begin{align}
U(x, y, z) = \mathcal{F}^{-1} \{ U(k_x, k_y, z) \} = \frac{1}{(2\pi)^2} \iint_{-\infty}^{\infty} U(k_x, k_y, z)\, e^{i(k_x x + k_y y)}\, dk_x\, dk_y.
\end{align}
An illustration of free-space propagation is shown in Fig. \ref{fig:asm_prop}.
This approach offers a highly accurate and computationally efficient numerical scheme, especially well-suited for the short to intermediate propagation distances typical in folded MPLC systems, where near-field and wide-angle effects are prominent. Here, short to intermediate refers to propagation distances that are on the order of or smaller than the Rayleigh length. The detailed structure and design of the folded MPLC will be discussed later.

While the ASM is mathematically accurate, its direct application to real-world MPLC systems is non-trivial. Practical setups often violate the method’s core assumptions such as normal incidence and matched sampling grids which will be detailed shortly. These limitations become especially relevant when modeling realistic optical setups that involve beam tilts or folded propagation paths, as we discuss next.

\begin{figure}[tb]
  \centering
  \begin{subfigure}[t]{0.04\textwidth}
    \textbf{(a)}
  \end{subfigure}
  \begin{subfigure}[t]{0.45\textwidth}
    \centering
    \includegraphics[width=0.15\linewidth, valign=t]{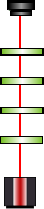}

  \end{subfigure}\hfill
  \begin{subfigure}[t]{0.04\textwidth}
    \textbf{(b)}
  \end{subfigure}
  \begin{subfigure}[t]{0.45\textwidth}
    \centering
    \includegraphics[width=\linewidth, valign=t]{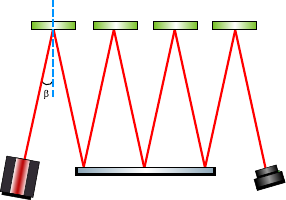}
  \end{subfigure}
  \caption[Transmissive vs folded MPLC]{Two MPLC architectures: (a) Transmissive configuration, where the beam passes through a sequence of distinct phase masks in free space; (b) Reflective (folded) configuration, where a single phase mask and a mirror enable multiple reflections to perform successive transformations within a compact footprint.}
  \label{fig:mplc_configurations}
\end{figure}

A basic MPLC can be built in a transmissive layout, where the light beam strikes each phase mask straight on, at right angles to the mask surfaces as can be seen in Fig. \ref{fig:mplc_configurations}a.  In practice, however, devices are usually made as a folded cavity, with one plane holding the phase mask and another a mirror as in Fig. \ref{fig:mplc_configurations}b.

To describe the reflective design, the ASM has two fundamental limitations:  First, it only works when the beam is exactly normal to both the input and output planes.  Second, the ASM requires that the numerical grids used to sample the field at the input and output planes have the same number and center. This constraint arises from the FFT-based convolution structure of the method and limits its applicability in scenarios where the beam is shifted, tilted, or undergoes magnification \cite{matsushima2020computerholography}. If the output plane is shifted sideways, the grid must be expanded so that it still contains both the incoming and outgoing fields, which greatly increases the computational effort.  Because of these restrictions, standard ASM is not well suited for simulating real MPLC systems where the beam may arrive at an angle. Despite these limitations, the ASM can still be applied in specific reflective configurations where the incidence angles remain small.
\begin{figure}[tb]
  \centering
  \begin{subfigure}[t]{0.04\textwidth}
    \textbf{(a)}
  \end{subfigure}
  \begin{subfigure}[t]{0.27\textwidth}
    \includegraphics[width=\linewidth, valign=t]{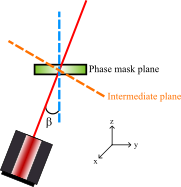}
    \label{fig:incidence_geom}
  \end{subfigure}\hfill
  \begin{subfigure}[t]{0.04\textwidth}
    \textbf{(b)}
  \end{subfigure}
  \begin{subfigure}[t]{0.27\textwidth}
    \includegraphics[width=\linewidth, valign=t]{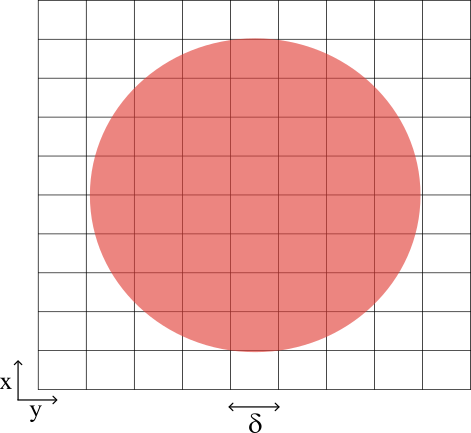}
    \label{fig:intermediate_grid}
  \end{subfigure}\hfill
  \begin{subfigure}[t]{0.04\textwidth}
    \textbf{(c)}
  \end{subfigure}
  \begin{subfigure}[t]{0.27\textwidth}
    \includegraphics[width=\linewidth, valign=t]{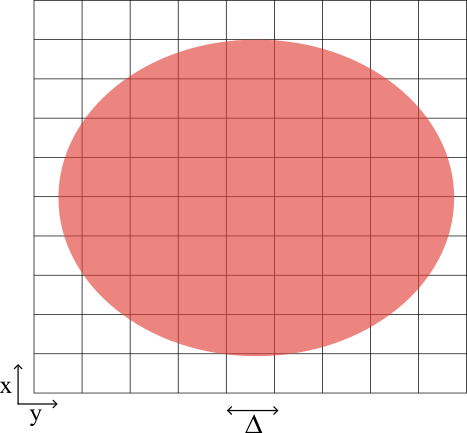}
    \label{fig:mask_grid}
  \end{subfigure}
  \caption[Sampling and alignment in folded MPLC]{(a) Beam incidence geometry showing the folded configuration. (b) Sampling grid on the intermediate plane (\(\delta\)) after propagation. (c) Sampling grid on the phase mask plane (\(\Delta\)) showing adjusted pixel alignment. These plots illustrate the geometric distortions that must be accounted for when aligning SLM pixels with the propagated field.}
  \label{fig:grid_sampling}
\end{figure}

In the reflective configuration, the incoming beam propagates along the optical axis, but the phase masks are physically aligned along the $y$-axis. For small incidence angles (typically under $10^\circ$), the ASM can  be used by modifying the spatial sampling. Specifically, the sampling grid in the $y$ direction is rescaled by a factor of $\cos(\beta)$, where $\beta$ is the incidence angle. The $x$-axis sampling remains unchanged, while the $y$-axis spacing is reduced to account for the projection of the beam path.

These geometric distortions must be taken into account when mapping fields between propagation planes and the physical pixel grid of the spatial light modulator (SLM), which we use in the laboratory to implement the required spatial transformations. As a result of this anisotropic scaling, a circular Gaussian profile in the intermediate plane becomes elliptical in the phase-mask plane. Since physical phase masks on the SLM are implemented on square-pixel grids, all calculations are performed on equidistant square grids defined in the phase-mask plane. When these square grids are mapped back to the intermediate propagation plane, they appear as rectangles due to the inverse projection. This grid distortion is illustrated in Fig. \ref{fig:grid_sampling}, which shows how a circular beam profile is mapped to an elliptical one due to the angled incidence.

\subsection{Sampling Requirements}

Accurate numerical modeling of free‐space propagation in multi‐plane light conversion relies critically on the choice of sampling parameters in both the spatial and spectral domains. In the angular spectrum framework, the pixel size $\Delta x$ (and $\Delta y$) sets the maximum spatial frequency that can be represented without aliasing, while the total number of grid points $(N_x, N_y)$ determines the sampling density and thus the ability to resolve the finest phase‐mask features. To ensure that the propagation kernel remains aliasing‐free these parameters must satisfy bounds derived from the Nyquist criterion. The constraints on $\Delta x$, $\Delta y$, $N_x$, $N_y$, and the propagation distance $z$ presented in the following section are derived from established formulations in \cite{tutorial,matsushima2020computerholography}. These references form the basis for understanding how to choose sampling parameters that avoid aliasing and ensure accurate, high-fidelity beam propagation using the angular spectrum method.

When using the angular spectrum method for free-space propagation, the system can be characterized by its transfer function where it can be seen in Eq. \ref{kernel}. This function acts as the kernel of ASM and can be rewritten as:
\begin{equation}
H(v_x,v_y)=e^{-iz \sqrt{k^2 - k_x^2 - k_y^2}}
=\exp\Bigl[-i\phi(v_x,v_y)\Bigr], \label{eq:transfer}
\end{equation}
where $\phi(v_x,v_y)=2\pi z\sqrt{\lambda^{-2}-v_x^2-v_y^2}$ and $v_{x,y}=\frac{k_{x,y}}{2\pi}$. By expressing the phase term $\phi(v_x,v_y)$ in normalized form, one obtains an ellipsoidal condition:
\begin{equation}
\frac{\phi(v_x,v_y)^2}{\left(\frac{2\pi z}{\lambda}\right)^2} + \frac{v_x^2}{\lambda^{-2}} + \frac{v_y^2}{\lambda^{-2}} = 1. \label{eq:ellipsoid}
\end{equation}
In a simulation, the spatial domain is sampled on a discrete grid. If we let
\begin{equation}
x = [-\frac{N_x}{2},\frac{N_x}{2}-1]\Delta x,
\quad
y = [-\frac{N_y}{2},\frac{N_y}{2}-1]\Delta y \label{eq:spatial_sampling}
\end{equation}
where $\Delta x$ and $\Delta y$ denote the grid sizes then the corresponding sampling points in the Fourier domain are given by:
\begin{equation}
v_x = [-\frac{N_x}{2},\frac{N_x}{2}-1]\Delta v_x, \quad
v_y =[-\frac{N_y}{2},\frac{N_y}{2}-1]\Delta v_y\label{eq:fourier_sampling}
\end{equation}
with $\Delta v_x=\frac{1}{N_x\Delta x},\quad \Delta v_y=\frac{1}{N_y\Delta y}.$

The range of spatial frequencies for \(v_x\) spans approximately from \(-1/(2\Delta x)\) to \(1/(2\Delta x)\), regardless of \(N_x\). To accurately resolve the finest features of the optical field and suppress aliasing, the number of pixels \(N_x\) and \(N_y\) must be chosen sufficiently large. Among the key parameters, pixel pitch \(\Delta x\) and wavelength \(\lambda\),  pixel numbers \(N_x\) and \(N_y\) are typically the most flexible in practice: \(\Delta x\) is set by the physical pixel size of the SLM, and \(\lambda\) is determined by the laser source. Therefore, the spatial resolution can be primarily adjusted by tuning \(N_x\) and \(N_y\), a strategy that will be discussed in more detail later.

Note that the ellipsoid’s width in Eq. \ref{eq:ellipsoid} remains fixed at $\lambda^{-1}$, whereas its height increases with propagation distance $z$. As $z$ grows, the local slope of the ellipsoid steepens, tightening the sampling requirements; for very large $z$, aliasing can occur even for a finite-bandwidth field.

The slope of the ellipsoid corresponds to the local spatial frequency. For each direction, one can define:
\begin{equation}
f_{v_x} = \frac{1}{2\pi}\frac{\partial\phi(v_x,v_y)}{\partial v_x}
=\frac{v_x\,z}{\sqrt{\lambda^{-2}-v_x^2-v_y^2}}, \quad
f_{v_y} = \frac{1}{2\pi}\frac{\partial\phi(v_x,v_y)}{\partial v_y}
=\frac{v_y\,z}{\sqrt{\lambda^{-2}-v_x^2-v_y^2}}. \label{eq:local_freq}
\end{equation}
For example, considering $f_{v_x}$, the maximum local spatial frequency is given by:
\begin{equation}
\max\bigl(f_{v_x}\bigr)=\frac{\max(v_x)\,z}{\sqrt{\lambda^{-2}-\max(v_x)^2-\max(v_y)^2}}. \label{eq:max_local_freq}
\end{equation}
If we approximate $\max(v_x)$ by $(N_x/2-1)\Delta v_x \approx 1/(2\Delta x)$ and similarly for $v_y$, then the Nyquist sampling condition which requires that the sampling frequency is at least twice the highest frequency of the signal imposes:
\begin{equation}
\frac{1}{\Delta v_x} > 2\,\max\bigl(f_{v_x}\bigr). \label{eq:nyquist}
\end{equation}
For a fixed $N_x$ inequality leads to a constraint on the propagation distance $z$, which can be written as:
\begin{equation}
z < \frac{\Delta x^2N_x}{2}\sqrt{\frac{4}{\lambda^2}-\frac{1}{\Delta x^2}-\frac{1}{\Delta y^2}}. \label{eq:constraint1}
\end{equation}
Applying this condition to the physical system that we will see in the next chapter, where the pixel pitch is fixed at \(8\,\mu\text{m}\) \((\Delta x = \Delta y)\) and the operating wavelength is \(\lambda = 633\,\text{nm}\), we can control the minimum allowable free-space propagation distance by appropriately selecting the number of pixels \(N_x = N_y\). For instance, choosing \(N_x = 512\) yields a lower bound of \(z = 51.7\,\text{mm}\) for the propagation distance, which serves as a critical design constraint in the subsequent system layout.

\subsection{Simulation Results and Performance Evaluation}
Although MPLC is in principle capable of implementing arbitrary unitary transformations on the spatial profile of light~\cite{Morizur2010}, in this work we focus exclusively on a practically motivated subclass: transformations from spatially separated Gaussian beams to co-located HG modes. This restriction is not due to a limitation of the MPLC architecture, but rather reflects the requirements of quantum-inspired superresolution imaging. In practical SPADE measurements, one seeks to extract information about sub-diffraction features by projecting the incoming optical field typically consisting of spatially overlapped point sources onto a basis of HG modes. However, in this thesis we consider the inverse transformation: we prepare well-defined HG modes and transform them into spatially separated Gaussian spots using MPLC. This reversed direction is experimentally advantageous, since it allows us to visually and spatially resolve the output spots corresponding to different input modes, thereby facilitating fidelity and crosstalk measurements. Importantly, the same MPLC design performs the transformation in either direction HG modes to spots, or spots to HG modes since it implements a unitary transformation. 

It is important to emphasize that while the experimental procedure operates in the inverse direction (mode to spot), all numerical simulations presented in this thesis follow the forward transformation (spot to mode) as originally motivated by SPADE’s requirement to project spatially localized signals onto an orthonormal mode basis. We now present numerical simulation results for a six-mode MPLC system using phase masks optimized via the WFM. We choose this example because it is representative of practical scenarios in optical mode demultiplexing, allowing us to analyze key performance metrics such as fidelity, spectral response, and robustness to misalignment and input parameter variation.

\subsubsection{Fidelity}
\begin{figure}[tb]
  \centering
  \begin{subfigure}[t]{0.04\textwidth}
    \textbf{(a)}
  \end{subfigure}
  \begin{subfigure}[t]{0.45\textwidth}
    \includegraphics[width=\linewidth, valign=t]{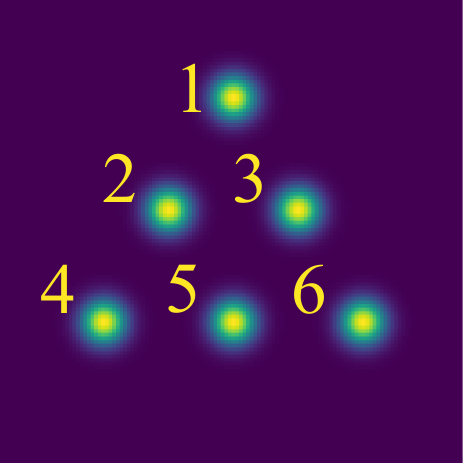}
  \end{subfigure}\hfill
  \begin{subfigure}[t]{0.04\textwidth}
    \textbf{(b)}
  \end{subfigure}
  \begin{subfigure}[t]{0.45\textwidth}
    \includegraphics[width=\linewidth, valign=t]{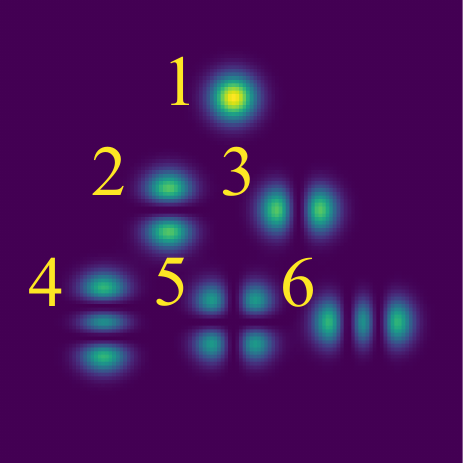}
  \end{subfigure}
  \caption[Six-beam HG-array simulation]{The simulation algorithm operates on (a) an input array of six Gaussian beams, each with a waist \(w_0 = 100~\mu\text{m}\), arranged in a triangular lattice. The target output (b) is a spatially co-located array of HG modes with \(w_0 = 200~\mu\text{m}\).}
  \label{mapping}
\end{figure}
\begin{figure}[H]
    \centering
    \begin{subfigure}[tbp]{0.59\textwidth}
        \centering
        \includegraphics[width=\textwidth]{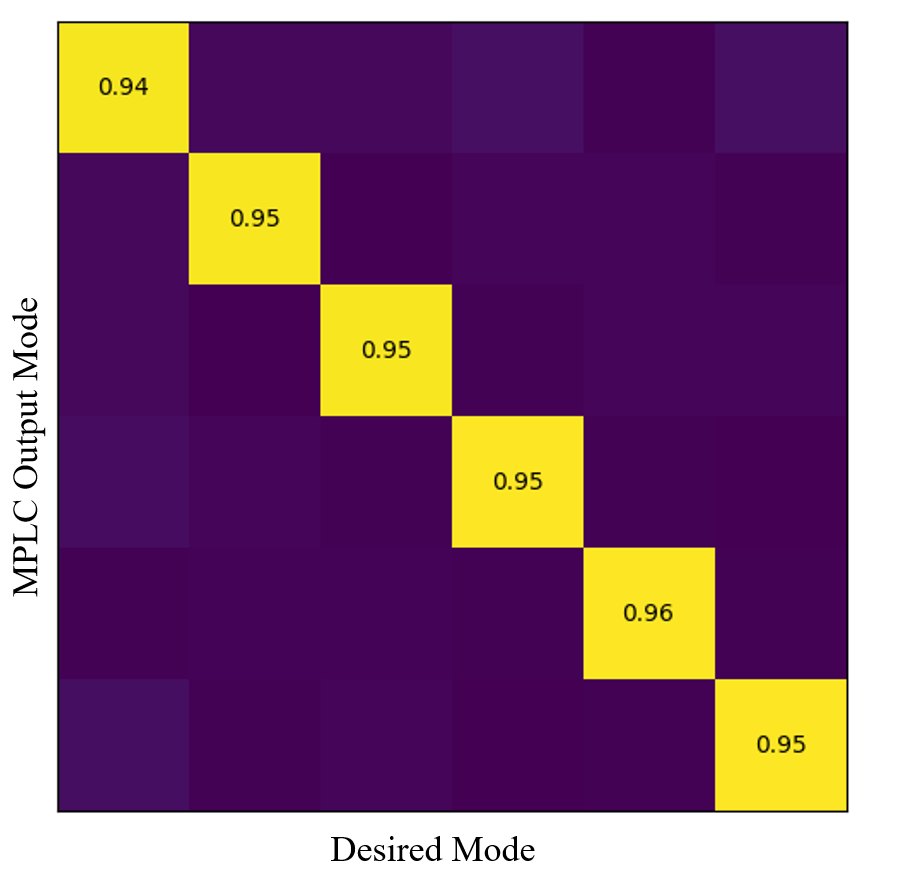}
    \end{subfigure}
    \caption[HG-array fidelity matrix]{Fidelity matrix for a six-mode MPLC transforming Gaussian beams ($w_0 =$ 100~$\mu$m) into co-located HG modes ($w_0 =$ 200~$\mu$m). }
    \label{fig:fidelty_matrix_1}
\end{figure}

We begin our evaluation with the core performance metric: the fidelity of the mode transformation implemented by the designed MPLC system. To assess the performance of our design, we simulated an MPLC system that maps six Gaussian beams into six co-located HG modes using 7 phase masks. At the input plane, each Gaussian beam had a waist of $w_0 = 100~\mu\mathrm{m}$, and their positions formed a triangular lattice. The output HG modes were designed to be co-located and have a waist of $w_0 = 200~\mu\mathrm{m}$. Fig.~\ref{mapping} illustrates this mapping: each numbered beam on the input corresponds to a numbered HG mode in the output, preserving their indexing. Triangular lattice geometry is especially relevant for SPADE implementations, where precise mode labeling and separation are critical.
The fidelity matrix for this design is presented in Fig. \ref{fig:fidelty_matrix_1}, where all diagonal elements lie between 0.94 and 0.96, indicating that at least 94\% of the input power is correctly routed into the intended HG mode and off-diagonal elements remain below 0.1, indicating low cross-talk. This high level of mode purity achieved with only seven phase masks demonstrates the effectiveness of the inverse‐design method for six-mode MPLC systems.

\subsubsection{Effect of Number of Phase Masks}
We next examine how the number of phase masks influences the achievable transformation fidelity, holding all other design parameters constant. In general, converting $N$ modes demands on the order of $N$ phase masks \cite{fontaine2022photonic}, yet no theory prescribes the absolute minimum number of planes for a given transformation. In practice, the required count depends on many variables, among them the total number of modes, the geometry of the input spot array, how the spatially separated input beams are assigned to target output modes, the specific output-mode set, the pixel pitch, and the spacing between mask planes.

\begin{figure}[h] 
  \centering
    \resizebox{1\textwidth}{!}{\input{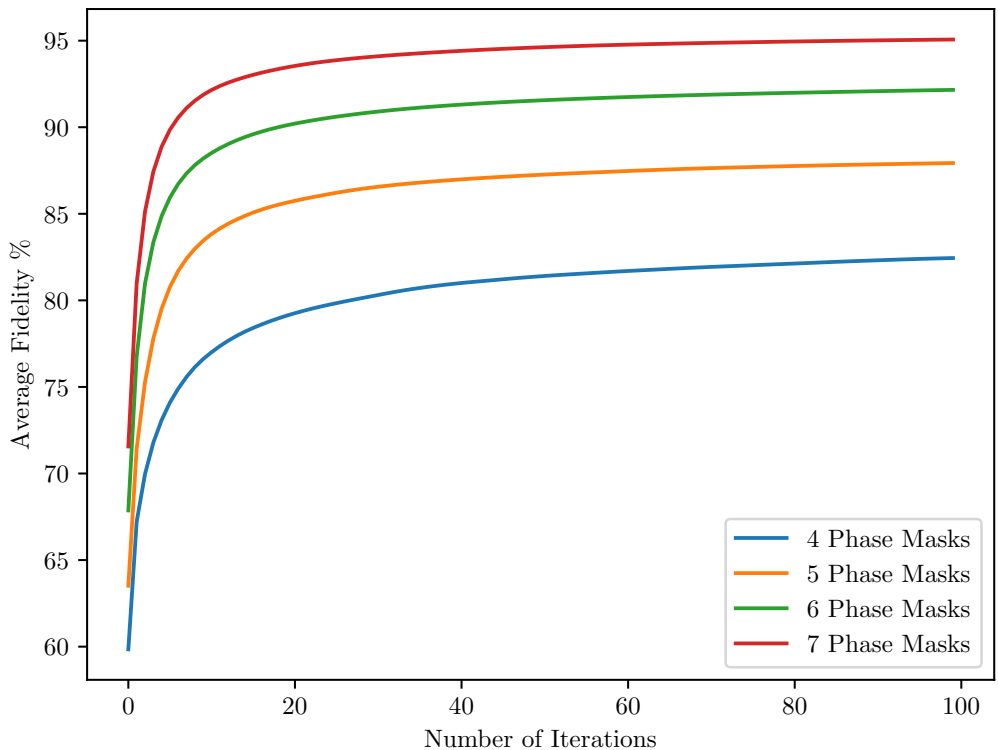}}
  \caption[Fidelity vs number of phase masks]{Average fidelity as a function of number of iterations for different number of phase masks. }
  \label{fig:num_masks_comp_1}
\end{figure}

Because each phase mask adds degrees of freedom for wavefront shaping, increasing the number of masks generally improves the achievable fidelity. To illustrate this effect, Fig.~\ref{fig:num_masks_comp_1} shows the evolution of the average fidelity $( \frac{1}{N}\sum_{i}F_{ii} )$ that is, the mean of overlap between each simulated output mode and its corresponding target HG mode over the course of 100 WFM iterations. 

The comparison includes configurations with 4, 5, 6, and 7 phase masks. In all cases, the input consisted of Gaussian beams with $w_0 = 100~\mu\mathrm{m}$, and the output consisted of HG modes with $w_0 = 200~\mu\mathrm{m}$. As expected, higher mask counts result in higher final fidelities, with the 7-mask system achieving over 95\% diagonal fidelity, compared to only 82\% for the 4-mask case. However, this improvement comes at a cost, as a higher number of masks increases the system's physical complexity, alignment sensitivity, and the computational time required for the WFM  algorithm.
\subsubsection{Tolerance to Input Misalignment}
\begin{figure}[H]
  \centering
  \begin{subfigure}[t]{0.04\textwidth}
    \textbf{(a)}
  \end{subfigure}
  \begin{subfigure}[t]{0.43\textwidth}
    \includegraphics[width=\linewidth, valign=t]{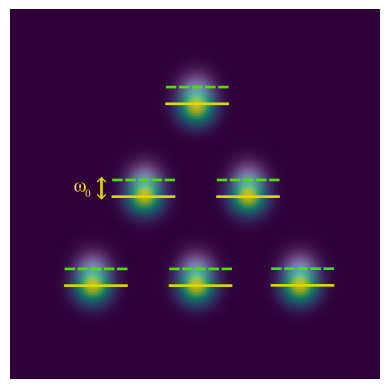}
  \end{subfigure}\hfill
  \begin{subfigure}[t]{0.04\textwidth}
    \textbf{(b)}
  \end{subfigure}
  \begin{subfigure}[t]{0.46\textwidth}
    \includegraphics[width=\linewidth, valign=t]{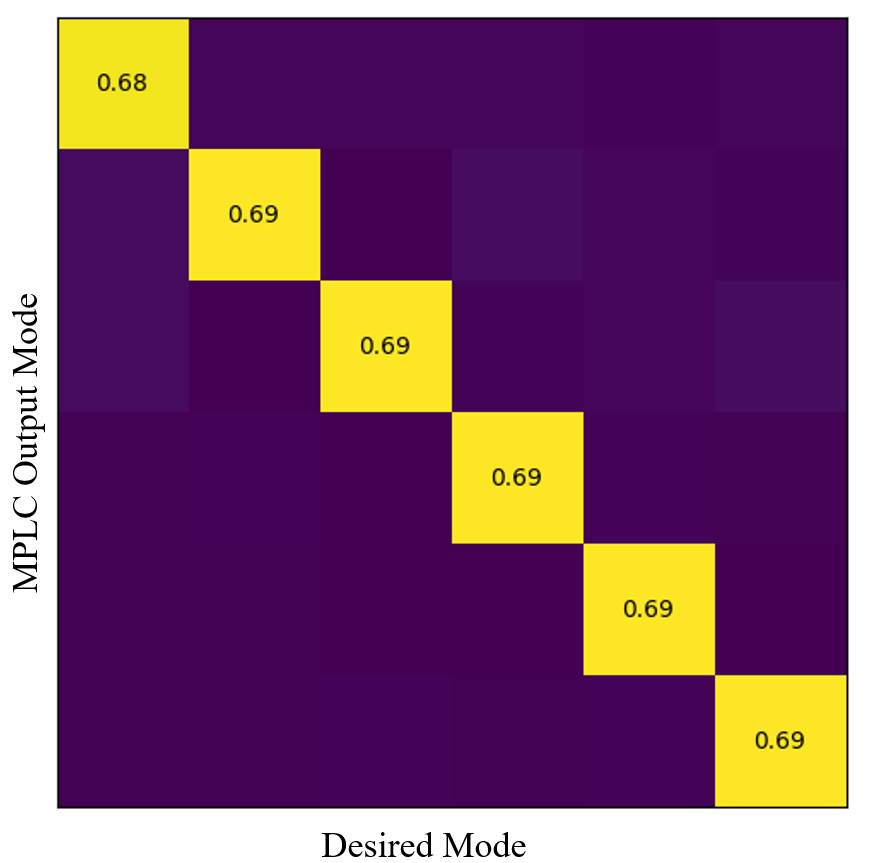}
  \end{subfigure}
  \caption[Beam-shift impact on fidelity]{Effect of input beam misalignment on fidelity. (a) shows the shifted Gaussian array (yellow) versus the original design positions (green). (b) presents the resulting fidelity matrix, illustrating degraded mode fidelity due to input misplacement.}
  \label{fig:beam_shift}
\end{figure}

\begin{figure}[H]
  \centering
  \begin{subfigure}[t]{0.03\textwidth}
    \textbf{(a)}
  \end{subfigure}
  \begin{subfigure}[t]{0.29\textwidth}
    \includegraphics[width=\linewidth, valign=t]{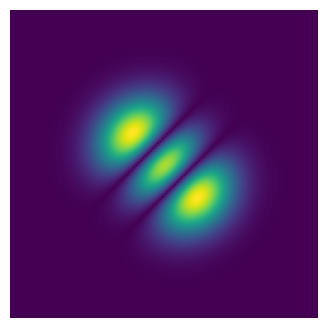}
    \label{fig:target}
  \end{subfigure}\hfill
  \begin{subfigure}[t]{0.03\textwidth}
    \textbf{(b)}
  \end{subfigure}
  \begin{subfigure}[t]{0.29\textwidth}
    \includegraphics[width=\linewidth, valign=t]{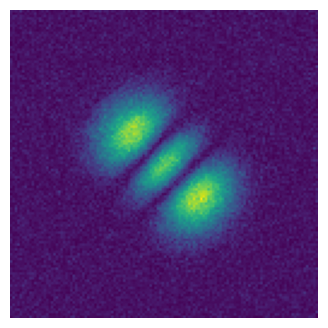}
    \label{fig:output}
  \end{subfigure}\hfill
  \begin{subfigure}[t]{0.03\textwidth}
    \textbf{(c)}
  \end{subfigure}
  \begin{subfigure}[t]{0.29\textwidth}
    \includegraphics[width=\linewidth, valign=t]{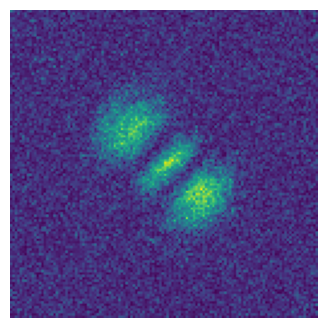}
    \label{fig:misoutput}
  \end{subfigure}
  \caption[HG$_{20}$ target vs misaligned output]{Comparison of the target HG$_{20}$ mode (a), the MPLC output for aligned inputs (b), and the distorted output resulting from a one-\(w_0\) downward shift of the input array (c). Misalignment causes contrast reduction, distorted lobes, and increased background intensity.}
  \label{fig:HG20_comparison}
\end{figure}

Precise alignment of the input beams is crucial for achieving high-fidelity mode conversion. To assess the sensitivity of the MPLC system to lateral misalignment, we simulated a scenario where the entire triangular array of six Gaussian beams was displaced downward by a distance equal to one beam waist ($w_0$). As can be seen in Fig.\ref{fig:beam_shift}, the original design coordinates are overlaid in green, while the shifted beam centers appear in yellow.
The resulting degradation in mode fidelity is visualized in Fig.~\ref{fig:HG20_comparison}, which shows (a) the ideal HG$_{20}$ mode, (b) the MPLC output for a perfectly aligned input, and (c) the distorted output after the downward shift. The misalignment leads to visibly distorted lobes, increased background intensity, and reduced contrast in the output mode.
Quantitatively, the fidelity matrix Fig.\ref{fig:beam_shift} reveals a significant drop in diagonal elements to approximately 0.68–0.69, compared to values around 0.94 in the aligned case. This substantial decrease indicates that even modest spatial displacement introduces mode mixing and cross-talk. These findings underscore the critical need for sub-beam-waist precision in beam placement to maintain high sorting fidelity in practical MPLC implementations.
\subsubsection {Sensitivity to Beam Size Variations}
\begin{figure}[H] 
  \centering
    \resizebox{0.8\textwidth}{!}{\input{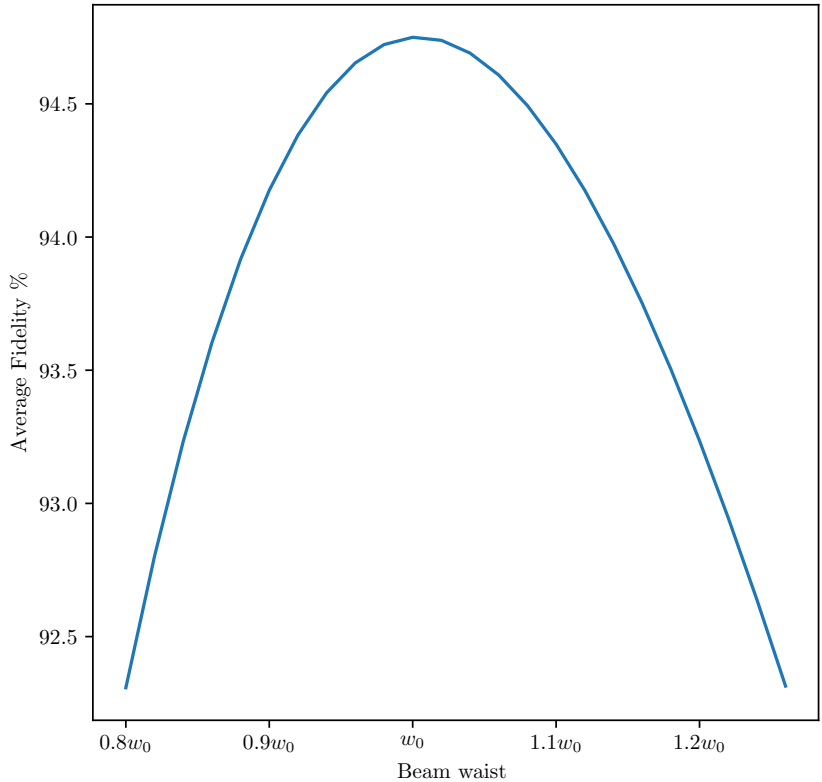}}
  \caption[Fidelity vs beam size]{Average fidelity of the MPLC system as a function of input beam waist.}
  \label{fig:beam_size_tolerance}
\end{figure}

Aside from lateral misalignment, errors in beam shaping can introduce discrepancies in waist size. We investigate how deviations from the design beam waist affect system performance. To evaluate the sensitivity of our seven‐mask MPLC design to the input‐beam size, we varied the $w_0$ of the triangular Gaussian array from 80 $\mu m$ to 125 $\mu m$, while keeping all other parameters fixed. Fig.~\ref{fig:beam_size_tolerance} shows the resulting average fidelity across all six modes as a function of input‐beam $w_0$. 
The conversion efficiency peaks at approximately 95\% when the input $w_0$ matches the design value of 100$\mu m$. As the beam size deviates from design value, the mode overlap degrades symmetrically, causing gradual drop in fidelity. Within a 10$\mu m$ window around the design $w_0$, the average fidelity remains above 94\%, indicating that the system tolerates moderate experimental beam‐shaping errors. Beyond this range, however, the fidelity declines more sharply, emphasizing the need for precise control of the input beam size in practical implementations.

\subsubsection{Sensitivity to Propagation Distance}
\begin{figure}[H] 
  \centering
    \resizebox{0.8\textwidth}{!}{\input{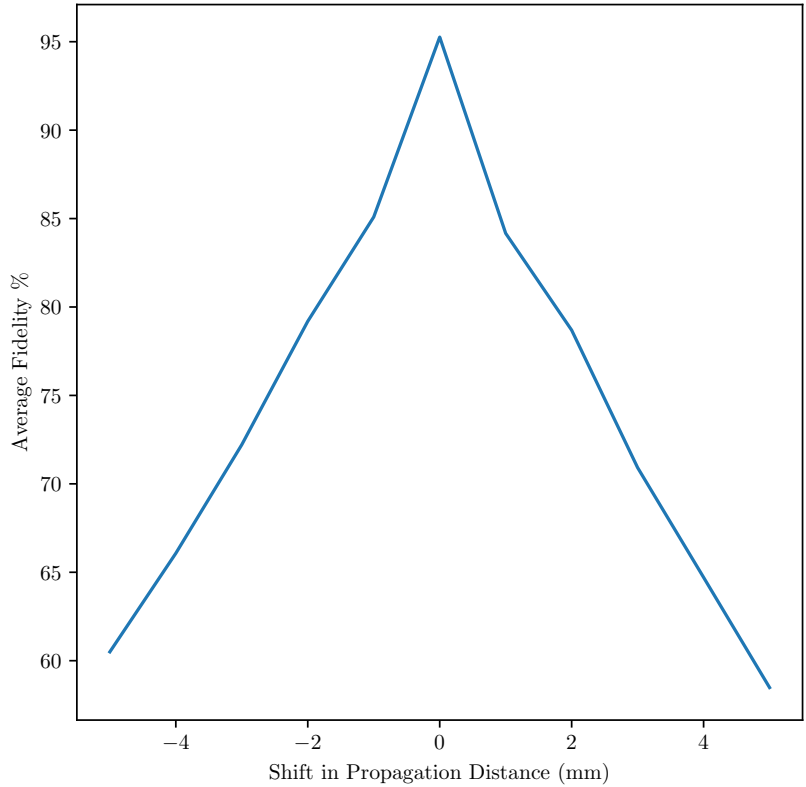}}
 \caption[Fidelity vs relative distance ]{Average fidelity as a function of deviation from the design phase mask–mirror spacing in the folded MPLC} 
    \label{fig:prop_distance}
\end{figure}

Accurate control over the physical distances between masks is critical. We now evaluate how axial deviations in propagation distance impacts conversion fidelity. To assess the sensitivity of the MPLC system to axial misalignment, we varied the propagation distance between the phase masks and the mirror in a folded configuration around its design value. Figure~\ref{fig:prop_distance} plots the average fidelity as a function of this deviation. The fidelity reaches its maximum when the propagation distance matches the design specification and falls off symmetrically as the distance increases or decreases. 
The simulation in Fig.~\ref{fig:prop_distance} models a folded configuration in which light propagates from the phase mask to a mirror placed $27\,\text{mm}$ away and then returns, yielding a total effective propagation distance of $54\,\text{mm}$.
This is remarkably close to the Rayleigh range of the beam:
\begin{align}
  z_R = \frac{\pi w_0^2}{\lambda} \approx \frac{\pi (100\,\mu\text{m})^2}{633\,\text{nm}} \approx 49.6\,\text{mm},
\end{align}
where we have used $w_0 = 100\,\mu\text{m}$ and $\lambda = 633\,\text{nm}$.
At this scale, even millimetre-level deviations lead to significant Gouy-phase shifts between the HG modes, which strongly impact the fidelity of the mode conversion. These results underscore the importance of precise axial alignment in MPLC systems.

\subsubsection{Spectral Performance Optimization}
\begin{figure}[H]
  \centering
  \begin{subfigure}[t]{0.04\textwidth}
    \textbf{(a)}
  \end{subfigure}
  \begin{subfigure}[t]{0.45\textwidth}
    \includegraphics[width=\linewidth, valign=t]{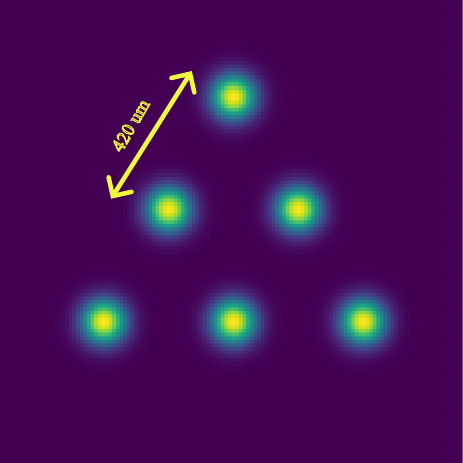}
  \end{subfigure}\hfill
  \begin{subfigure}[t]{0.04\textwidth}
    \textbf{(b)}
  \end{subfigure}
  \begin{subfigure}[t]{0.45\textwidth}
    \includegraphics[width=\linewidth, valign=t]{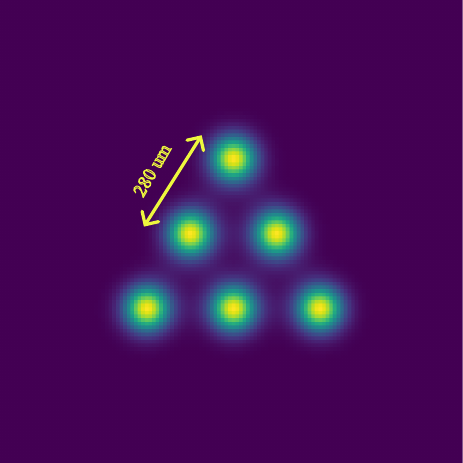}
  \end{subfigure}
  \caption[Reduced beam-spacing design]{Input configuration before (a) and after (b) reducing the beam spacing from 420~\(\mu\)m to 280~\(\mu\)m. This adjustment allows for better utilization of the spatial light modulator’s active area and improves mode separation.}
  \label{more_space}
\end{figure}

Lastly, we assess the wavelength sensitivity of our MPLC design, which was optimized for 633 nm, by examining fidelity over a range of operating wavelengths. To explore this, we simulated the fidelity of mode conversion as a function of wavelength for two different input-beam spacings: $420~\mu\mathrm{m}$ and $280~\mu\mathrm{m}$. This adjustment is visually represented in Fig. \ref{more_space}, which compares the original and reduced beam spacing configurations.

\begin{figure}[tbp] 
  \centering
    \resizebox{1\textwidth}{!}{\input{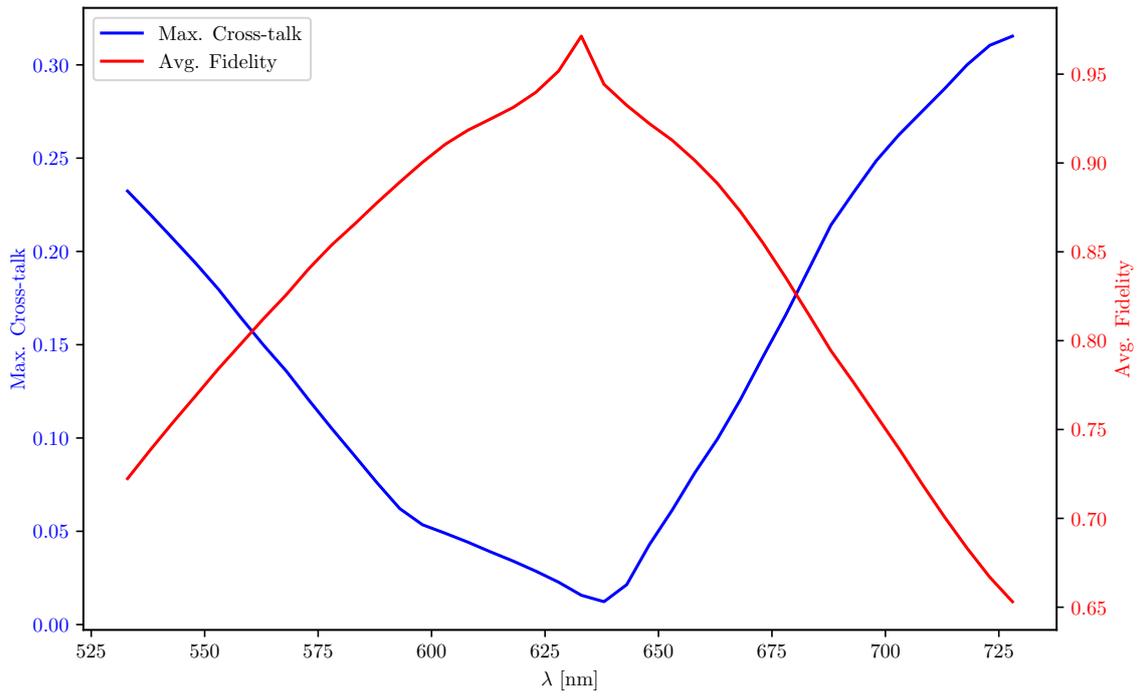}}
  \caption[Spectral fidelity and cross-talk]{Spectral performance of the six-mode MPLC for a 280~$\mu$m. beam spacing. Red: average fidelity; blue: maximum cross-talk. Fidelity degrades at off-design wavelengths, indicating sensitivity to spectral detuning.}
  \label{fig:broad}
\end{figure}

With the original spacing of $420~\mu\mathrm{m}$, fidelity declined rapidly at off-design wavelengths, and cross-talk increased. By contrast, as can be seen in Fig. \ref{fig:broad} the reduced spacing of $280~\mu\mathrm{m}$ yielded a modest improvement in spectral stability: the average fidelity remained above 90\% across the 600–650 nm range, and the maximum cross-talk remained below 10\%. The underlying mechanism for this improvement is not definitively established; it may relate to a more spatially uniform phase pattern across the input aperture or to reduced sensitivity to wavelength-dependent phase distortions due to overlapping mode fields, but further analysis is required to isolate the contributing factors. 

Although this example shows that spectral robustness can be improved through geometric adjustment of the input array, the result is design-specific and does not generalize to arbitrary configurations. Moreover, spectral performance is highly sensitive to the mode structure and beam overlap. As such, additional simulations would be needed to draw broader conclusions. This case study suggests that high-fidelity conversion is achievable across a moderate spectral bandwidth, provided careful attention is paid to input geometry and wavelength detuning.

In summary, our simulations demonstrate that MPLC systems designed with the WFM algorithm can achieve high-fidelity spatial mode transformations using a modest number of phase masks. While the system is sensitive to limitations of alignment, beam waist, and propagation distance, the robustness observed within tolerable experimental ranges suggests practical viability. Moreover, the observed improvements in spectral performance through beam-spacing optimization highlight the potential for further refinement in future designs.

\chapter{Experimental Results}
This chapter details the experimental realization and performance characterization of the MPLC system introduced in previous chapters. While earlier sections focused on theoretical design and numerical simulation, here we implement the proposed configuration using a spatial light modulator and assess its mode transformation capabilities in practice. The aim is to validate the simulated wavefront-matching strategy under controlled laboratory conditions and to experimentally quantify the fidelity of spatial-mode sorting in a realistic optical setup.

\section{Experimental Setup}
\begin{figure}[ht]
    \centering
    \begin{subfigure}[h]{1\textwidth}
        \centering
        \includegraphics[width=\textwidth]{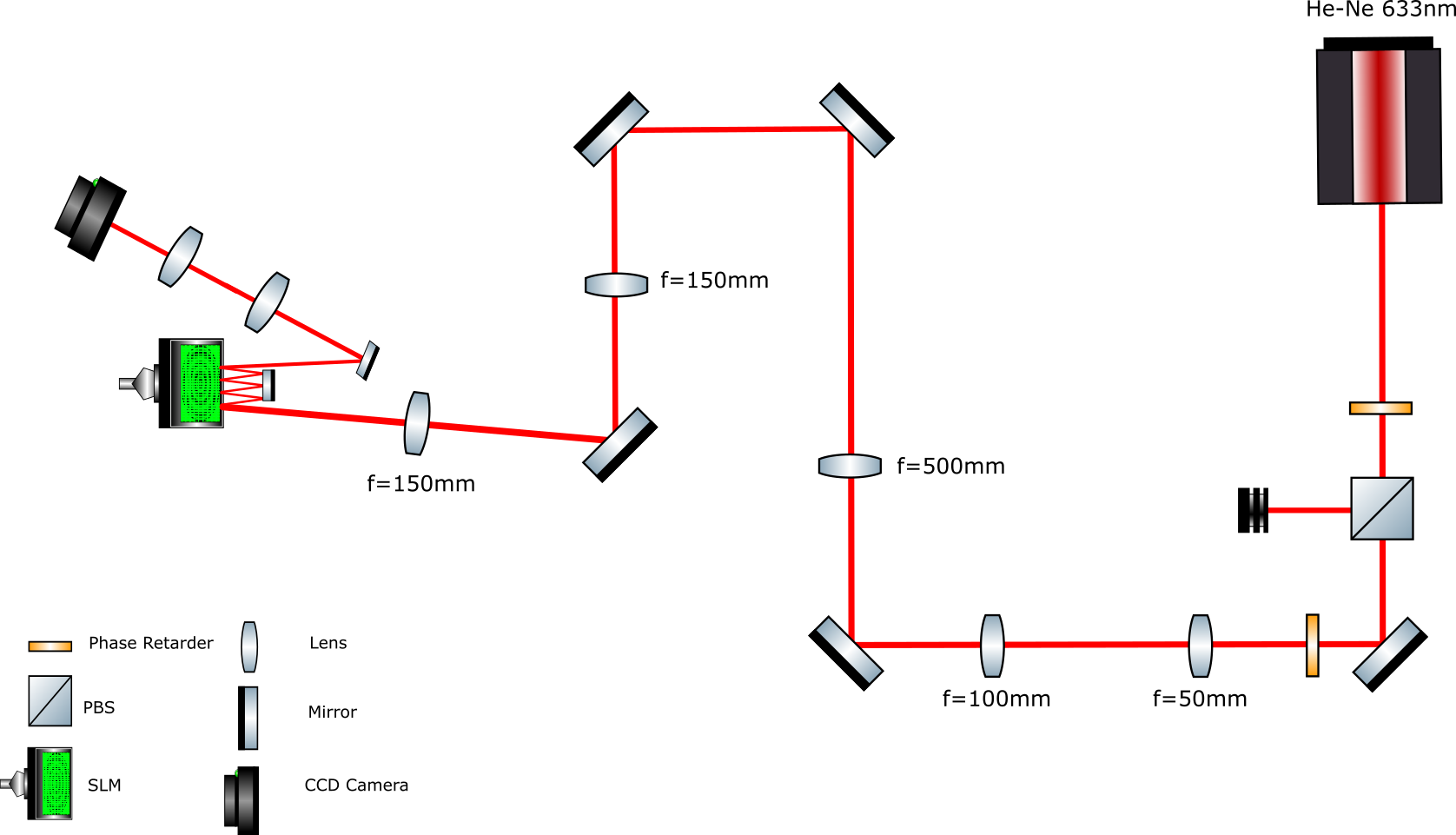}
    \end{subfigure}  
    \caption[Experimental MPLC setup (reflective design)]{Schematic of the experimental MPLC setup using a reflective configuration. }
    \label{setup}
\end{figure}

The design of the MPLC setup used in this work is depicted in Fig.\ref{setup}: To illuminate the system, we used a He-Ne (633 nm) coherent laser source (10 mW Linos Optics). In order to ensure that the incident beam was linearly polarized, we implemented a polarization control scheme consisting of two adjustable half-wave plates with a polarizing beam splitter (PBS) positioned between them. This configuration serves two primary purposes. First, it guarantees linear and uniform polarization of the beam, which is critical for the effective operation of a spatial light modulator (Holoeye PLUTO 1). Such polarization ensures a consistent phase response across the beam profile, because the spatial light modulator (SLM) imparts phase modulation effectively only to light polarized along its active axis typically linear polarization aligned with the liquid crystal director. Any deviation from this polarization leads to reduced phase modulation depth and spatial non-uniformities. Second, the setup enables convenient control of the beam intensity by rotating the half-wave plate placed before PBS, allowing for precise tuning of optical power without altering the beam path or coherence properties.

Once the light is linearly and uniformly polarized the beam is subsequently directed through a telescope setup designed to reduce its beam size. The telescope comprises a pair of lenses with focal lengths of 50 mm and 100 mm, respectively, configured in a 4$f$ arrangement. This configuration results in a demagnification factor of 2, effectively reducing the beam diameter by half while preserving its collimation.

After spatial demagnification, the beam is focused using a lens with a focal length of 500 mm. To finely control the angle and position of the input beam at the first phase mask of the MPLC system, two mirrors are placed at the image and Fourier planes of the optical path, forming a standard 4f configuration with a pair of 150 mm focal length lenses. This arrangement enables precise independent adjustment of  angular incidence and lateral displacement, which are critical for mode alignment and optimal transformation efficiency. The MPLC mirror itself is positioned in front of the SLM at a distance of 27 mm, a value determined through numerical simulations to be optimal for the spatial mode profiles and transformations targeted in this experiment.
\begin{figure}[H]
    \centering
    \begin{subfigure}[h]{0.6\textwidth}
        \centering
        \includegraphics[width=1\textwidth]{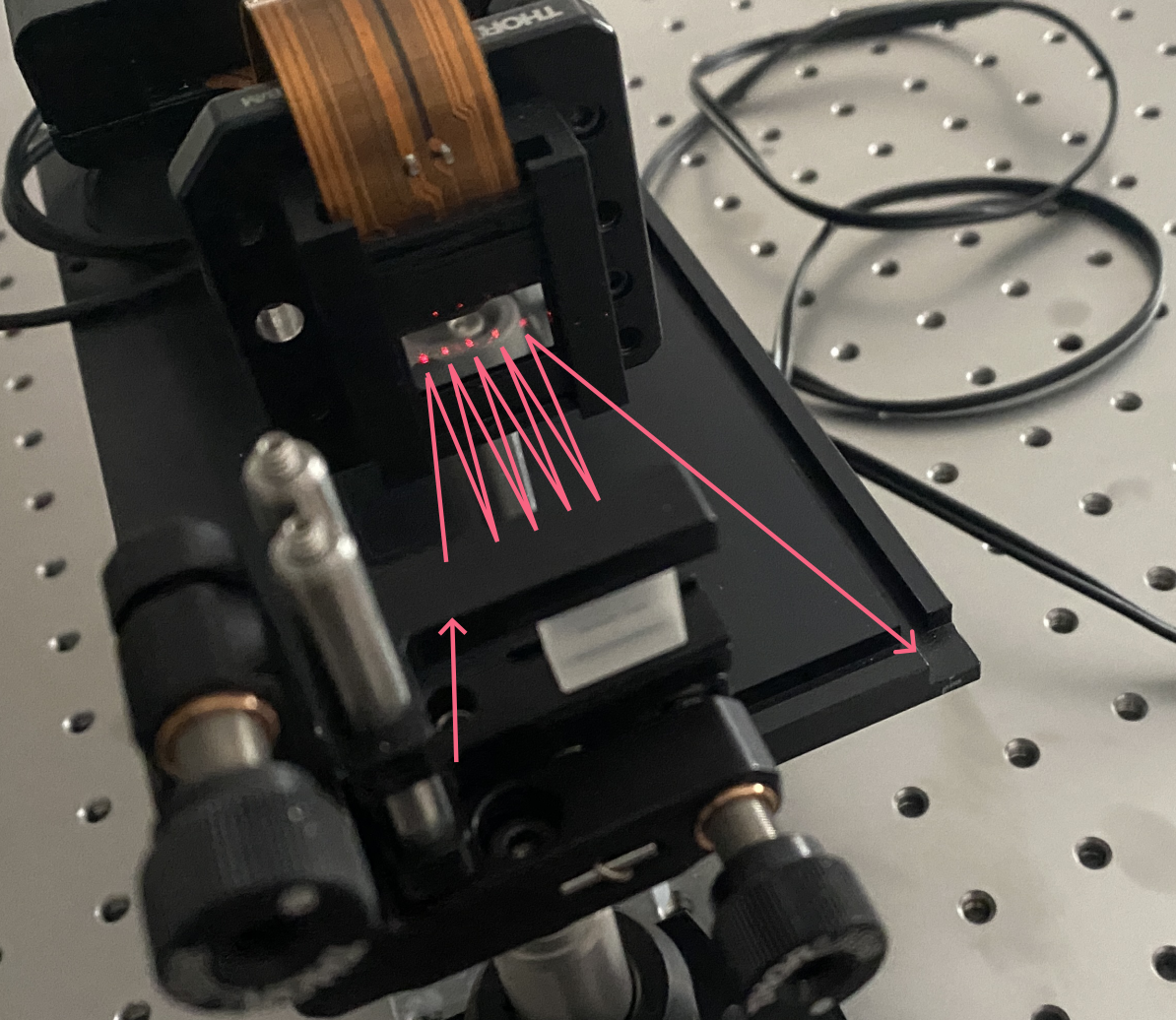}
    \end{subfigure}  
    \caption[Five SLM beam reflections]{Five successive beam reflections from the spatial light modulator}
    \label{5_bounce}
\end{figure}
A key challenge during the implementation of the folded cavity design is to avoid clipping the beam at the edges of the SLM or mirrors. Because the setup is extremely compact and the inter-component spacing is minimal, the beam’s path runs very close to the mounts and optical elements. Ensuring that the entire beam reaches the CCD camera is essential for reliable spatial-mode sorting. By adjusting the beam’s angular incidence and lateral displacement, we achieved five successive beam reflections from the SLM. After the fifth bounce, the output beam was focused onto a CCD camera for spatial intensity measurements and mode analysis. The successful alignment of five successive reflections is shown in
Fig. \ref{5_bounce}.

\subsection{Locating Beams Positions}

After implementing five successive beam reflections on the SLM, the next critical step was to determine the precise pixel coordinates at which each reflected beam strikes the SLM surface. As shown in simulations, accurate beam localization is essential for centering the phase masks and ensuring high-fidelity mode transformations. As discussed, even a displacement of the order of the beam waist leads to a significant fidelity drop, underscoring the need for precise spatial alignment.
\begin{figure}[h!]
    \centering
    \begin{subfigure}[h]{1\textwidth}
        \centering
        \includegraphics[width=\textwidth]{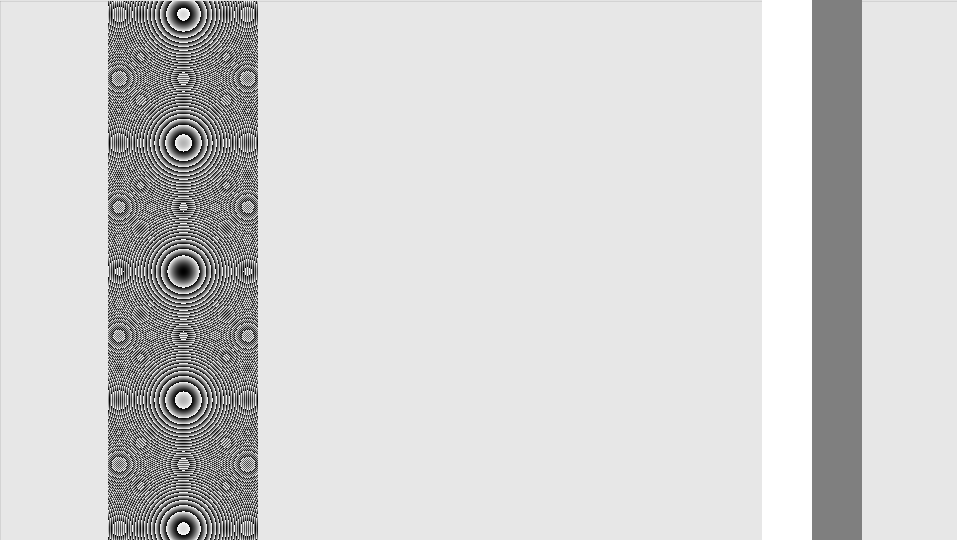}
    \end{subfigure}  
    \caption[Phase-step mask for beam-centroid alignment]{Phase‐strap mask along with fresnel lens displayed on the SLM for beam‐centroid alignment. The central vertical band shows the binary $\pi$–step pattern.}
    \label{phase_strip}
\end{figure}

To simplify real-time adjustment of the SLM phase patterns and facilitate the beam-centering procedure, we developed a custom graphical user interface (GUI). The SLM (Holoeye PLUTO 1), with a pixel pitch of 8 $\mu m$, was treated as a secondary monitor, and the GUI enabled interactive control of the phase mask image in pixel steps. Live feedback from the imaging camera was displayed on a computer screen alongside the GUI controls, allowing rapid iterations.
\begin{figure}[tb]
  \centering
  \begin{subfigure}[t]{0.04\textwidth}
    \textbf{(a)}
  \end{subfigure}
  \begin{subfigure}[t]{0.45\textwidth}
    \includegraphics[width=\linewidth, valign=t]{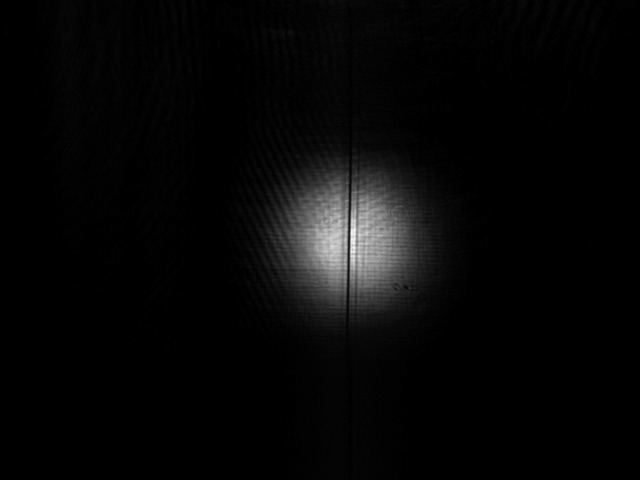}
  \end{subfigure}\hfill
  \begin{subfigure}[t]{0.04\textwidth}
    \textbf{(b)}
  \end{subfigure}
  \begin{subfigure}[t]{0.45\textwidth}
    \includegraphics[width=\linewidth, valign=t]{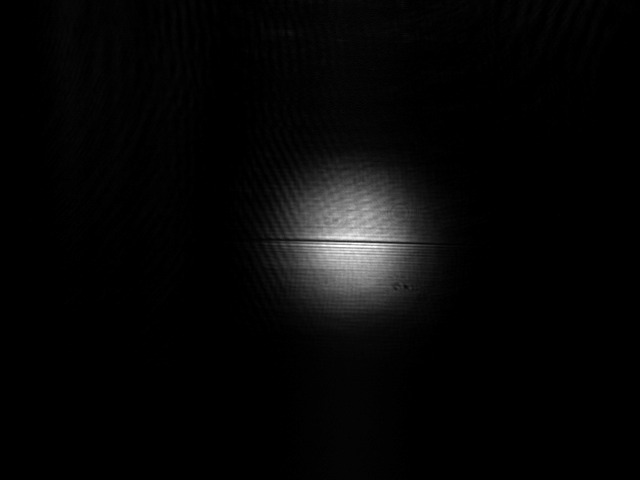}
  \end{subfigure}
  \caption[Beam-centroid alignment using bisection mask]{Camera images showing the Gaussian input beam intersected by the holographic mask in (a) horizontal and (b) vertical orientations. The dark fringe at the phase discontinuity reveals the beam centroid, enabling precise alignment along the \(x\)- and \(y\)-axes for accurate phase modulation and mode preparation.}
  \label{bisect}
\end{figure}
We employed a binary $\pi$–step phase mask to locate the beam centroid in both transverse directions. A sharp intensity null is created when the $\pi$–step discontinuity intersects the beam, providing a high-contrast marker.  An example of $\pi$–step phase mask, in conjunction with a Fresnel lens, is shown in Fig. \ref{phase_strip}. This pattern was first aligned horizontally and then vertically to identify the x and y centroids. As can be seen in Fig.\ref{bisect} the optimal position was defined as the pixel row or column where the dark fringe symmetrically bisected the Gaussian envelope and minimized the central intensity.

Although alternative methods such as centroid fitting or 2D Gaussian fitting could be employed for beam localization, the $\pi$–step technique proved effective in practice due to its visual immediacy and interactive control. This method enabled the localization of beam centers with an accuracy of a few pixels, where each pixel corresponds to $8~\mu\text{m}$, which is crucial for centering the computed phase masks at each reflection point.

\subsection{Generating HG modes}
Before attempting full mode sorting, we first validated the functionality of the WFM algorithm through a minimal experimental test. The goal was to verify whether WFM can generate a target HG mode from a Gaussian input using the minimal configuration of two phase masks, which is the smallest setup that allows beam shaping via WFM.

\begin{figure}[H]
  \centering
  \begin{subfigure}[t]{0.03\textwidth}
    \textbf{(a)}
  \end{subfigure}
  \begin{subfigure}[t]{0.27\textwidth}
    \includegraphics[width=\linewidth, valign=t]{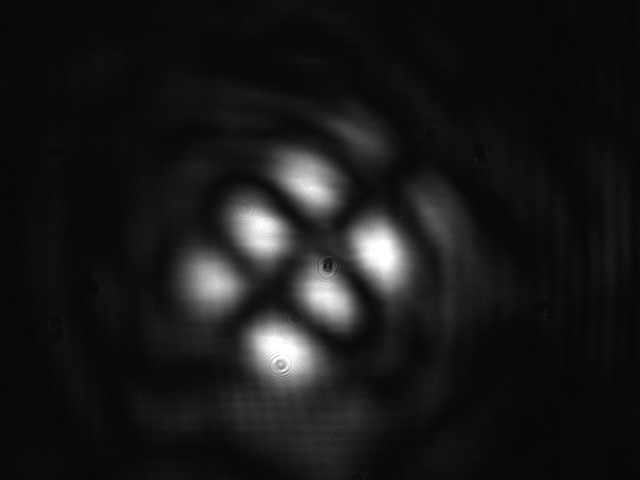}
    \label{fig:HG32_raw}
  \end{subfigure}\hfill
  \begin{subfigure}[t]{0.04\textwidth}
    \textbf{(b)}
  \end{subfigure}
  \begin{subfigure}[t]{0.27\textwidth}
    \includegraphics[width=\linewidth, valign=t]{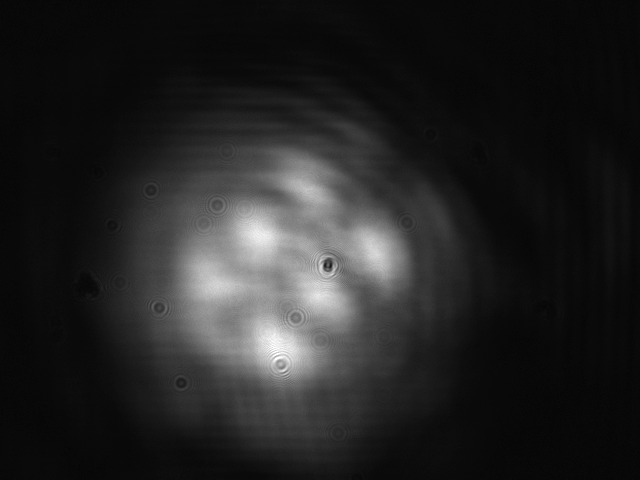}
    \label{fig:HG32_with_bg}
  \end{subfigure}\hfill
  \begin{subfigure}[t]{0.04\textwidth}
    \textbf{(c)}
  \end{subfigure}
  \begin{subfigure}[t]{0.27\textwidth}
    \includegraphics[width=\linewidth, valign=t]{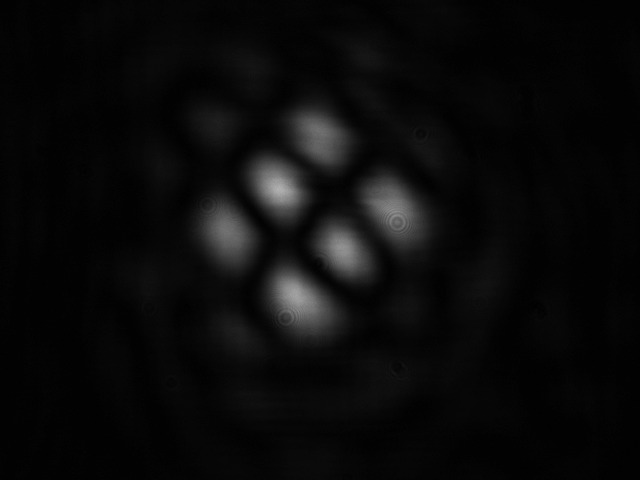}
    \label{fig:HG32_final}
  \end{subfigure}
  \caption[HG$_{32}$ generation: before and after blazing]{Generation of the HG$_{32}$ mode before and after polarization-order separation: (a) WFM output showing the HG$_{32}$ structure with a bright zeroth-order Gaussian background; (b) overlay of the modulated mode and the unmodulated Gaussian beam, highlighting contrast degradation; (c) improved HG$_{32}$ profile after isolating the first diffraction order via a linear grating and reapplying the second phase mask.}
  \label{fig:HG32_comparison}
\end{figure}

As a proof-of-concept, we selected the HG$_{32}$ mode due to its  spatial structure and asymmetry, which makes it visually distinct and sensitive to phase errors. Using the procedure outlined in Section \ref{WFM}, we applied the WFM algorithm to compute a pair of phase profiles that transform Gaussian beam into the desired higher-order mode. The input beam was modeled with a waist of $100~\mu\text{m}$ to match the experimental conditions, while the output mode was designed to have a waist of $200~\mu\text{m}$

The two generated phase masks were sequentially displayed at the corresponding pixel positions on the SLM. Upon illumination with the Gaussian beam, the resulting intensity pattern qualitatively matched the expected HG$_{32}$  mode structure (see Fig.\ref{fig:HG32_comparison}a), confirming the feasibility of mode shaping via WFM with only two phase masks.

However, the observed output also included a strong zeroth-order background, arising from imperfect phase modulation by the SLM (see Fig.\ref{fig:HG32_comparison}b). This unmodulated component is characteristic of spatial light modulators and can interfere with the mode structure, especially for higher-order patterns.
\begin{figure}[h!]
    \centering
    \begin{subfigure}[h]{1\textwidth}
        \centering
        \includegraphics[width=\textwidth]{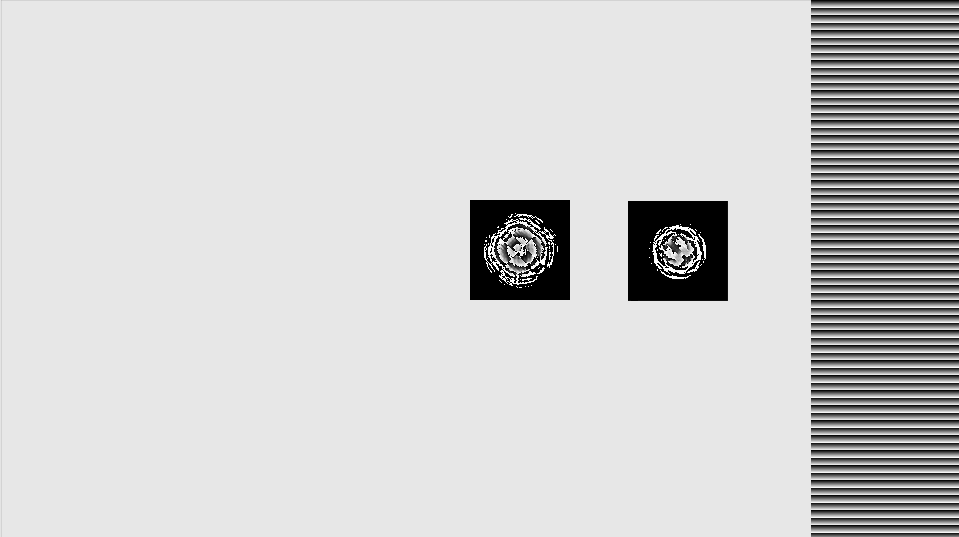}
    \end{subfigure}  
    \caption[Blazed grating and phase masks for HG$_{32}$ mode generation]{Blazed grating along with calculated phase-masks displayed on the SLM to produce HG$_{32}$ mode.}
    \label{blazed}
\end{figure}
To suppress the unwanted, unmodulated zeroth‐order reflection from the SLM, we first placed a linear blazed grating onto the region where the incoming Gaussian beam initially hits the SLM. This grating steers the modulated light into the first diffraction order, spatially separating it from the residual zeroth order. The diffracted, now well‑isolated beam is then redirected onto the subsequent phase‑mask regions. Fig.~\ref{blazed} shows a representative phase mask containing both the blazed grating and the phase masks pattern. As a result, the final output (Fig.~\ref{fig:HG32_comparison}c) exhibits markedly cleaner lobes and higher contrast, more closely matching the ideal HG$_{32}$ field distribution.

This minimal configuration served as a practical check for the WFM method and confirmed that even with only two phase planes, high-order HG modes can be generated with reasonable fidelity using numerically optimized phase masks. These results justify the more complex configurations used for full spatial-mode sorting in the subsequent section.

\subsection{Mode Sorting}

Having verified the basic functionality of WFM algorithm through single-mode transformation, we proceeded to test its ability to perform SPADE in a three-mode case. While the simulations in Chapter 4 considered a six-mode configuration of Gaussian arrays, here we deliberately start from a minimal and more controllable example to validate the MPLC design under experimental conditions.

\begin{figure}[h!]
    \centering
    \begin{subfigure}[h]{0.6\textwidth}
        \centering
        \includegraphics[width=\textwidth]{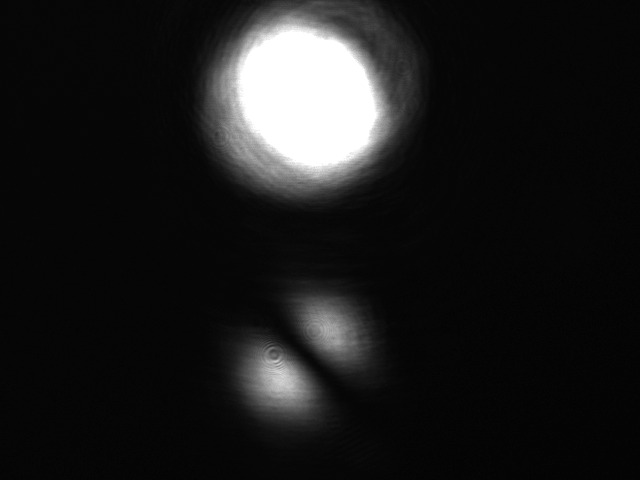}
    \end{subfigure} 
   
    \caption[Computer-generated hologram with unmodulated beam]{HG$_{01}$ mode produced with computer generated hologram (bottom) and unmodulated beam on top}
     \label{holo_unmodulated}
\end{figure}

To precisely control the input modes and maximize the utility of our SLM platform, we choose to generate the input HG modes themselves using a computer-generated hologram (CGH) at the first beam location, rather than WFM method. This method allows for efficient generation of complex optical fields with independently defined amplitude and phase profiles using phase-only modulation.

To implement CGH, we use a method as described in~\cite{Arrizon2007}. In our implementation, the goal is to generate an arbitrary complex optical field
\begin{align}
s(x, y) = a(x, y) \, e^{i\phi(x, y)}.
\end{align}
where \( a(x, y) \) is the amplitude and \( \phi(x, y) \) is the phase. Since our SLM is phase-only, we use a CGH to encode this complex field into a phase modulation function.
The transmittance of the CGH is defined as:
\begin{align}
h(x, y) = \exp[i\Theta(a, \phi)].
\end{align}
where \( \Theta(a, \phi) \) is a phase function specifically designed so that the first-order diffraction term of the hologram contains the desired field \( s(x, y) \).
To achieve this, we express the CGH transmittance as a Fourier series:
\begin{align}
 h(x, y)= \sum_{q=-\infty}^{\infty} h_q(x, y),\quad h_q(x, y) = c_q(a) \, e^{i q \phi(x, y)}.
\end{align}
where \( c_q(a) \) are the Fourier coefficients. The desired field appears in the first-order diffraction term \( h_1(x, y) \) if the coefficient \( c_1(a) \) satisfies:
\begin{align}
c_1(a) = A \, a(x, y),
\end{align}
where \( A \leq 1 \) is a constant encoding efficiency factor.
In our system, we use a Type-3 CGH encoding scheme as described in~\cite{Arrizon2007}, where the phase modulation is given by:
\begin{align}
\Theta(a, \phi) = f(a) \, \sin(\phi),
\end{align}
leading to Fourier coefficients defined in terms of Bessel functions:
\begin{align}
c_q(a) = J_q[f(a)],
\end{align}
where \( J_q \) is the Bessel function of the first kind of order \( q \). The condition
\begin{align}
J_1[f(a)] = A \, a(x, y).
\end{align}
is met by numerically inverting the Bessel function to determine \( f(a) \). This ensures that the first-order diffraction term reconstructs the complex field with high fidelity. For our design, the optimal efficiency is \( A \approx 0.5819 \), which corresponds to a maximum phase modulation depth of approximately \( 1.17\pi \) within the range of standard SLM.

An additional benefit of this encoding is that the Bessel function structure naturally suppresses higher diffraction orders, improving the purity of the generated mode. As a result, CGHs are particularly well-suited for initializing HG modes in our MPLC-based mode-sorting experiments, even when using SLMs with limited spatial resolution. An image of a generated mode alongside the residual unmodulated beam can be seen in Fig. \ref{holo_unmodulated}.

\begin{figure}[tb]
  \centering
  \begin{subfigure}[t]{0.04\textwidth}
    \textbf{(a)}
  \end{subfigure}
  \begin{subfigure}[t]{0.28\textwidth}
    \includegraphics[width=\linewidth, valign=t]{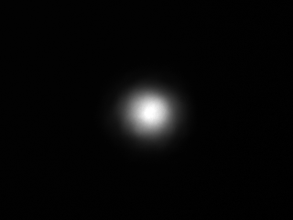}
  \end{subfigure}\hfill
  \begin{subfigure}[t]{0.04\textwidth}
    \textbf{(b)}
  \end{subfigure}
  \begin{subfigure}[t]{0.28\textwidth}
    \includegraphics[width=\linewidth, valign=t]{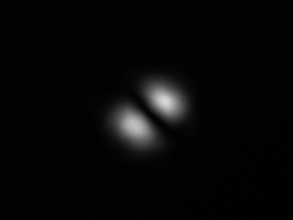}
  \end{subfigure}\hfill
  \begin{subfigure}[t]{0.04\textwidth}
    \textbf{(c)}
  \end{subfigure}
  \begin{subfigure}[t]{0.28\textwidth}
    \includegraphics[width=\linewidth, valign=t]{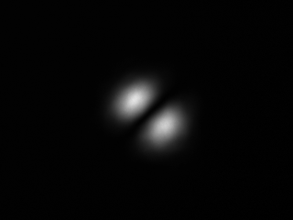}
  \end{subfigure}
  \caption[HG$_{00}$, HG$_{10}$, HG$_{01}$ modes generated via CGH]{HG modes generated via CGH displayed on the SLM: (a) HG$_{00}$, (b) HG$_{10}$, and (c) HG$_{01}$. The beam waist in each case was approximately 150~\(\mu\)m.}
  \label{holo_modes}
\end{figure}

\begin{figure}[H]
    \centering
    \begin{subfigure}[h]{1\textwidth}
        \centering
        \includegraphics[width=\textwidth]{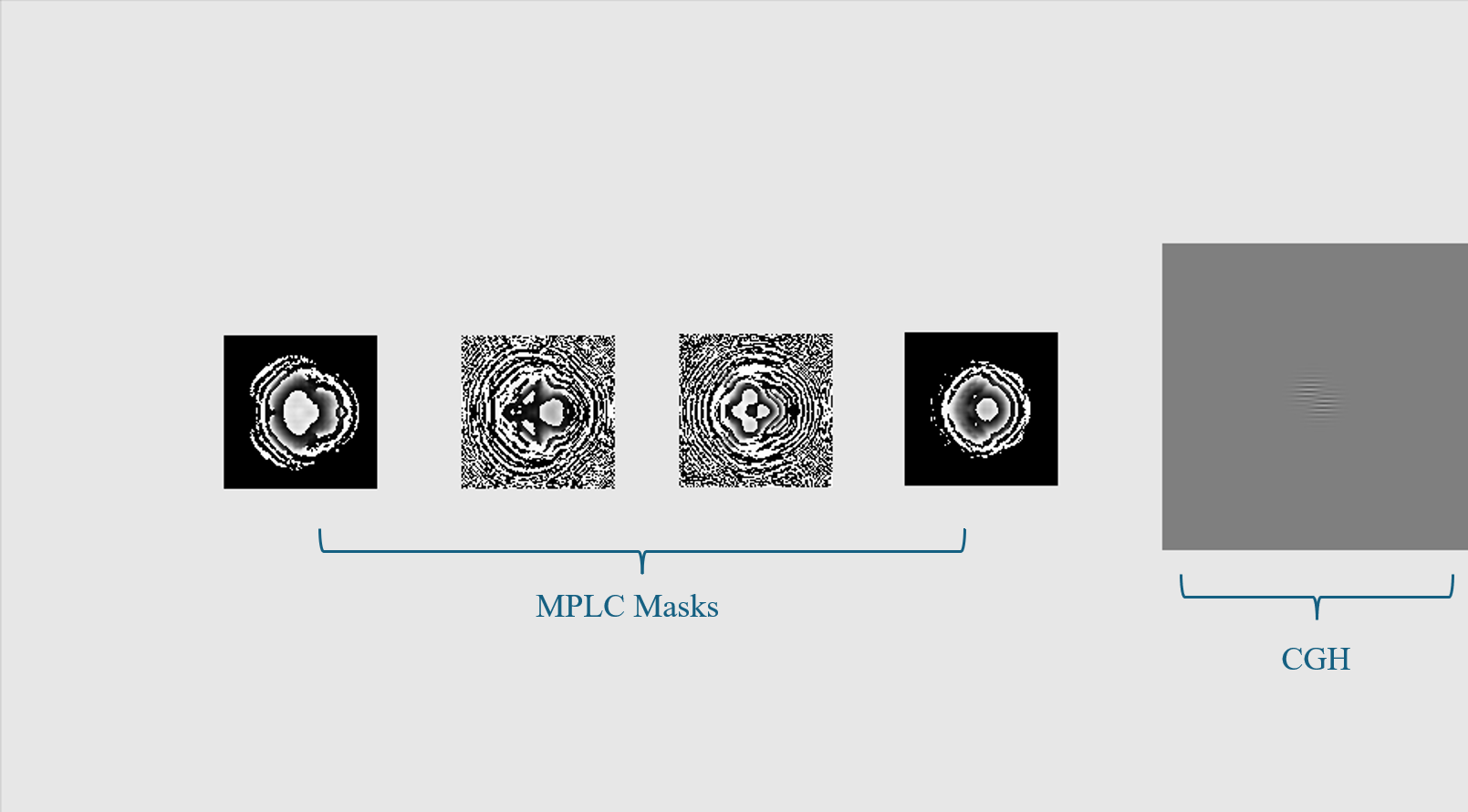}
    \end{subfigure}  
    \caption[Four-mask SLM layout for mode sorting]{HG mode generation CGH and 4 phase masks displayed on the SLM for mode sorting}
    \label{4mask_setup}
\end{figure}
In this experiment, the input beam was a Gaussian beam with a waist of 100~$\mu$m, achieved through careful alignment of the optical setup using a He-Ne laser as described in Section~5.1. To prepare the desired HG modes, we applied the CGH at the initial reflection point on the SLM. This allowed us to generate higher-order modes (HG$_{00}$, HG$_{10}$, HG$_{01}$) with a beam waist of 150 $\mu m$, chosen for its spatial compatibility with the MPLC. Fig. \ref{holo_modes} displays the HG$_{00}$, HG$_{10}$, HG$_{01}$ modes successfully generated using this CGH technique.

For this experiment, we designed a four-phase-mask MPLC configuration using the WFM algorithm, with parameters chosen to match the physical setup. The input mode was assumed to have a beam waist of 150$\mu m$, consistent with the CGH HG modes, while the output was targeted to a beam waist of 80$\mu m$, to enable tighter spatial separation of the sorted modes. As shown in Fig. \ref{4mask_setup}, the experimental setup implements a four phase mask design along with single CGH mask on a single SLM. The distance between successive phase planes in the simulation was set to 27 mm, aligning with the actual spacing between the SLM and the mirror in the folded experimental geometry, as described earlier.
\begin{figure}[tb]
  \centering
  \begin{subfigure}[t]{0.04\textwidth}
    \textbf{(a)}
  \end{subfigure}
  \begin{subfigure}[t]{0.27\textwidth}
    \includegraphics[width=\linewidth, valign=t]{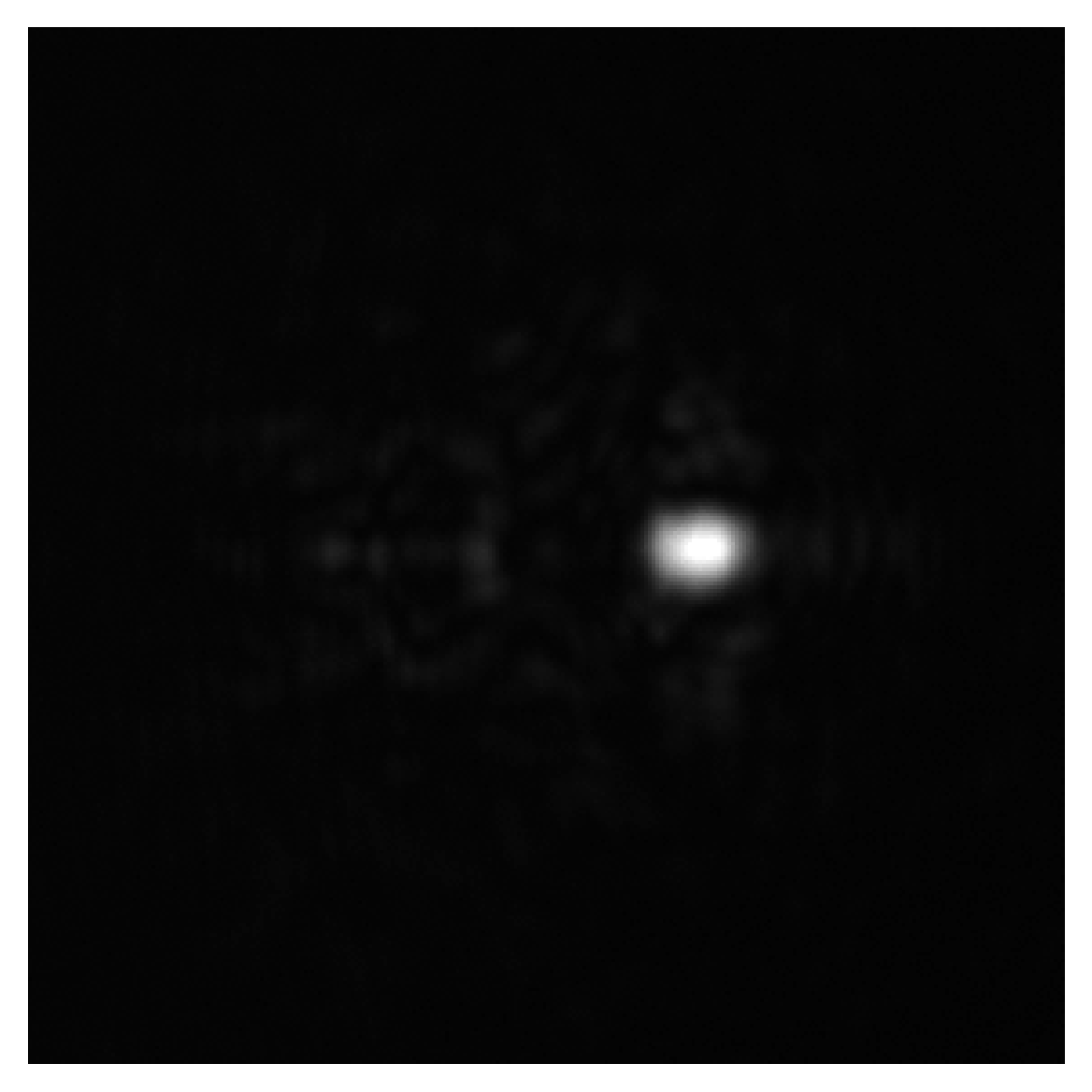}
  \end{subfigure}\hfill
  \begin{subfigure}[t]{0.04\textwidth}
    \textbf{(b)}
  \end{subfigure}
  \begin{subfigure}[t]{0.27\textwidth}
    \includegraphics[width=\linewidth, valign=t]{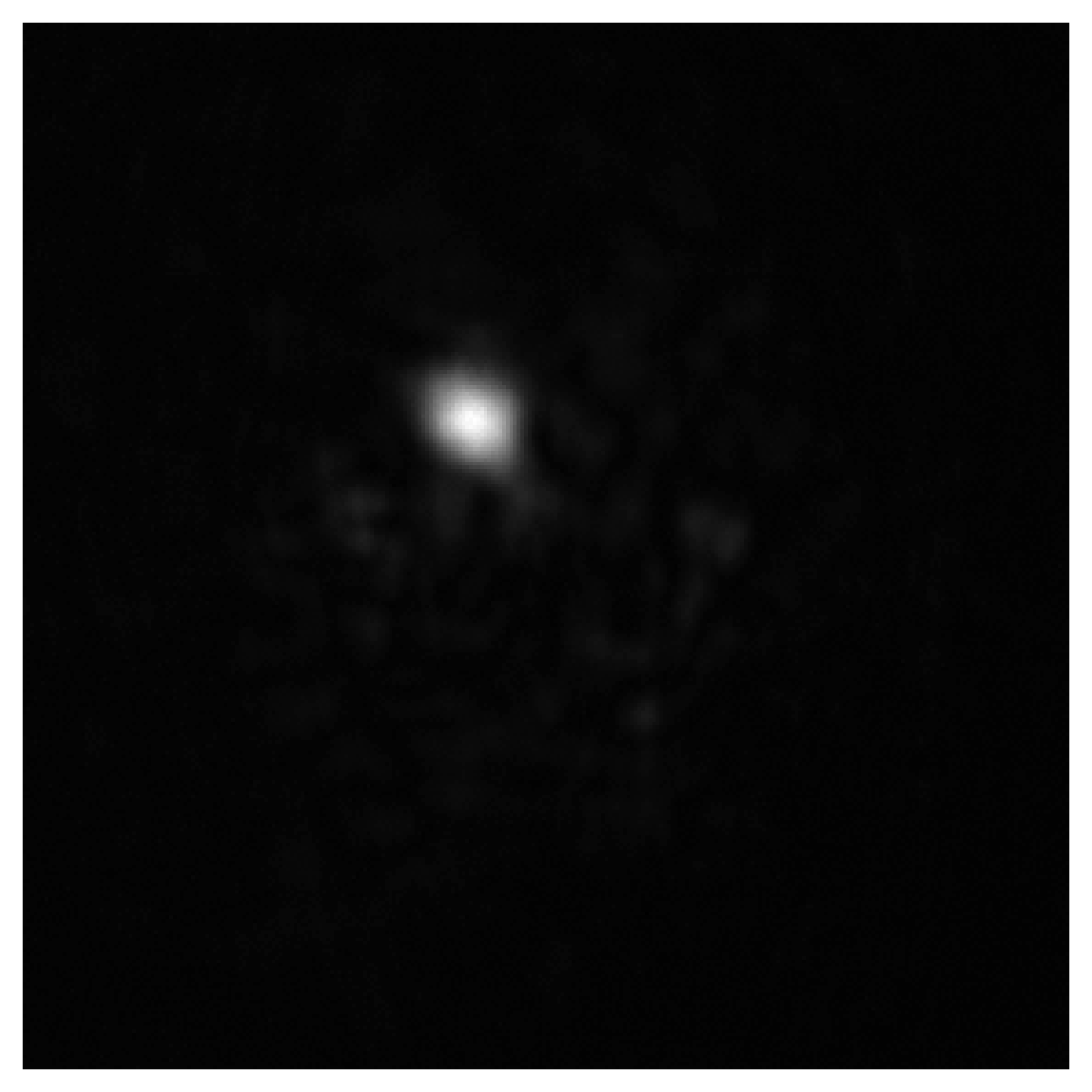}
  \end{subfigure}\hfill
  \begin{subfigure}[t]{0.04\textwidth}
    \textbf{(c)}
  \end{subfigure}
  \begin{subfigure}[t]{0.27\textwidth}
    \includegraphics[width=\linewidth, valign=t]{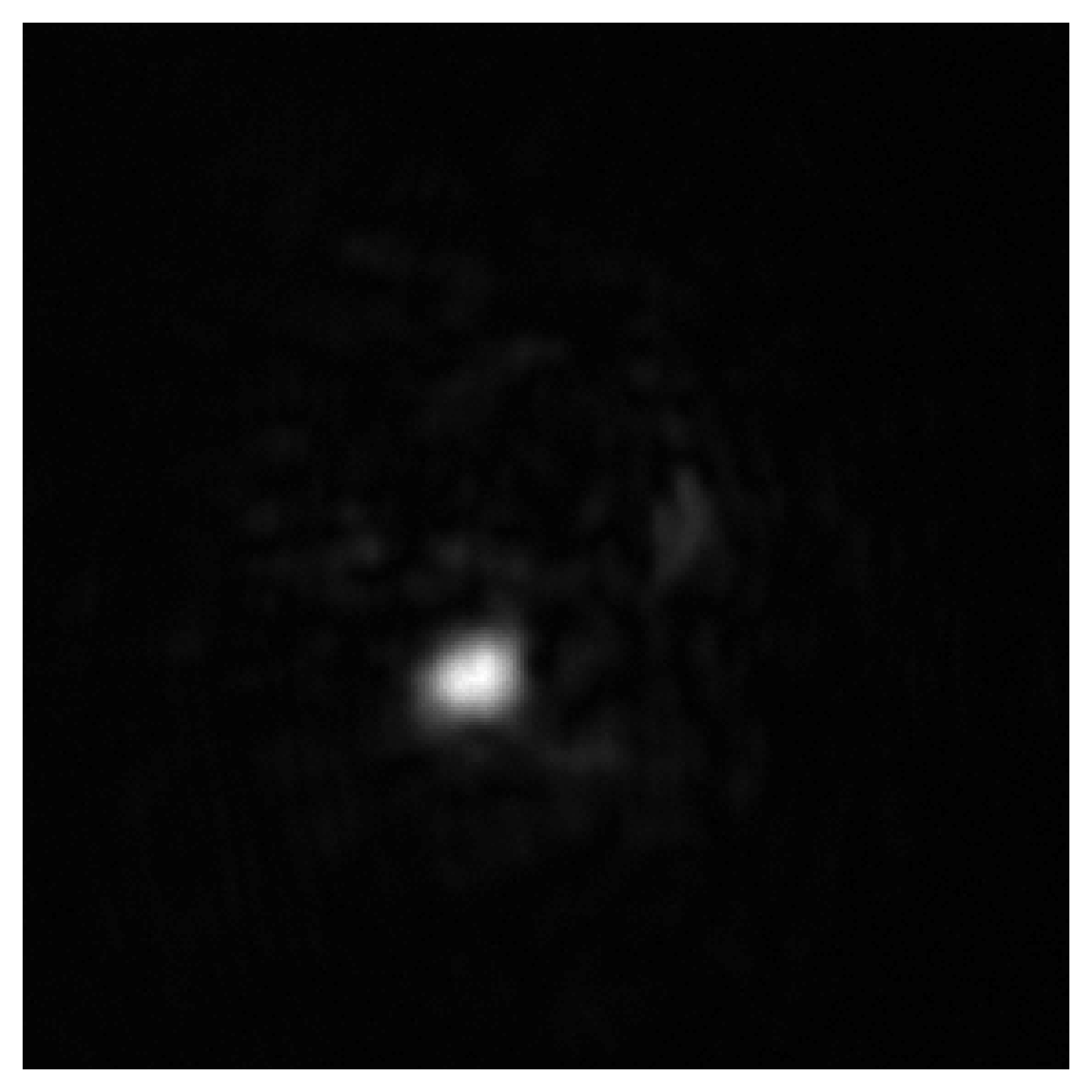}
  \end{subfigure}
  \caption[CCD images of sorted HG$_{00}$, HG$_{10}$, HG$_{01}$ output spots]{Output spots corresponding to sorted HG modes after passing through the MPLC system: (a) HG$_{00}$, (b) HG$_{10}$, and (c) HG$_{01}$. Each mode is directed to a distinct spatial location on the CCD, verifying successful mode sorting.}
  \label{mode_sort}
\end{figure}

\begin{figure}[tb]
  \centering
  \begin{subfigure}[t]{0.04\textwidth}
    \textbf{(a)}
  \end{subfigure}
  \begin{subfigure}[t]{0.45\textwidth}
    \includegraphics[width=\linewidth, valign=t]{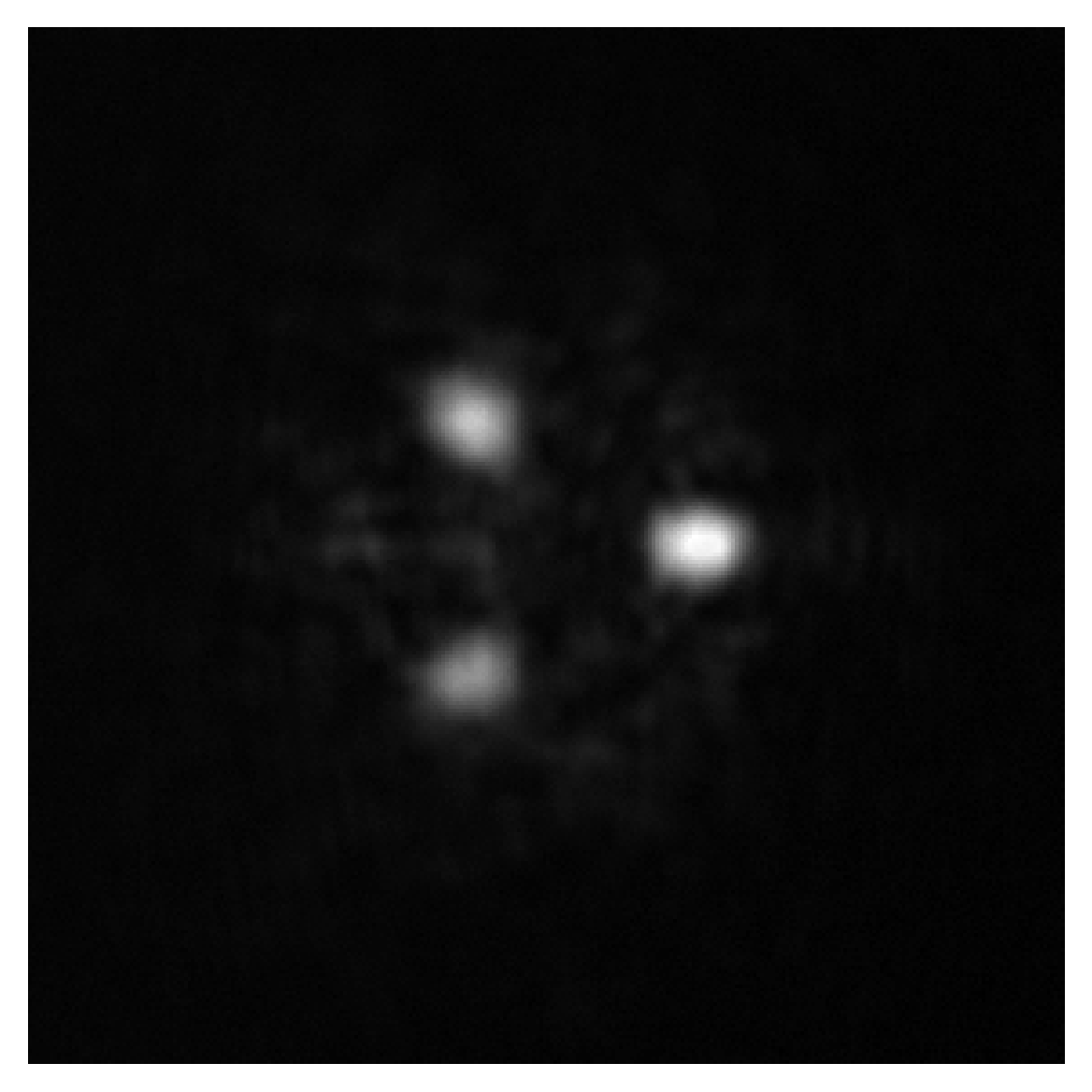}
  \end{subfigure}\hfill
  \begin{subfigure}[t]{0.04\textwidth}
    \textbf{(b)}
  \end{subfigure}
  \begin{subfigure}[t]{0.45\textwidth}
    \includegraphics[width=\linewidth, valign=t]{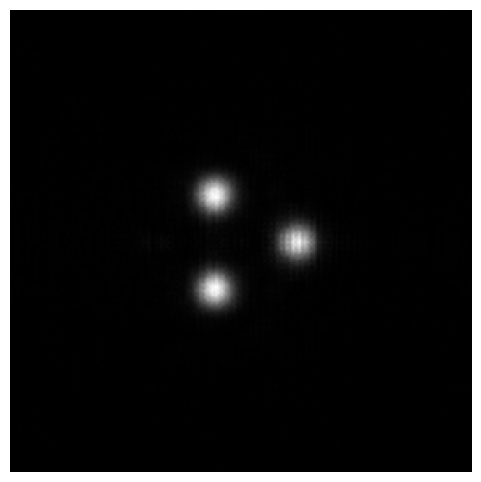}
  \end{subfigure}
  \caption[Experimental vs simulated HG mode sorting]{Overlap of output spots corresponding to the sorted HG modes: (a) experimental result, (b) simulated profile. The comparison highlights the spatial match between measured and expected mode positions.}
  \label{exp_sim}
\end{figure}

Each CGH-generated mode was sent through the MPLC system, and the resulting output field was captured by a CCD camera. As shown in Fig. \ref{mode_sort}, the three modes were clearly directed to distinct spatial positions as expected. A visual comparison of the experimental and simulated output spot overlaps is presented in Fig. \ref{exp_sim}.

\subsubsection{Quantification of Crosstalk Between Spatial Modes}

To assess the degree of mode sorting and the spatial quality of the output modes, we recorded the intensity profiles of each sorted beam and fitted them to a two-dimensional Gaussian model. From the fit parameters we extracted both the beam waist and the goodness‐of‐fit. In simulations, the output profiles of all three modes exhibited a 98\% Gaussian overlap, indicating good agreement between the simulated intensity distribution and a fitted two-dimensional Gaussian function. This overlap quantifies the goodness of fit and serves as a proxy for how well the beam resembles a spatially localized Gaussian spot, as expected at the output of the mode sorter. Experimentally, we measured an average overlap of 88\%, reflecting some degradation due to alignment imperfections, imperfect calibration of SLM phase profiles, thermal drift or vibrations in the setup and SLM quantization error. 

Nonetheless, the close correspondence between the simulated and measured beam waist is noteworthy: while the simulations predicted an output waist of 80~$\mu$m for all three modes, the experimentally measured values were 82~$\mu$m, 91~$\mu$m, and 92~$\mu$m for the HG$_{00}$, HG$_{10}$, and HG$_{01}$ modes, respectively. The high Gaussian overlap further confirms that our four-plane MPLC design successfully separates the HG$_{00}$, HG$_{10}$, and HG$_{01}$ modes into distinct, spatially localized output spots.

\begin{figure}[tb]
  \centering
  \begin{subfigure}[t]{0.04\textwidth}
    \textbf{(a)}
  \end{subfigure}
  \begin{subfigure}[t]{0.27\textwidth}
    \includegraphics[width=\linewidth, valign=t]{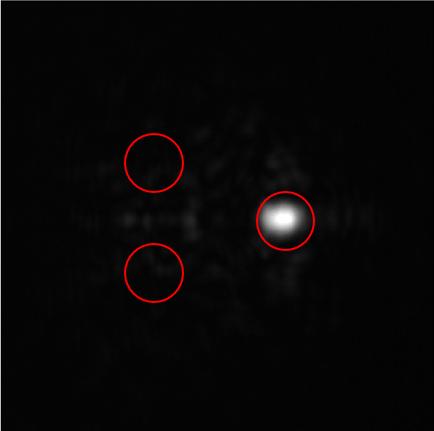}
  \end{subfigure}\hfill
  \begin{subfigure}[t]{0.04\textwidth}
    \textbf{(b)}
  \end{subfigure}
  \begin{subfigure}[t]{0.27\textwidth}
    \includegraphics[width=\linewidth, valign=t]{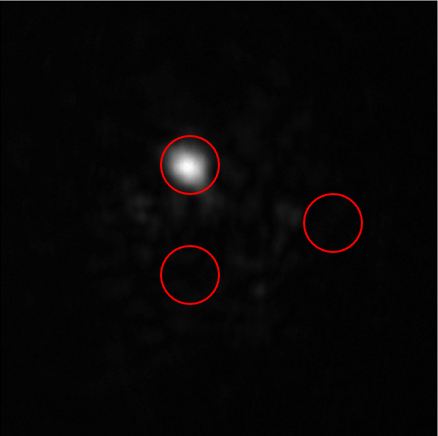}
  \end{subfigure}\hfill
  \begin{subfigure}[t]{0.04\textwidth}
    \textbf{(c)}
  \end{subfigure}
  \begin{subfigure}[t]{0.27\textwidth}
    \includegraphics[width=\linewidth, valign=t]{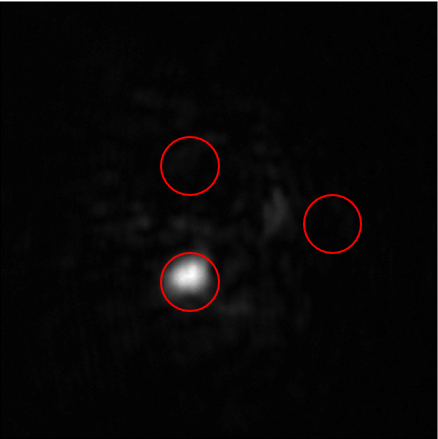}
  \end{subfigure}
  \caption[Gaussian fitting of output spots with red circle overlay]{Output modes corresponding to (a) HG$_{00}$, (b) HG$_{10}$, and (c) HG$_{01}$ with red circles indicating the position and size obtained from a 2D Gaussian fit. These fits were used for fidelity and crosstalk analysis.}
  \label{red_circle}
\end{figure}

In addition to spatial mode quality, we quantified the crosstalk between the output modes to assess the sorting fidelity. After identifying the beam centers via the Gaussian fitting routine, we computed the integrated intensity $I_i$ for each spot by summing the pixel values within a radius equal to the $1/e^2$ waist determined from the fit. 
\begin{align}
I_{\mathrm{total}} &= \sum_{i=1}^{3} I_i \,.
\end{align}
The experiment fidelity of the mode sorting process is defined as
\begin{align}
F &= \frac{I_{\mathrm{correct}}}{I_{\mathrm{total}}} \,,
\end{align}
where \(I_{\mathrm{correct}}\) is the summed intensity at the expected (correct) positions. Crosstalk is quantified by
\begin{align}
C &= \frac{I_{\mathrm{incorrect}}}{I_{\mathrm{total}}} \,,
\end{align}
where \(I_{\mathrm{incorrect}}\) denotes the summed intensity at the non‐target positions. In Fig.~\ref{red_circle}, the output modes corresponding to different HG inputs are displayed, with red circles indicating the position and size of the two-dimensional Gaussian fits. These circles were used to define the integration regions for the crosstalk analysis.

This allowed us to construct a fidelity matrix indicating the fraction of input power routed to the correct output mode, and thereby quantify both sorting efficiency and residual mode overlap.

\begin{figure}[H]
  \centering
  \begin{subfigure}[t]{0.04\textwidth}
    \textbf{(a)}
  \end{subfigure}
  \begin{subfigure}[t]{0.45\textwidth}
    \includegraphics[width=\linewidth, valign=t]{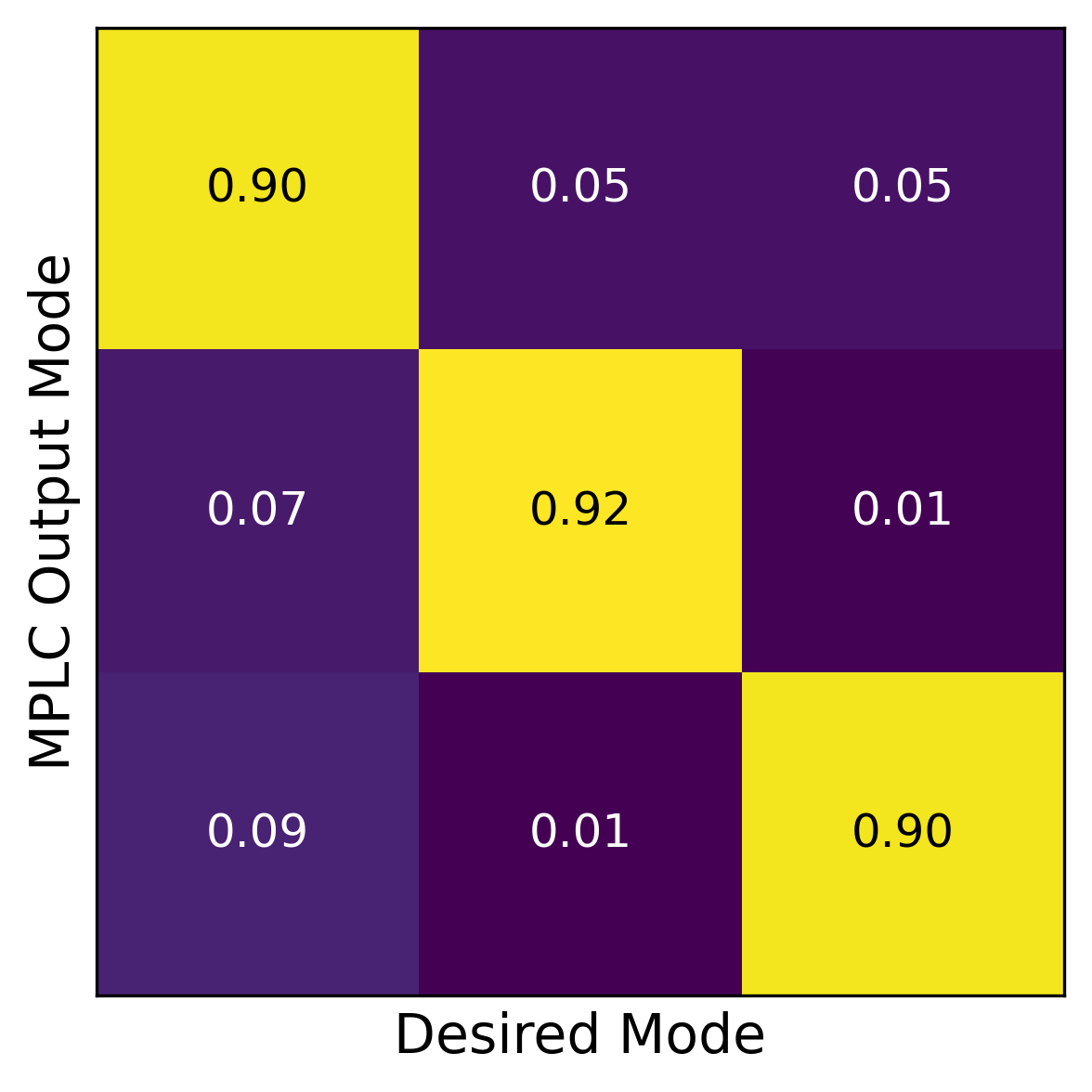}
    \label{fig:3mod_fid_exp}
  \end{subfigure}\hfill
  \begin{subfigure}[t]{0.04\textwidth}
    \textbf{(b)}
  \end{subfigure}
  \begin{subfigure}[t]{0.45\textwidth}
    \includegraphics[width=\linewidth, valign=t]{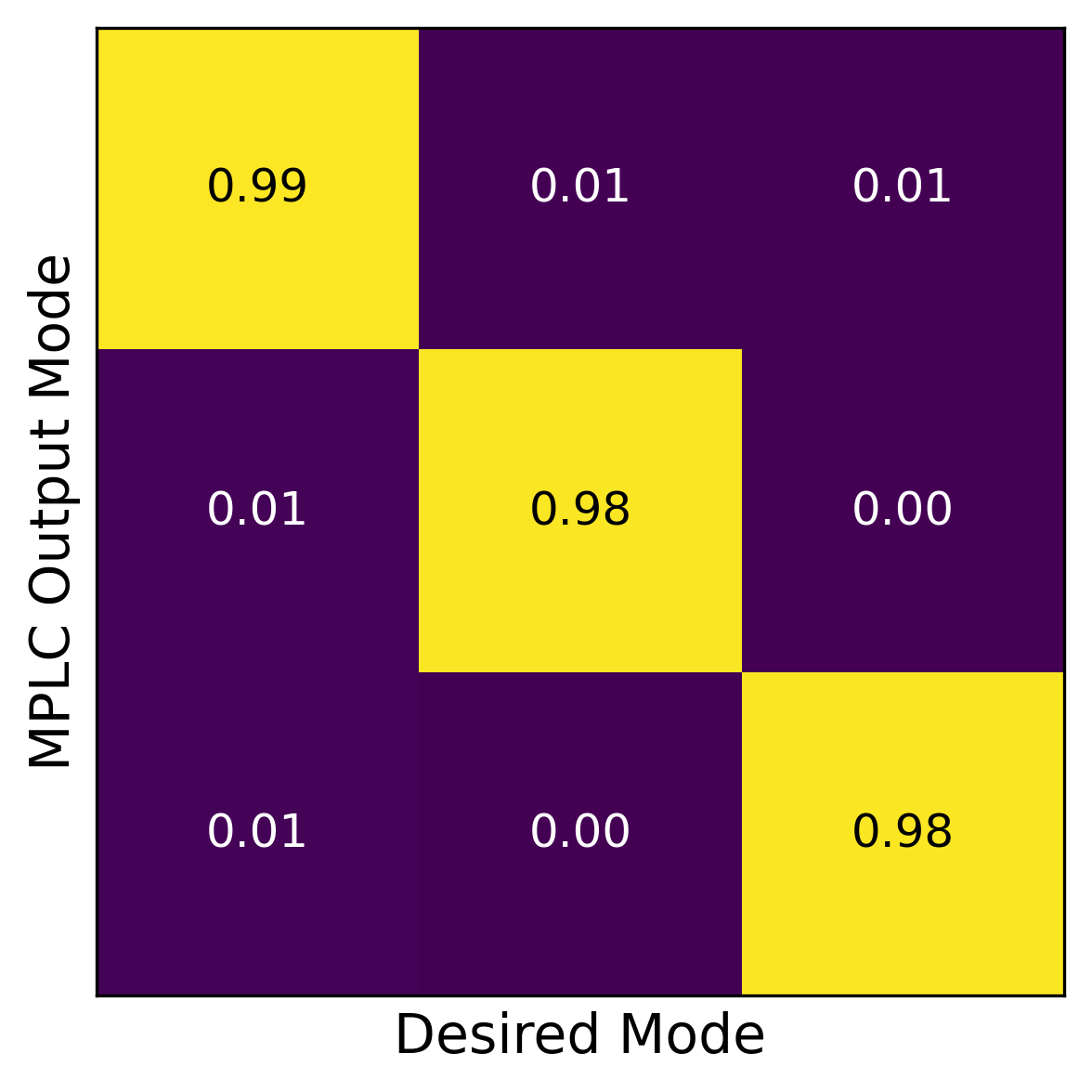}
    \label{fig:3mod_fid_sim}
  \end{subfigure}
  \caption[Fidelity matrix comparison: experimental vs simulated]{Comparison of fidelity matrices for sorting HG$_{00}$, HG$_{10}$, and HG$_{01}$ modes. (a) shows the matrix derived from experimental CCD measurements, while (b) presents the idealized simulated result. Diagonal dominance indicates successful mode separation.}
  \label{fig:3mod_fid}
\end{figure}

To complete our performance evaluation, we present a side-by-side comparison of the experimental and simulated fidelity matrices for the three-mode spatial sorter, as shown in Fig.~\ref{fig:3mod_fid}. The left figure displays the matrix obtained from CCD camera measurements, while the right panel shows the corresponding simulation result computed for four-mask MPLC configuration.

The experimental fidelity matrix exhibits a dominant diagonal structure, confirming that the MPLC system successfully directs each input mode to its intended output mode and effectively performs spatial-mode sorting. Both the simulated and experimental fidelity matrices were computed by integrating the output intensity within predefined spatial regions, corresponding to the expected locations of the sorted modes. Since the same method was used in both cases, the resulting matrices are directly comparable. The experimental matrix showed an average fidelity of 90.6\%,  matched by the simulation result of 98.3\%. The close quantitative agreement between them demonstrates that the MPLC design reliably implements the desired mode sorting and performs robustly under experimental conditions.

Having presented the implementation and performance of the MPLC system for spatial mode generation and sorting, we now turn to its practical limitations and underlying constraints. The source of the discrepancy between the simulation and experiment can be partially attributed to alignment sensitivity, as predicted in Chapter 4. While our experimental alignment was precise, a residual misalignment of even a fraction of the beam waist, combined with the SLM's finite pixel resolution, can possibly account for the observed reduction in fidelity. Moreover, the current design is limited to three spatial modes (HG\textsubscript{00}, HG\textsubscript{10}, HG\textsubscript{01}). Scaling this approach to support more HG mode would require additional phase planes. This would also increase the alignment complexity and computational overhead in WFM.

\chapter{Conclusion and Outlook}

\section{Conclusion}

This thesis has demonstrated, from theoretical design through to experimental validation, that MPLC is a viable and high-fidelity method for implementing SPADE, a quantum-inspired technique designed to surpass the classical diffraction limit in imaging. The key objective was to implement a practical mode sorter that projects incoming light onto an orthonormal mode basis, to be used in future experiments to extract more information than direct imaging and overcome Rayleigh’s criterion.

We perform theoretical and numerical investigations in the design of a six-mode MPLC system with seven phase masks. This configuration was engineered to transform an input array of six laterally displaced Gaussian beams into a co-located set of HG modes. Simulations predicted high-fidelity mode conversion, with diagonal elements of the resulting fidelity matrix ranging from 0.94 to 0.96. These values indicate that at least 94\,\% of the optical power in each input mode was correctly mapped to its intended HG mode. Additionally, numerical robustness analyses revealed a strong dependence on alignment precision: a lateral misalignment of one beam waist ($100~\mu$m) led to a drop in fidelity to approximately 69\,\%, emphasizing the critical role of accurate optical alignment in maintaining performance. The system showed moderate tolerance to variations in the input beam size, with fidelity remaining above 94\% within a 10 $\mu$m window of the design waist. However, it was highly sensitive to the axial propagation distance, where even millimeter-level deviations caused significant fidelity degradation.

Building upon these numerical results, a physical MPLC system was constructed and experimentally validated. To ensure quantifiable benchmarking, a three-mode sorting experiment was performed using the HG$_{00}$, HG$_{10}$, and HG$_{01}$ modes. The experimental setup successfully sorted these modes into distinct and spatially separated output spots, achieving a measured average sorting fidelity of 90.6\,\%. This value  matched the simulated fidelity of 98.3\,\% for the same configuration, thereby validating both the accuracy of the computational design and the reliability of the physical implementation. Moreover,  crosstalk was measured to be as low as 1\,\% for specific mode pairs, offering  evidence of MPLC’s efficacy in high-fidelity spatial-mode demultiplexing.

\section{Outlook}

While the results are promising, there remain several parts to improve the system’s performance and to broaden its impact. In the following, we outline practical improvements for the current setup as well as broader conceptual and application-based extensions:

\begin{itemize}
    \item \textbf{Scalability to More and Higher-Order Modes:} The present demonstrator validates MPLC for three fundamental HG modes, yet future SPADE experiments will likely demand tens or even hundreds of modes. Achieving such scale depends on the number of phase planes: in practice, roughly one carefully designed plane per mode is enough to keep fidelities above 90\%, but this quickly increases alignment effort and computation time. The WFM algorithm used throughout this thesis has already proven highly capable indeed, it underpins recent demonstrations with up to 1035 spatial modes so it remains a solid baseline\cite{Fontaine2021}. At the same time, complementary optimization strategies such as gradient‑based adjoint methods have begun to show advantages in convergence speed and design‑space exploration for high‑dimensional MPLCs \cite{Kupianskyi2023}. Combining these techniques with the existing WFM for instance, by using a few adjoint‑optimized planes as an initial seed might offer a pragmatic route to expanding the sorter’s dimensionality without abandoning the robust design philosophy developed here.

    \item \textbf{Dynamic reconfigurable MPLC:}  
        While our prototype is static, real‑time mask‑update architectures based on MEMS mirrors or ferroelectric SLMs have reached kilohertz refresh rates across multi‑plane systems \cite{Rocha2025}.  
        Incorporating similar hardware could potentially extend the SPADE concept to scenarios that demand adaptive mode sorting, such as turbulence compensation or real‑time system calibration.

    \item \textbf{Photon‑number‑resolved detection:}  
        The current validation used CCD intensity measurements.  
        Photon‑number‑resolving detectors (PNR) are increasingly compatible with free‑space and fiber‑coupled optics \cite{Marsili2009}. Combining such detectors with the MPLC output could allow direct tests of quantum‑Fisher information predictions that motivated this work.

\end{itemize}

% Appendix - - - - - - - - - - - - - - - - - - - - - - - - - - - - -
\appendix
\chapter{Derivations }

\section{Quantization of Electromagnetic Waves}
\label{appendix_qem}
To simplify the classical Hamiltonian, we use Parseval’s theorem. First we define the Fourier transform of \(f(\vec r)\) by
\begin{align}
\tilde f(\vec k)
  \;=\;\mathcal{F}\bigl[f(\vec r)\bigr]
  \;=\;\frac{1}{\sqrt{V}}\int_V f(\vec r)\,e^{-i\vec k\cdot\vec r}\,d\vec{r}.
\end{align}
Then Parseval’s theorem reads
\begin{align}
\int_V \lvert f(\vec r)\rvert^2\,d\vec{r}
  &= \sum_{\vec k} \lvert \tilde f(\vec k)\rvert^2.
\end{align}
Then ignoring prefactors, we get:
\begin{align}  
\int \left| \frac{\partial \vec{A}(\vec{r},t)}{\partial t} \right|^2 d\vec{r} &= \sum_{\vec{k}} \left| \mathcal{F} \left( \frac{\partial \vec{A}(\vec{r},t)}{\partial t} \right) \right|^2, \quad 
\int |\nabla \times \vec{A}(\vec{r},t)|^2 d\vec{r} = \sum_{\vec{k}} \left| \mathcal{F} (\nabla \times \vec{A}(\vec{r},t)) \right|^2.
\end{align}
Computing the Fourier transform gives:
\begin{align}
\mathcal{F}(\nabla \times \vec{A}(\vec{r},t))(\vec{k}) 
&= (i\vec{k} \times \vec{\epsilon}) 
\left( A_{\vec{k}} e^{-i\omega_k t}
- A^*_{\vec{k}} e^{i\omega_k t}\right). \\
\mathcal{F} \left( \frac{\partial \vec{A}(\vec{r},t)}{\partial t} \right)(\vec{k}) 
&=  (i\omega) 
\left( A_{\vec{k}} e^{-i\omega_k t}  
- A^*_{\vec{k}} e^{i\omega_k t}\right) \vec{\epsilon}.
\end{align}
so that
\begin{align}
\left| \mathcal{F}(\nabla \times \vec{A}(\vec{r},t)) \right|^2 
&=
(i\vec{k} \times \vec{\epsilon}) \cdot (-i\vec{k} \times \vec{\epsilon}) 
\left( A_{\vec{k}} e^{-i\omega_{\vec{k}} t}  
- A^*_{\vec{k}} e^{i\omega_{\vec{k}} t} \right)
\\& \times\left( A^*_{\vec{k}} e^{i\omega_{\vec{k}} t}  
- A_{\vec{k}} e^{-i\omega_{\vec{k}} t}  \right)\\
&= 2 |\vec{k}|^2 |A_{\vec{k}}|^2.
\end{align}
and similarly:
\begin{align}
\left| \mathcal{F} \left( \frac{\partial \vec{A}(\vec{r},t)}{\partial t} \right) \right|^2 
&= 
(i\omega)(-i\omega) 
\left( A_{\vec{k}} e^{-i\omega_{\vec{k}} t} 
- A^*_{\vec{k}} e^{i\omega_{\vec{k}} t} \right) \\
&\quad \times \left( A_{\vec{k}} e^{-i\omega_{\vec{k}} t} 
- A^*_{\vec{k}} e^{i\omega_{\vec{k}} t} \right) 
\vec{\epsilon} \\
&= 2 \omega_{\vec{k}^2} |A_{\vec{k}}|^2.
\end{align}
Hence we get $H_{\text{class}} = 2 \varepsilon_0 \omega_{\vec{k}^2} |A_{\vec{k}}|^2 $

\section{Quantum Fisher Information}

Eq. \ref{QCRB_EQ} can be proved as follows \cite{PezzeSmerzi2014}:
\begin{align}
    \mathcal{F}[\hat{\rho}(\theta), E_y(\vec{x})] = \int d\vec{x} \frac{\operatorname{Tr}[E_y(\vec{x})\, \partial_{\theta} \hat{\rho}(\theta)]^2}{\operatorname{Tr}[E_y(\vec{x})\, \hat{\rho}(\theta)]}.
\end{align}
Using the definition of \( \hat{L}_\theta \), Eq.~\ref{eq:sld}, and the identity:
\begin{align}
\operatorname{Tr}[(\hat{\rho}(\theta)\hat{L}_\theta\, \, E_y(\vec{x}))^\dagger] = \operatorname{Tr}[\hat{\rho}(\theta)\, \hat{L}_\theta\, E_y(\vec{x})]^*,
\end{align}
We get:
\begin{align}
    \operatorname{Tr}[E_y(\vec{x})\, \partial_{\theta} \hat{\rho}(\theta)] = \Re\left(\operatorname{Tr}[E_y(\vec{x})\hat{\rho}(\theta)\, \hat{L}_\theta\, ]\right),
    \label{eq:real}
\end{align}
where \( \Re(z) \) and \( \Im(z) \) denote the real and imaginary parts of the complex number \( z \), respectively. The bound on the Fisher information is obtained from the following chain of inequalities:
\begin{align}
    \Re\left(\operatorname{Tr}[E_y(\vec{x})\hat{\rho}(\theta)\, \hat{L}_\theta\, ]\right)^2 
    &\leq \left|\operatorname{Tr}[E_y(\vec{x})\hat{\rho}(\theta)\, \hat{L}_\theta\, ]\right|^2 \nonumber \\
    &\leq \operatorname{Tr}[E_y(\vec{x})\hat{\rho}(\theta)\,]\, \operatorname{Tr}[E_y(\vec{x})\, \hat{\rho}(\theta)\, \hat{L}_\theta^2],
    \label{eq:inequality}
\end{align}
where the second inequality follows from the Cauchy-Schwarz inequality. Combining Eq.\ref{eq:real} and \ref{eq:inequality}, we obtain:
\begin{align}
    \frac{\operatorname{Tr}[E_y(\vec{x})\, \partial_{\theta} \hat{\rho}(\theta)]^2}{\operatorname{Tr}[E_y(\vec{x})\, \hat{\rho}(\theta)]}
    \leq \operatorname{Tr}[E_y(\vec{x})\, \hat{\rho}(\theta)\, \hat{L}_\theta^2].
\end{align}
Summing over \( \vec{x} \), and using the fact that \( \int d\vec{x}    E_Y(\vec{x}) = \mathbb{1} \), we find:
\begin{align}
    \mathcal{F}[\hat{\rho}(\theta), E_y(\vec{x})] 
    &\leq \int d\vec{x}  \operatorname{Tr}[E_y(\vec{x})\, \hat{\rho}(\theta)\, \hat{L}_\theta^2] 
    = \operatorname{Tr}[\hat{\rho}(\theta)\, \hat{L}_\theta^2] = \mathcal{F}_Q[\hat{\rho}(\theta)].
\end{align}
% Bibliography - - - - - - - - - - - - - - - - - - - - - - 
\bibliographystyle{IEEEtran}
\bibliography{refs}
% - - - - - - - - - - - - - - - - - - - - - - - - - - - - -

\end{document}